\documentclass[a4paper,11pt]{article}
\pdfoutput=1 

\usepackage{jheppub} 
\pdfoutput=1
\usepackage{amsmath}
\usepackage{xcolor}
\usepackage{subcaption}

\usepackage{graphicx}
\usepackage{amssymb,xfrac,framed,verbatim,amsthm}
\usepackage{simplewick}
\usepackage{graphicx}
\usepackage{here}
\usepackage{slashed,mathtools}
\usepackage{graphicx,enumerate,bm}
\usepackage{tikz}
\usetikzlibrary{angles, quotes}
\usetikzlibrary{decorations.markings}
\usetikzlibrary{decorations.pathmorphing}

\usepackage{jheppub} 

\usepackage[T1]{fontenc} 

\preprint{TIFR/TH/21-2} \title{\bf Bounds on Regge growth of flat space scattering from bounds on chaos}

\author[b,1]{Deeksha Chandorkar,\note{deeksha.chandorkar@gmail.com}}
\author[a,2]{Subham Dutta Chowdhury,\note{subham@theory.tifr.res.in}}
\author[a,3]{Suman Kundu,\note{suman@theory.tifr.res.in}}
\author[a,4]{ Shiraz Minwalla \note{minwalla@theory.tifr.res.in}}
\affiliation[a]{Department of Theoretical Physics, Tata Institute of Fundamental Research, Homi Bhabha Rd, Mumbai 400005, India}
\affiliation[b]{Department of Physics, Jai Hind College (Autonomous), A Rd, Churchgate, Mumbai 400020, India}

\abstract{
	We study four-point functions of scalars, conserved currents, and stress tensors in a conformal field theory,  generated by a local contact term in the bulk dual description,   in two different causal configurations. The first of these is the standard Regge configuration in which the chaos bound applies. The second is the `causally scattering configuration' in which the correlator develops a bulk point singularity. We find an expression for the coefficient of the bulk point singularity in terms of the bulk S matrix of the bulk dual metric, gauge fields and scalars, and use it to determine the Regge scaling of the correlator on the causally scattering sheet in terms of the Regge growth of this S matrix. We then demonstrate that the Regge scaling on this sheet is governed by the same power as  in the standard  Regge configuration, and so is constrained by the chaos bound, which turns out to be violated unless the bulk flat space S matrix grows no faster than $s^2$ in the Regge limit. It follows that in the context of the AdS/CFT correspondence, the chaos bound applied to the boundary field theory implies that the S matrices of the dual bulk scalars, gauge fields, and gravitons obey the Classical Regge Growth (CRG) conjecture.
}

\begin{document} 
\maketitle
\flushbottom

\section{Introduction}


It has recently been conjectured  that the tree level S matrix of any `consistent'
\footnote{That is a causal classical theory whose energy is bounded from below (and perhaps is also required to obey other constraints of a similar general nature). See \cite{Chowdhury:2019kaq, Chakraborty:2020rxf} for a discussion and \cite{Chowdhury:2020ddc} for a generalization to gluons.} classical theory  never grows faster with $s$ than $s^2$ in the Regge limit. Arguments supporting this so-called Classical Regge Growth (CRG) conjecture were advanced in the CEMZ paper \cite{Camanho:2014apa}, the chaos bound paper \cite{Maldacena:2015waa} and also in  discussions of the inversion formula \cite{Caron-Huot:2017vep} in large $N$ theories. (see e.g. \cite{Turiaci:2018dht}).

The authors of \cite{Chowdhury:2019kaq}
demonstrated that the classical Einstein S matrix is the only CRG consistent local tree-level \footnote{That is an S matrix that is the sum of  a polynomial in momenta plus a finite number of physical particle exchange poles.} gravitational S matrix in $D\leq 6$. This striking result suggests that if the CRG conjecture indeed holds then the set of consistent classical gravitational S matrices is highly constrained by simple general considerations \footnote{See in particular the introduction of  \cite{Chowdhury:2019kaq}, 
	and in particular Conjectures 1-3 presented in that introduction, for a detailed discussion.}, and motivates a more detailed investigation of the CRG conjecture. 

We are aware of three arguments in support of the CRG conjecture. The first (and weakest) of these arguments is the simple observation that the CRG conjecture is indeed obeyed in every classical theory that is  known for certain to be consistent.  For instance, all two derivative classical theories always obey the CRG conjecture \cite{Chowdhury:2019kaq}. Moreover, the classical Einstein S matrix saturates CRG growth \footnote{Demonstrating the impossibility of a tighter than $s^2$ bound on Regge growth.} and the Type II and Heterotic analogues of the  Virasoro Shapiro amplitude of string theory temper this growth to a power that is strictly less than two at every physical finite value of $\alpha't$ (see \cite{Chowdhury:2019kaq} for some more details). 

A second reason to believe the CRG conjecture is contained in the analysis of 
\cite{Camanho:2014apa}. The authors of \cite{Camanho:2014apa} used a `signal model' (see Appendix D of that paper) to argue that the the function 
\begin{equation} \label{argen}
{\cal T}(\delta , s)= 1+i \frac{S(\delta, s)}{s} 
\end{equation} 
should obey the inequality 
\begin{equation}\label{inequ}
|{\cal T}(\delta,s)| \leq 1
\end{equation}
everywhere in the upper half complex $s$ plane for every nonzero physical value of $\delta$. Here $S(\delta,s)$ is the transition amplitude - atleast roughly speaking-  for a particle passing through a shock at impact parameter $\delta$. 
Roughly, $S(\delta,s)$ can be thought of as the invariant amplitude for the scattering in impact parameter space 
(see \cite{Kologlu:2019bco} for related discussions).
In the classical limit the tree amplitude is parametrically small \footnote{This is why loops are sub dominant compared to trees.} and \eqref{inequ} reduces to the condition  
\begin{equation}\label{inequuu}
{\rm Im} \left( \frac{S(\delta, s)}{s} \right) \geq 0
\end{equation}  
If the large $|s|$ fixed $\delta$ expression for the classical S matrix is  
$$S(\delta, s)= B(\delta)s^{A(\delta)}$$
and $s=|s|e^{i\phi}$  then  \eqref{inequuu} implies
\begin{equation}\label{Tt}
T(\delta) \sin \left( (A(\delta)-1) \phi \right)  \geq 0, ~~~~~0 \leq \phi \leq \pi
\end{equation} 
When $A(\delta)>2$ the LHS of \eqref{Tt} switches sign as $\phi$ varies over the range $[0,\pi]$ so \eqref{Tt} cannot be obeyed for all values of $\phi$ in this range. \footnote{When $|A(\delta)|\leq 2$, the $\sin$ function that appears in \eqref{Tt} is always positive, and  \eqref{Tt} is obeyed when $B(\delta)>0$.}. It follows that the condition \eqref{inequ} can only be satisfied if $A(\delta)\leq 2$, i.e if the CRG conjecture is obeyed. As the passage from the impact space S matrix $S(\delta,s)$ to the usual momentum space S matrix $S(t, s)$ is, roughly speaking a 
Fourier Transform in $t$ at fixed $s$, this argument suggests that $S(t,s)$ obeys the CRG conjecture.  The weakness of this beautiful argument lies in the fact it contains elements that are intuitive rather than completely precise. For instance the transition from `impact parameter space' to `momentum space' may have subtleties related to $\delta$ function terms in impact parameter space \footnote{We thank S. Caron Huot for very useful discussions on this point.}

The  third (and possibly the strongest) reason to believe in the correctness of the CRG conjecture is its connection to the chaos bound in holographic theories. A relationship between the chaos bound and the CRG conjecture was already suggested in the 
original paper \cite{Maldacena:2015waa} 
(see  Appendix \ref{cb} for a discussion and of the relevant comments in that paper) and has been somewhat elaborated upon since 
(see e.g. \cite{Turiaci:2018dht}). The key strength of the chaos-bound based argument for the CRG conjecture is that the starting point for this argument is a theorem, at least by physicists' standards. One weakness of this argument lies in the fact that the connection between the chaos bound and the CRG conjecture has never (to our knowledge) been carefully argued through 
\footnote{See however the papers \cite{Costa:2017twz, Afkhami-Jeddi:2016ntf,  Afkhami-Jeddi:2017rmx, Afkhami-Jeddi:2018apj, Afkhami-Jeddi:2018own, Kologlu:2019bco, Kulaxizi:2018dxo, Kulaxizi:2017ixa, Meltzer:2017rtf} which put use the chaos bound (both the sign constraint and the growth constraint part of this bound) or closely related causality constraints \cite{Hofman:2008ar, Hartman:2015lfa, Hartman:2016dxc, Hofman:2016awc, Komargodski:2016gci, Hartman:2016lgu, Chowdhury:2017vel, Cordova:2017zej, Cordova:2017dhq, Meltzer:2018tnm, Cordova:2017dhq, Chowdhury:2018uyv, Kologlu:2019mfz, Belin:2019mnx} to obtain many interesting results that are in the same broad universality class as this paper. In particular the analysis of \cite{Afkhami-Jeddi:2018apj, Perlmutter:2016pkf} use such arguments to demonstrate the unphysicality of dual bulk theories with a finite number of higher spin fields. Analogous causality bounds for CFTs (not necessarily large $N$) have been studied in \cite{Kundu:2020gkz, Hartman:2016lgu, Afkhami-Jeddi:2018own}.}
(see Appendix \ref{cb} for some discussion). \footnote{Another weakness is the fact that it uses an elaborate theoretical framework - namely that of holography and consistency of the quantum structure of the boundary field theory - to establish a fact about scattering in the simpler theoretical structure of classical field theories.} The goal of this  paper is to fill in this gap for the case that bulk particles are either scalars, gauge fields, or the metric (i.e. all particles relevant  to the analysis of \cite{Chowdhury:2019kaq}). 

In particular, in this paper we demonstrate that the four-point function generated holographically by a local bulk contact term for scalars, gauge fields, or the metric necessarily violates the chaos bound whenever the flat space S matrix generated by the same contact term violates the CRG conjecture. In other words, the classical holographic dual of  a consistent unitary boundary field theory necessarily obeys the CRG conjecture. 

Our argument for a sharp connection between the chaos bound and the CRG conjecture consists largely of stitching together well known results from the now classic  papers \cite{Heemskerk:2009pn} and \cite{Gary:2009ae, Penedones:2010ue, Okuda:2010ym, Maldacena:2015iua} (see also \cite{ Paulos:2016fap} \footnote{See \cite{Komatsu:2020sag, Hijano:2019qmi} for recent progress regarding external massive scalar states.}).
In section \ref{kinem} we study time-ordered four-point functions in a holographic conformal field theory of four operators of arbitrary spin inserted on the two-parameter family of points \eqref{fpscatnnnn}. \footnote{The same kinematic configurations were studied in section 6.1 of \cite{Maldacena:2015iua}.} Our four-point function is normalized by dividing with a product of two two-point functions (see \eqref{normfp}).
As the insertion parameters $\theta$ and $\tau$ run over the range of study \eqref{rangepar}, the conformal cross ratios $\sigma$ and $\rho$ (or equivalently $z$ and ${\bar z}$ -see around \eqref{sigrhodef} for definitions) range over three different sheets in the complex cross ratio space corresponding to three distinct causal configurations \eqref{causalrel} of the boundary points. The first of these is the principal sheet which we refer to as  the Causally Euclidean sheet through this paper. The second sheet -which we refer to as the Causally Regge sheet through this paper - is reached starting from the Causally Euclidean sheet by circling the branch point at ${\bar z}=1$ in a counter-clockwise manner. The third sheet - which we refer to as the Causally Scattering sheet through this paper- is obtained starting from the Causally Regge sheet by circling counter-clockwise around the branch point at $z=0$. 

Our argument proceeds by focusing attention on two special one parameter limits of our two-parameter set of insertion points \eqref{fpscatnnnn}.  The first of these is the much-studied `Regge limit' \eqref{newlim} in which $\sigma$ is taken to be small but $\rho$ is allowed to be  arbitrary. Here and throughout this paper, $\rho$ and $\sigma$ are conformal cross ratios, related by the more familiar cross ratios $z$ and ${\bar z}$ by the relations $z=\sigma e^{\rho}$ and ${\bar z}= \sigma e^{-\rho}$. In several studies of conformal field theory, the Regge limit is studied on the Causally Regge sheet \eqref{causalrel} of cross ratio space. The Regge limit studied in section \ref{sgzl} of this paper, however,  straddles across both the Causally Regge and the Causally scattering configuration
\footnote{One reason for this is that our `coordinate patch' \eqref{fpscatnnnn} is more flexible than the $\rho$ coordinate patch \eqref{labn} often employed in CFT studies. As we explain in Appendix \ref{stdconfig}, that part of the two-parameter insertion space \eqref{fpscatnnnn} which lies on the Causally Scattering sheet has no image into time-ordered correlators on the $\rho$ plane \eqref{labn}. }
In section \ref{sgzl} below we use use a small variant of the analysis of section  5.2 of 
\cite{Heemskerk:2009pn} to demonstrate that to leading order in this limit (i.e. to leading order in small $\sigma$ ) the $\sigma$ and $\rho$ dependence of our correlator takes the form 
\begin{equation}\label{parse}
\frac{g_{CS}(e^{2\rho})}{\sigma^{A'-1}} ~~~~~{\rm and}~~~~~~
\frac{g_{CR}(e^{2\rho})}{\sigma^{A'-1}}
\end{equation}
respectively in the Causally Scattering and Causally Regge regimes, where $A'$ is a fixed but as yet unknown number. $g_{CS}(e^{2\rho})$ and 
$g_{CR}(e^{2\rho})$ are as yet unknown functions of the cross ratio $\rho$. The crucial point here, however, is that they are not completely independent of each other. There exists a function ${\tilde H}(z)$ which is analytic apart away from the branch cut at $z=0$. $g_{CR}(e^{2\rho})$ equals ${\tilde H}(z)$ evaluated on $z =e^{2 \rho}$ on one sheet of this function, while $g_{CR}(e^{2\rho})$
equals ${\tilde H}(z)$ evaluated on $z =e^{2 \rho}$ on a second sheet. \footnote{On the configurations studied in this paper $e^{2 \rho}$ is real and lies in the interval  $(0,1)$.} This fact makes it impossible for $g(e^{2 \rho})$ to vanish identically in the Causally Regge branch if it is nontrivial on the Causally Scattering branch.
\footnote{The point $\tau=\theta$ in the configuration \eqref{fpscatnnnn} maps to $e^{2 \rho}=z=0$, the branch point of the correlator.
	As a consequence the functions $g_{CS}(\rho)$ and 
	$g_{CR}(e^{2\rho})$ are not directly smoothly connected but are more indirectly related, as explained in this paragraph.} 
It follows, in other words, that in the small $\sigma$ limit, correlators scale with the same power of $\sigma$ in the Causally Scattering and Causally Regge sheets of cross ratio space. 

In section \eqref{rgzl} we turn to the study the limit $\rho \to 0$ on the Causally Scattering sheet; this is the bulk point limit of  \cite{Gary:2009ae,Penedones:2010ue, Okuda:2010ym, Maldacena:2015iua}.
 In section \eqref{rgzl} we generalize the discussion  of section 6 of \cite{Maldacena:2015iua} and \cite{Penedones:2010ue, Okuda:2010ym} to demonstrate that the four point function is given,  to leading order in this limit, by the expression 
\begin{equation}\label{nntrevint}
\begin{split} 
G_{\rm sing}& \propto-\frac{1}{4 \sqrt{\sigma (1-\sigma)}}
\int_{\mathbb{H}_{D-2}} d^{D-2} X \int d\omega \omega^{\Delta -4}
e^{i\omega P.X } {\cal S}\\
\end{split} 
\end{equation}
(see \eqref{nntrespin} for the proportionality factor that makes this equation precise) 
where ${\cal S}$ is the classical flat space S matrix of 
four specified waves with momenta \eqref{forvec},  polarizations \eqref{polphograv},  $\omega$ is the energy of each of these waves, $\Delta$ is the sum of the scaling dimensions of the four operators and the $\mathbb{H}_{D-2}$ is the part of $AdS_{D+1}$ space that is orthogonal 
to the four boundary points $P_i$ at $\tau=0$ (see subsection \ref{ibl}).\footnote{In other words it consists of the points $X$ in the embedding space $\mathbb{R}^{D,2}$ which obey the equations $P_i.X=0$ for all $i$ together with $X^2=-1$. See Appendix \ref{leb} for details.}  

Whenever the bulk $S$ matrix is generated by a local contact term, ${\cal S}$ grows like a non-negative power of $\omega$. In this situation the integral over $\omega$ in \eqref{nntrevint} receives its dominant contributions from large values of $\omega$. The phase factor $e^{i \omega P.X}$ cuts off a would-be large $\omega$ divergence in this integral and turns it into a power-law singularity, 
of the form $\frac{1}{\rho^a}$, in $\rho$ \footnote{When 
	${\cal S}$ grows like $\omega^r$, 
	$$a=\Delta+r-3.$$ }; this is the famous bulk point singularity of \cite{Maldacena:2015iua}(see Section \ref{rgzl} for details). \eqref{nntrevint} allows one to compute the coefficient of this singularity; we find that it is proportional to a simple known function of $\sigma$ times an integral of  flat space S matrix over $\mathbb{H}_{D-2}$, with the scattering angle $\theta$ and scattering polarizations determined in terms of boundary cross ratios and polarization in a simple way.
    \footnote{Specifically,  $\theta$ is identified with the conformal cross ratio $\sigma$ according $\sin^2 \frac{\theta}{2} = \sigma$, while the transverse polarizations are identified with the parallel transport or boundary polarizations to the bulk point $X$ on $\mathbb{H}_{D-2}$ along the unique null geodesic that connects the boundary points $P_a$ to $X$.}  Our final result for the coefficient of the singularity $1/\rho^a$ in terms of the flat space S matrix, presented in \eqref{gsingnngen}, is  a generalization of the results of \cite{Gary:2009ae} to the study of holographic correlators of arbitrary spin, and reduces to the results of 
	\cite{Gary:2009ae} in the special case that all particles have spin-zero. 

Recall that \eqref{parse} applies when $\sigma$ is small while \eqref{nntrevint} applies when 
$\rho$ is small. We will now extract information from the fact that these two expressions must hold simultaneously when $\rho$ and $\sigma$ are both small. 
When we specialize the small $\rho$ discussion (\eqref{nntrevint} and surrounding) to small values of $\sigma$ it turns out that we find the following simple universal result. If the flat space $S$ matrix ${\cal S}$ scales at large $s$ but fixed $t$ like 
\begin{equation}\label{lsft}
S \propto s^A 
\end{equation} 
then the small $\sigma$ limit of the $\rho \to 0$ limit of the  Greens function - normalized as in \eqref{normfp} - is proportional to 
\begin{equation} \label{ssigp}
G \propto \frac{1}{\rho^a \sigma^{A-1}}
\end{equation} 

\eqref{ssigp} tells us how fast our correlator grows with $\sigma$ in the $\rho \to 0 $ limit. On the other hand  \eqref{parse} captures the fastest growth of the correlator at any value of $\rho$. In subsection \ref{ire} we present a detailed study of  the inter relation between the small $\sigma$ and small $\rho$ expansions, and establish in particular that
\begin{equation}\label{alphaa}
A' \geq A
\end{equation} 

Recall that the chaos bound theorem of \cite{Maldacena:2015waa} implies (see the Appendix of that paper and section \ref{crgcha} for a brief review) that the correlator for a well behaved (unitary etc) boundary theory cannot grow faster than $\frac{1}{\sigma}$ in the small $\sigma$ limit on the causally Regge sheet. As $A'$ in \eqref{parse} equals $A$ even on the causally Regge sheet, it follows as a consequence of the chaos bound as 
\begin{equation}\label{aineq} 
A \leq 2.
\end{equation}
In words, the flat space $S$ matrix  cannot grow faster than $s^2$ in the Regge limit. 
Restated,  the chaos bound applied to  boundary correlators of a unitary theory implies that the bulk dual of that theory obeys the CRG conjecture.




\section{Kinematics} \label{kinem} 

\subsection{Insertion locations}

In this paper we study the four point function of boundary operators in a holographic field theory. Following section 6 of \cite{Maldacena:2015iua}, we study correlators of operators inserted at the following two parameter set of boundary points  
of $AdS_{D+1}$: 
\begin{equation}\label{fpscatnnnn} 
\begin{split}
&P_1=(\cos\tau , \sin\tau, 1,0, {\vec 0})\\
&P_3=(\cos \tau, \sin \tau ,-1,0, {\vec 0})\\
&P_2=(-1,0, -\cos \theta, -\sin \theta, {\vec 0})\\
&P_4=(-1,0, \cos \theta, \sin \theta, {\vec 0})\\
\end{split}
\end{equation}

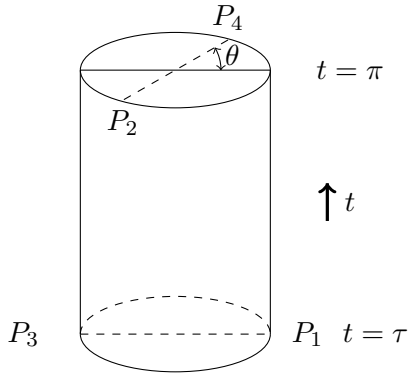
\begin{figure}
\begin{center}
\begin{tikzpicture}[
my angle/.style={draw, <->, angle eccentricity=1.3, angle radius=6mm}]
\draw (0,0) ellipse (1.25 and 0.5);
\draw (-1.25,0) -- (-1.25,-3.5);
\draw (-1.25,-3.5) arc (180:360:1.25 and 0.5);
\draw [dashed] (-1.25,-3.5) arc (180:360:1.25 and -0.5);
\draw (1.25,-3.5) -- (1.25,0);
\draw (2.3,-3.5) node {$P_1~~ t=\tau$};
\draw (-2.0,-3.5) node {$P_3$};  
\draw (2.3,0) node {$t=\pi$};
\draw (2.3,-1.75) node {$t$};
\draw [dashed] (-1.25,-3.5) -- (1.25,-3.5);
\draw 
		[
		very thick,
		decoration={markings, mark=at position 1.0 with {\arrow[black,line width=0.5mm]{>}}},
		postaction={decorate}
		]
		(2.0,-2.0)--(2.0,-1.5);
\coordinate (A)  at (1.25,0);
\coordinate (O)  at (0,0);
\coordinate (B)  at (-1.25,0);
\coordinate[label=above:$P_4$]   (M) at (30:0.8);
\coordinate [label=below:$P_2$]  (N) at (210:0.8);
\draw[dashed]    (M) -- (N);
\draw (-1.25,0) -- (1.25,0);
\pic[my angle, "$\theta$"] {angle = A--O--M};
\end{tikzpicture}
\caption{Insertion points in global AdS}
\end{center}
 \label{fig:adscylinderins}
\end{figure}

where we parameterize boundary points in $AdS_{D+1}$ by null
rays in embedding space with the first two coordinates timelike and the remaining $D$ coordinates space-like (see Appendix \eqref{ncbm} for notations and conventions and a brief review of the embedding space formalism), ${\vec 0}$ is the zero vector in $D-2$ dimensional Euclidean space. Points 
$P_1$ and $P_3$ are inserted at global time $\tau$, 
and points $P_2$ and $P_4$ are inserted at global time $\pi$. We will focus our attention on the range of parameters
\begin{equation}\label{rangepar} 
0 \leq \tau \leq \pi, ~~~~0 \leq \theta \leq \frac{\pi}{2}
\end{equation} 
\footnote{The restriction to $\theta$ in the range 
	$0 \leq \theta \leq \frac{\pi}{2}$ rather than 
	$0 \leq \theta \leq \pi$ is a matter of convenience; the flip $2 \leftrightarrow 4$ maps the smaller to the bigger range.}

\subsection{Causal relations and cross ratio sheets}

The causal relations between the points $P_a$ are given by \footnote{See around Fig. 8 of\cite{Maldacena:2015iua}. The argument is as follows. The pairs $(P_1, P_3)$ and $(P_2, P_4)$ 
	are each inserted at equal global times; it follows that elements of the same pair are always space-like separated from each other. 
	The time difference between the insertion points of the second and first pair is  $\pi -\tau$. The angular separation between $1$ and $3$ (or $2$ and $4$) is $\pi -\theta$, and the angular separation between $1$ and $4$ (or $2$ and $3$) is $\theta$. Using the fact that two  points are time-like separated only if the time difference between them exceeds their angular separation gives \eqref{causalrel}. } 
\begin{equation}\label{causalrel}
\begin{split}
&\tau  > \pi -\theta ~~~~~~~~~~~~~{\rm Causally ~ Euclidean}, \\
&\pi -\theta >\tau>\theta ~~~~~~ {\rm  ~~ Causally ~Regge}~~~~~~~~~~~(P_4>P_1, ~{\rm and}~P_2> P_3) \\
&\tau < \theta ~~~~~~~~~~~~~~~~~~{\rm ~ Causally ~ Scattering} 
~~~~~~~~~ (P_4,P_2)> (P_1, P_3)\\
\end{split}
\end{equation} 
where we have used the notation $A>B$ to denote that $A$ is in the causal future of $B$ and it is understood that two points are space-like separated with respect to each other if their causal ordering is not specified\footnote{For complete clarity, all four points are space-like separated from each other in the Euclidean configuration,  $P_4$ is in the future light-cone of $P_1$, $P_2$ is the the future light-cone of $P_3$ while each of the pair $P_1, P_4$ is space-like separated with each of the pair $P_3, P_2$ 
	in the Regge configuration. On the other hand $P_1$ and $P_3$ are space-like separated with respect to each other, $P_2$, and $P_4$ are also space-like separated with respect to each other, but each of $P_4, P_2$ lie in the future light-cones of each of $P_1, P_3$ in the causally scattering regions.} 
\footnote{We use the name Causally Regge because this is 
	the causal relationship between points in the well studied Regge limit of CFT on the cross ratio sheet on which the chaos bound constrains the small $\sigma$ behavior of correlators.} Note that the three distinct causal relations \eqref{causalrel} correspond to three distinct sheets in complex cross ratio space.

We end this subsection with an aside. We see from \eqref{causalrel} that when $\tau > \theta$, $\tau=\theta$ or $\tau <\theta$, the pairs of points $(P_2, P_1)$ and $(P_4, P_3)$  are, respectively both space-like, null or time-like  separated. The fact that the switch of causal relations for the pairs $(P_2, P_1)$ and $(P_4, P_3)$ happens simultaneously indicates that the two parameter set of coordinates  \eqref{fpscatnnnn} are rather special especially in the neighbourhood of $\tau =\theta$.   If we allowed for general variations of all four boundary points, in the  neighbourhood of the configuration \eqref{fpscatnnnn} with $\tau=\theta$ 
we would find  points with $P_2>P_1$ but $(P_4,P_3)$ space-like and vice versa. The configurations \eqref{fpscatnnnn} do not include any such causal configurations, and so are non generic in the neighbourhood of $\tau=\theta$. See 
Appendix \ref{reggeblowup} for a detailed analysis in the case that $\tau$ and $\theta$ are both small. 

\subsection{Conformal cross ratios} \label{ccrr}

The conformal cross ratios (see Appendix \ref{ccra}) associated with these four points are given by 
\begin{equation} \label{crrnnn}
\begin{split} 
&z {\bar z} \equiv \frac{(P_2.P_1) (P_3.P_4)}{(P_2.P_4) (P_3.P_1)}=\frac{(\cos \tau -\cos \theta )^2}{4} \equiv A\\
&(1-z)(1-{\bar z}) \equiv \frac{(P_4.P_1) (P_2.P_3)}{(P_2.P_4) (P_3.P_1)}=\frac{(\cos \tau + \cos \theta )^2}{4} \\
& \implies z + {\bar z} = 1-\cos \tau \cos \theta \equiv 2 B \\
& z^2-2 B z+A=0\\
& z= B \pm \sqrt{B^2-A}\\
&z=\frac{1-\cos \left( \tau - \theta \right)}{2},  \\
& {\bar z}= \frac{1-\cos \left( \tau + \theta \right)}{2},
\end{split} 
\end{equation}
Note that both $z$ and ${\bar z}$ lie in the interval 
\begin{equation}\label{zineq}
z \in [0.1], ~~~~{\bar z} \in [0,1]
\end{equation} 

As usual we define the cross ratios $\sigma$ and $\rho$ by the relations 
\footnote{See Appendix \ref{ccra} for more about various cross ratios.}
\begin{equation}\label{sigrhodef}
z=\sigma e^\rho, ~~~~{\bar z}= \sigma e^{-\rho}
\end{equation} 
With these definitions 
\begin{equation} \label{crrnnnr}
\begin{split} 
& \sigma^2 = \frac{(\cos\theta-\cos\tau)^2}{4} \\
& \sinh^2\rho = \frac{\sin^2\theta\sin^2\tau}{(\cos\theta-\cos\tau)^2}
\end{split} 
\end{equation}

To end this subsection we specialize \eqref{crrnnn} in the two parametric limits that are of particular interest in this paper.

\subsubsection{The small $\tau$ limit} 

 In the small $\tau$ limit at fixed $\theta$, \eqref{crrnnn} simplifies to  
\begin{equation}\label{crst} 
\begin{split} 
z&= \sin \frac{\theta}{2} \left( \sin \frac{\theta}{2} 
-\tau \cos \frac{\theta}{2} \right) +{\cal O}(\tau^2) \\
{\bar z}&=\sin \frac{\theta}{2} \left( \sin \frac{\theta}{2} 
+\tau \cos \frac{\theta}{2} \right) +{\cal O}(\tau^2) \\
\sigma&=\sin^2 \frac{\theta}{2}+ {\cal O}(\tau^2) \\
\rho & = -\tau \cot \frac{\theta}{2} +{\cal O}(\tau^3)  \\
\end{split}
\end{equation}
Note, in particular, that $\rho$ approaches zero (from the negative side) as $\tau \to 0$.

\subsubsection{The Regge limit} 

On the other hand in the Regge limit 
\begin{equation}\label{newlim} 
\tau \to 0, ~~~~\theta \to 0, ~~~~ \frac{\tau}{\theta}=
a ={\rm fixed}
\end{equation}  
\eqref{crrnnn1} simplifies to  
\begin{equation}\label{nun}
\begin{split} 
&z= \frac{(\theta -\tau - i \epsilon)^2}{4}= \frac{\theta^2}{4} (1-a -i \epsilon)^2 +{\cal O}(\theta^4) \\
&{\bar z}= \frac{(\theta +\tau + i \epsilon)^2}{4}  
= \frac{\theta^2}{4} (1+a +i \epsilon)^2+{\cal O}(\theta^4) \\
& \sigma^2 = \frac{\theta^4(1-a^2)}{16}+{\cal O}(\theta^6)\\
& e^{2 \rho} = \left( \frac{1-a-i\epsilon}{1+a +i\epsilon} \right)^2 +{\cal O}(\theta^2)
\end{split} 
\end{equation}
(in \eqref{newlim} we have presented $i\epsilon$ corrected 
formulae using the method of subsection \ref{ccr} below).

Note that the Regge limit explores the neighbourhood of the boundary point \eqref{fpscatnnnn} obtained by setting $\tau=\theta=0$, i.e. the point 
\begin{equation}\label{reggepoints} \begin{split}
&P_1=(1 , 0, 1,0, {\vec 0})\\
&P_3=(1,0,-1,0, {\vec 0})\\
&P_2=(-1,0, -1, 0, {\vec 0})\\
&P_4=(-1,0, 1, 0, {\vec 0})\\
\end{split}
\end{equation}
In this paper we study only that part of the neighbourhood of this point that we can be reached by turning on small values of $\theta$ and $\tau$ in \eqref{fpscatnnnn}. As an aside we note that this two parameter set of points do not give a complete cover of the neighbourhood  of \eqref{reggepoints} modulo conformal transformations. In addition to the points obtained from \eqref{fpscatnnnn} at small values of $\theta$ and $\tau$, there are additional infinitesimal deformations of \eqref{reggepoints} in which the insertion points enjoy different causal relations from any of those listed in \eqref{causalrel}. 
This fact (which we will never use anywhere else in this paper) is explained in some detail in Appendix 
\ref{reggeblowup}.

\subsubsection{Overlap between small $\tau$ and Regge}

Note that the small $\theta$ limit of the small $\tau$ limit \eqref{crst} overlaps with the small $a$ limit of the Regge limit \eqref{nun}. In particular if we expand the RHS of  \eqref{crst} at leading order in $\theta$ we obtain 
\begin{equation}\label{expansion} 
\begin{split} 
\sigma &= \frac{\theta^2}{4} \\
\rho&=-\frac{2\tau}{\theta}\\
\end{split} 
\end{equation}
On the other hand if we expand \eqref{nun} to leading order in $a$ we obtain 
\begin{equation}\label{expansionn} 
\begin{split} 
\sigma &= \frac{\theta^2}{4} \\
\rho&=-2 a\\
\end{split} 
\end{equation}
It follows from \eqref{newlim} that \eqref{nun} and \eqref{crst} are equivalent. 

Notice that the $\tau \to 0$ limit places us in the Causally Scattering regime of \eqref{causalrel}. On the other hand, the Regge limit lies in the causally scattering regime when $a<1$ but in the Causally Regge regime for $a>1$. The fact that the Regge limit straddles two distinct causal regimes will be of central importance to this paper.

\subsection{A path in cross ratio space} \label{pcrs}

It is useful to track the evolution of $z$, ${\bar z}$, $\sigma$ and $\rho$ as we keep $\theta$ fixed and adiabatically decrease $\tau$ from $\tau=\pi$ to $\tau=0$. When $\tau=\pi$, $z={\bar z}=\frac{1+\cos \theta}{2}$ and our configuration is Euclidean. 
As we decrease $\tau$, $z$ decreases while ${\bar z}$ increases. ${\bar z}$ reaches its maximum value, namely unity when $\tau=\pi -\theta$  (i.e. at the boundary between the Causally Euclidean and Causally Regge regimes) . As $\tau$ is further 
decreased, $z$ continues to decrease, but now 
${\bar z}$ also begins to decrease. Once $\tau$ reaches $\theta$ (the boundary between the causally Regge and causally scattering regime) $z$ reaches its minimum value namely zero. As 
$\tau$ is decreased even further ${\bar z}$ continues 
to decrease but $z$ now starts to increase. Finally, at $\tau=0$ we have $z={\bar z}= \frac{1-\cos \theta}{2}$. \footnote{It follows (see \eqref{qeq}) that the starting point $\tau=\pi$ and the end point 
	$\tau=0$ of our paths each have $\rho=0$.}

\begin{figure}[H]
	\begin{center}
		\includegraphics[width=13cm]{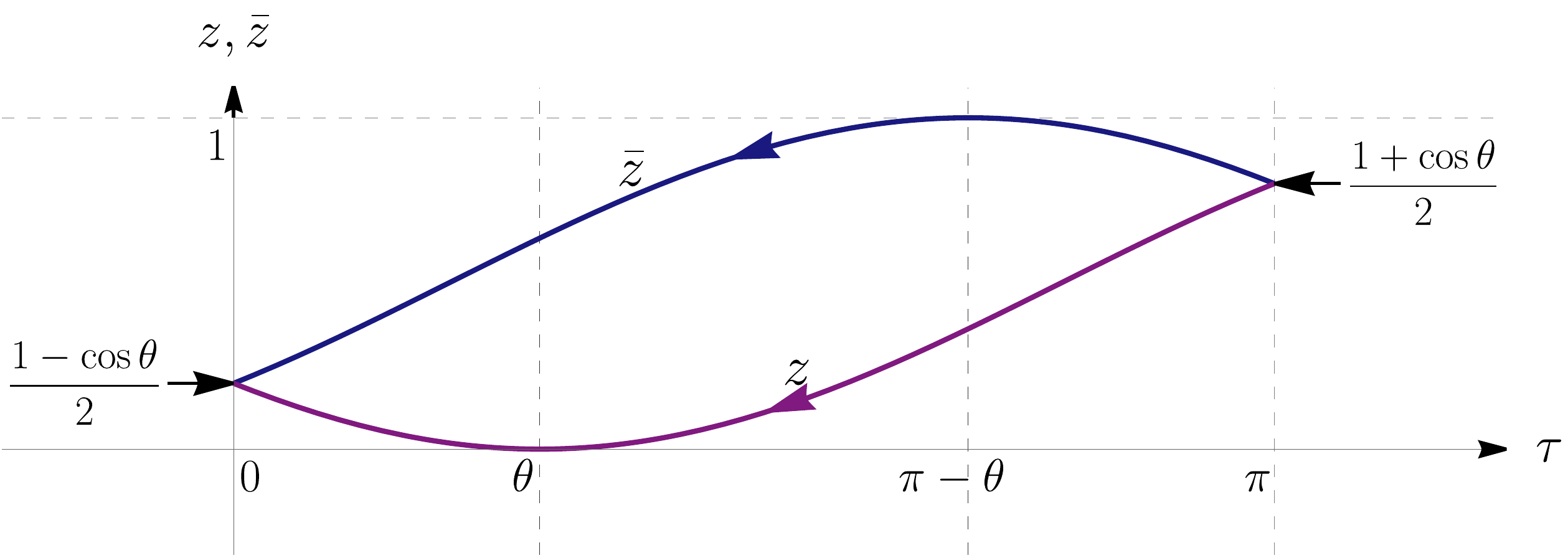}
		\caption{The evolution of $z$ and ${\bar z}$ as $\tau$ is decreased from $\pi$ down to $0$ at fixed $\theta$. Note ${\bar z}$ touches its maximum value unity at $\tau=\pi -\theta$ and $z$ touches its minimum value, $z=0$ at $\tau=\theta$.}
		\label{zphh}
	\end{center}
\end{figure}

\begin{figure}[H]
	\begin{center}
		\includegraphics[width=11cm]{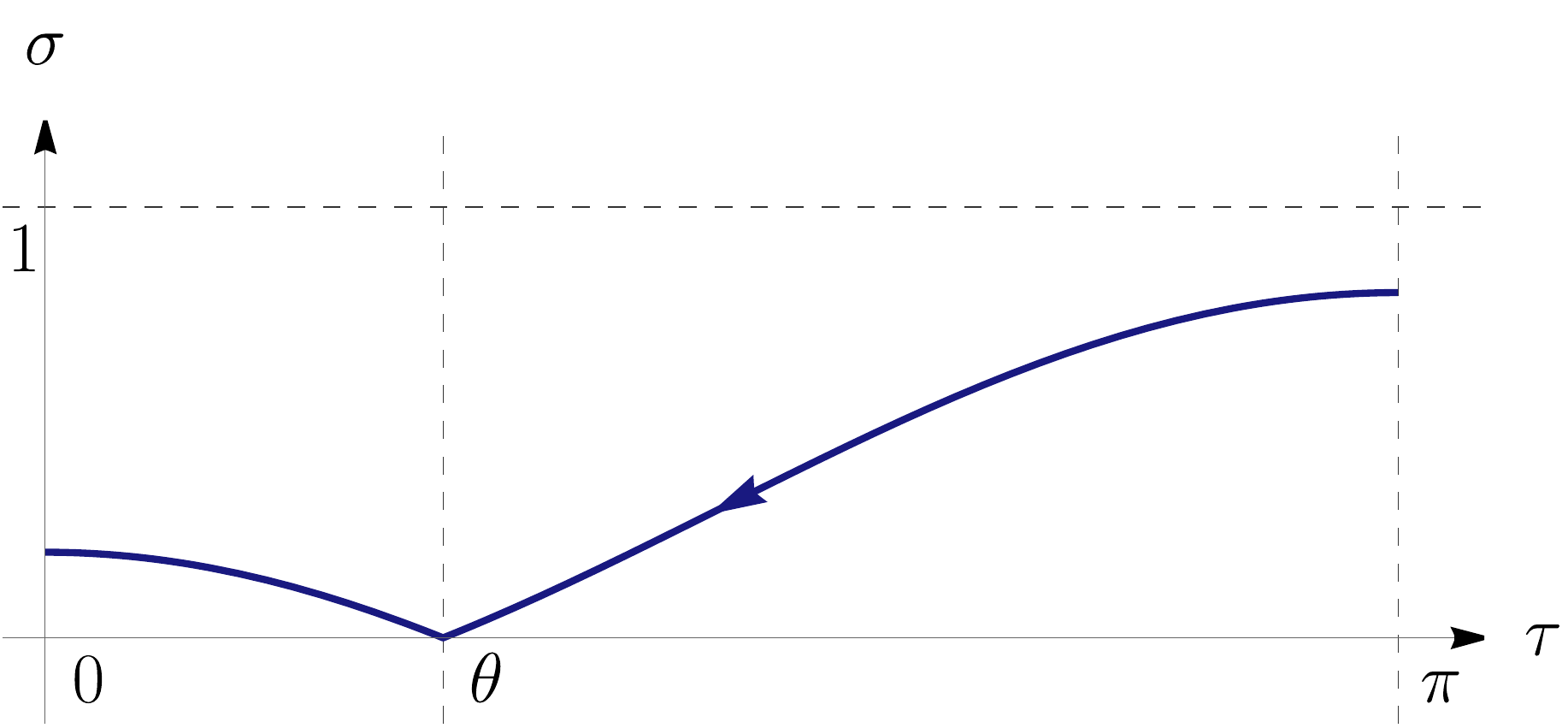}
		\caption{The evolution of the cross ratio $\sigma$ as $\tau$ is decreased from $\pi$ down to $0$ at fixed $\theta$. Note that $\sigma$ touches its minimum value, $\sigma=0$ at $\tau=\theta$, when $z$ vanishes.}
		\label{sp}
	\end{center}
\end{figure}

\begin{figure}[H]
	\begin{center}
		\includegraphics[width=11cm]{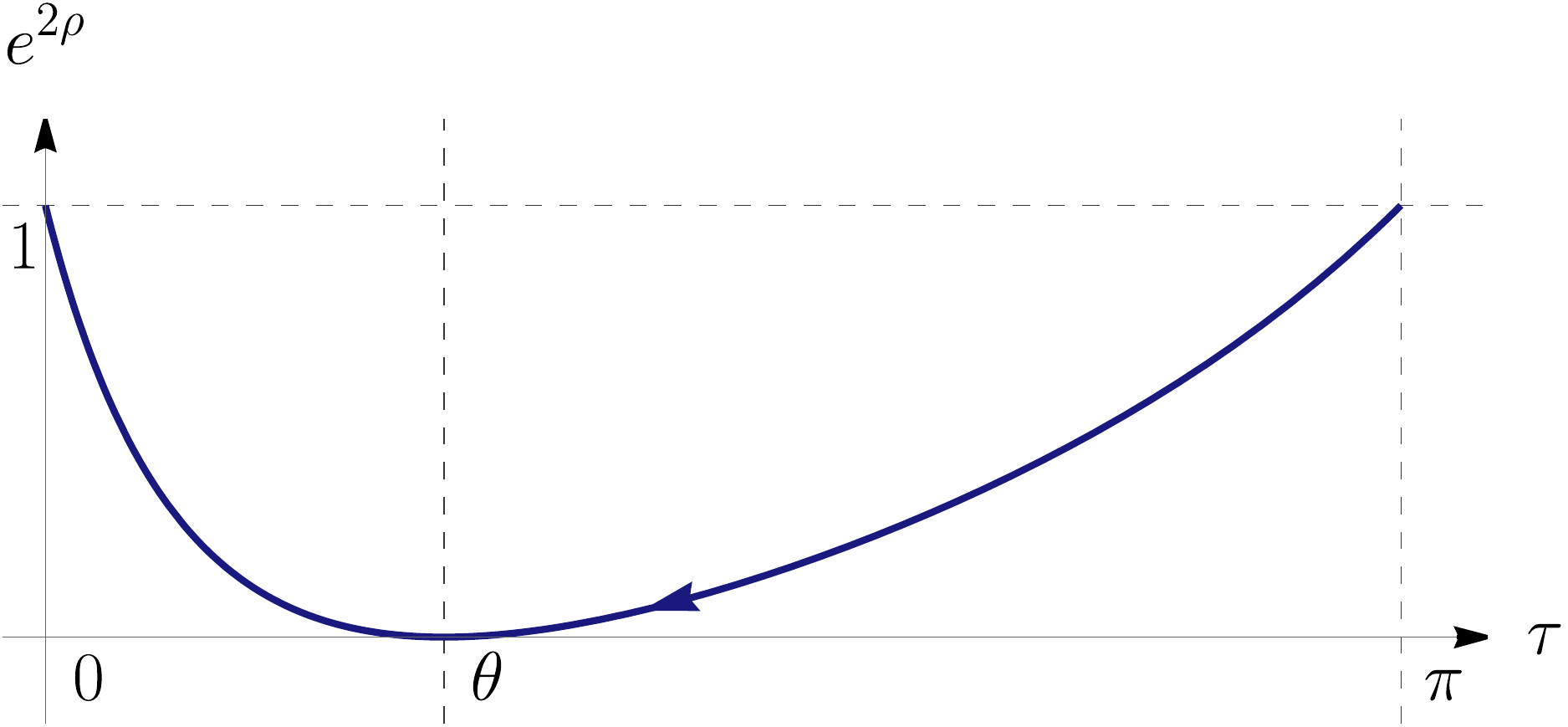}
		\caption{The evolution of the cross ratio $e^{2 \rho}= \frac{z}{\bar z}$ as $\tau$ is decreased from $\pi$ down to $0$ at fixed $\theta$. Note that $e^{2 \rho}$ touches its minimum value, $e^{2 \rho}=0$ at $\tau=\theta$, when $z$ vanishes.}
		\label{rp}
	\end{center}
\end{figure}

In the previous paragraph, we have described a one parameter path in configuration space. The evolutions of various conformal cross ratios along this path is depicted in the graphs Figs \ref{zphh},  \ref{sp} and \ref{rp} 
below in which we have displayed 
graphs of the cross ratios $z$, ${\bar z}$, $\sigma$  $\rho$ versus $\tau$ (note the arrows in those graphs track the journey from Causally Euclidean configurations, starting at $\tau=\pi$, to Causally Scattering configurations, ending at 
$\tau=0$.)

We end this subsection with a remark. Several investigations of CFT work in the so-called `$\rho$ plane'. \footnote{To forestall confusion we emphasize that in this paragraph and in parts of Appendix \ref{stdconfig} - but nowhere else in this paper - $\rho$ refers to the insertion coordinate in \eqref{labn} and not to the conformal cross ratio paired with $\sigma$.} In this plane the four CFT operators are inserted in $\mathbb{R}^{1,1}$ at the locations \eqref{labn}. A question that might occur to the reader is the following: how does the path described in this subsection (fixed $\theta$, $\tau$ lowered from $\pi$ to zero) map to the $\rho$ plane? This question is addressed in detail in Appendix \ref{stdconfig}. Here we only make a simple qualitative point. In the $\rho$ coordinate system 
\eqref{labn} the insertion points $P_3$ and $P_4$ are fixed and unmoving; in particular, these two points are spacelike separated from each other. It follows immediately that the part of 
the trajectory of this subsection that lies in the Causally scattering region ($\tau < \theta$) has no image in the $\rho$ plane. On the other hand, the part of the trajectory of this subsection that lies on the Causally Euclidean and Causally Regge sheets has a faithful map onto the $\rho$ plane, as we describe in Appendix \ref{stdconfig}.

\subsection{The same path on the complex cross ratio sheets}
\label{ccr}

As conformal correlators have branch points at ${\bar z}= 1$ and $z=0$, and as the trajectories described in the previous subsection all 
`touch' these branch points, it follows that the description of the paths presented in the previous subsection is ambiguous and needs to be improved. \footnote{Along the path of the previous subsection, for instance, ${\bar z}$ 
	increases to unity and then retreats back. We need to know whether the retreat occurs on a different sheet from the onward trajectory, and if so which one. A similar question arises for the part of the trajectory that touches $z=0$.}
The true paths traversed in cross ratio space are given by making the replacements $\tau \rightarrow \tau - i \epsilon \tau$ (see Appendix \eqref{reprule}). The $i \epsilon$ corrected insertion points are
\begin{equation}\label{fpscatnnnn1} \begin{split}
&P_1=(\cos(\tau - i\epsilon \tau) , \sin(\tau - i\epsilon\tau), 1,0)\\
&P_3=(\cos (\tau - i\epsilon\tau), \sin (\tau - i\epsilon\tau) ,-1,0)\\
&P_2=(\cos(\pi - i\pi \epsilon),\sin(\pi - i\pi \epsilon), -\cos \theta, -\sin \theta)\\
&P_4=(\cos(\pi - i\pi \epsilon),\sin(\pi - i\pi \epsilon), \cos \theta, \sin \theta)\\
\end{split}
\end{equation}
with $\epsilon>0$. The $i \epsilon$ corrected cross ratios are given by 
\begin{equation} \label{crrnnn1}
\begin{split}
&z {\bar z} \equiv \frac{(P_2.P_1) (P_3.P_4)}{(P_2.P_4) (P_3.P_1)}=\frac{(\cos \theta -\cos \left[\tau + i \epsilon(\pi-\tau) \right] )^2}{4} \equiv A\\
&(1-z)(1-{\bar z}) \equiv \frac{(P_4.P_1) (P_2.P_3)}{(P_2.P_4) (P_3.P_1)}=\frac{(\cos \theta +\cos \left[\tau + i \epsilon(\pi-\tau) \right])^2}{4} \\
& \implies z + {\bar z} = 1-\cos \theta \cos \left[\tau + i \epsilon(\pi-\tau) \right] \equiv 2 B \\
& z^2-2 B z+A=0\\
& z= B \pm \sqrt{B^2-A}\\
&z=\frac{1}{2} (1-\cos (\theta -\tau -i (\pi -\tau ) \epsilon ) ),  \\
& {\bar z}=\frac{1}{2} (1-\cos (\theta +\tau +i (\pi -\tau ) \epsilon )), \\
& \sigma^2 = \frac{(\cos\theta-\cos[\tau+i\epsilon (\pi-\tau)])^2}{4} \\
& \sinh^2\rho = \frac{\sin^2\theta\sin[\tau+i\epsilon (\pi-\tau)]^2}{(\cos\theta-\cos[\tau+i\epsilon (\pi-\tau)])^2}
\end{split} 
\end{equation}

\subsubsection{The neighbourhood of ${\bar z}=1$}

To examine the route traced by the path of subsection \ref{pcrs} in complex cross ratio space, we first examine 
\eqref{crrnnn1} in the neighbourhood of ${\bar z}=1$  by setting $\tau=\pi -\theta +\delta \tau$, assuming $\delta \tau = {\cal O}(\epsilon)$ and expanding to second order in $\epsilon$. We find that the formula for ${\bar z}$ reduces to 
\begin{equation}\label{barzexp}\begin{split}
&{\bar z}-1= - \frac{(\delta \tau +i {\tilde \epsilon})^2}{4} \\
&|{\bar z}-1|=
\frac{\delta \tau^2+ {\tilde \epsilon}^2}{4} \\
&
{\rm Arg}({\bar z-1})=-\pi +2 \tan^{-1} \left( 
\frac{{\tilde \epsilon}}{\delta \tau } \right), ~~~~~~
\tan^{-1}(x) \in [0, \pi) \\
\end{split} 
\end{equation} 
where ${\tilde \epsilon}= (\pi -\tau) \epsilon$

It follows immediately from \eqref{barzexp} that 
as $\tau$ is lowered from just above $\pi -\theta$ to just below $\pi -\theta$, our path in cross ratio space circles around the branch point at ${\bar z}=1$ in a counter-clockwise manner. The much-studied passage from the Euclidean to standard `Regge' behavior - involves traversing precisely the same path in cross ratio space (see e.g. subsection 5.1 of \cite{Kravchuk:2018htv} ). 
It follows that the Causally Regge sheet of 
\eqref{causalrel} is the sheet encountered in studies of the CFT in the  Regge limit\footnote{As far as we are aware, this limit was first studied in the context of $AdS/CFT$ in \cite{Cornalba:2006xm,Cornalba:2006xk} and later extended to more general CFTs in \cite{Cornalba:2007fs, Costa:2012cb}. See \cite{Afkhami-Jeddi:2016ntf, Caron-Huot:2017vep, Simmons-Duffin:2017nub, Kravchuk:2018htv} for more recent use of this kinematic limit in recent times.},  i.e. the sheet on which CFT correlators are constrained by the chaos bound of 
\cite{Maldacena:2015waa}.

\subsubsection{The neighbourhood of $z=0$}

In the similar fashion we examine the behaviour of our path in the neighbourhood of $z=0$ by setting $\tau =  \theta + \delta \tau$. Once again we take ${\cal O}(\delta \tau)={\cal O}(\epsilon)$ and work to second order in $\epsilon$ to obtain 
\begin{equation}\label{zexp}\begin{split}
&{z}= \frac{(\delta \tau +i {\tilde \epsilon})^2}{4} \\
&|z|= 
\frac{\delta \tau^2+ {\tilde \epsilon}^2}{4} \\
&
{\rm Arg}(z)= 2 \tan^{-1} \left( 
\frac{{\tilde \epsilon}}{\delta \tau } \right), ~~~~~~
\tan^{-1}(x) \in [0, \pi) \\
\end{split} 
\end{equation}
It follows that as $\tau$ is lowered from just above $\theta$ to just below $\theta$, our path circles round the branch point at $z=0$ in a counter-clockwise manner (note ${\tilde \epsilon}>0)$ (see Fig. 6 of \cite{Heemskerk:2009pn}). 

\subsubsection{Summary of the trajectory in the complex plane}

\begin{figure}[H]
	\begin{center}
		\includegraphics[width=10cm]{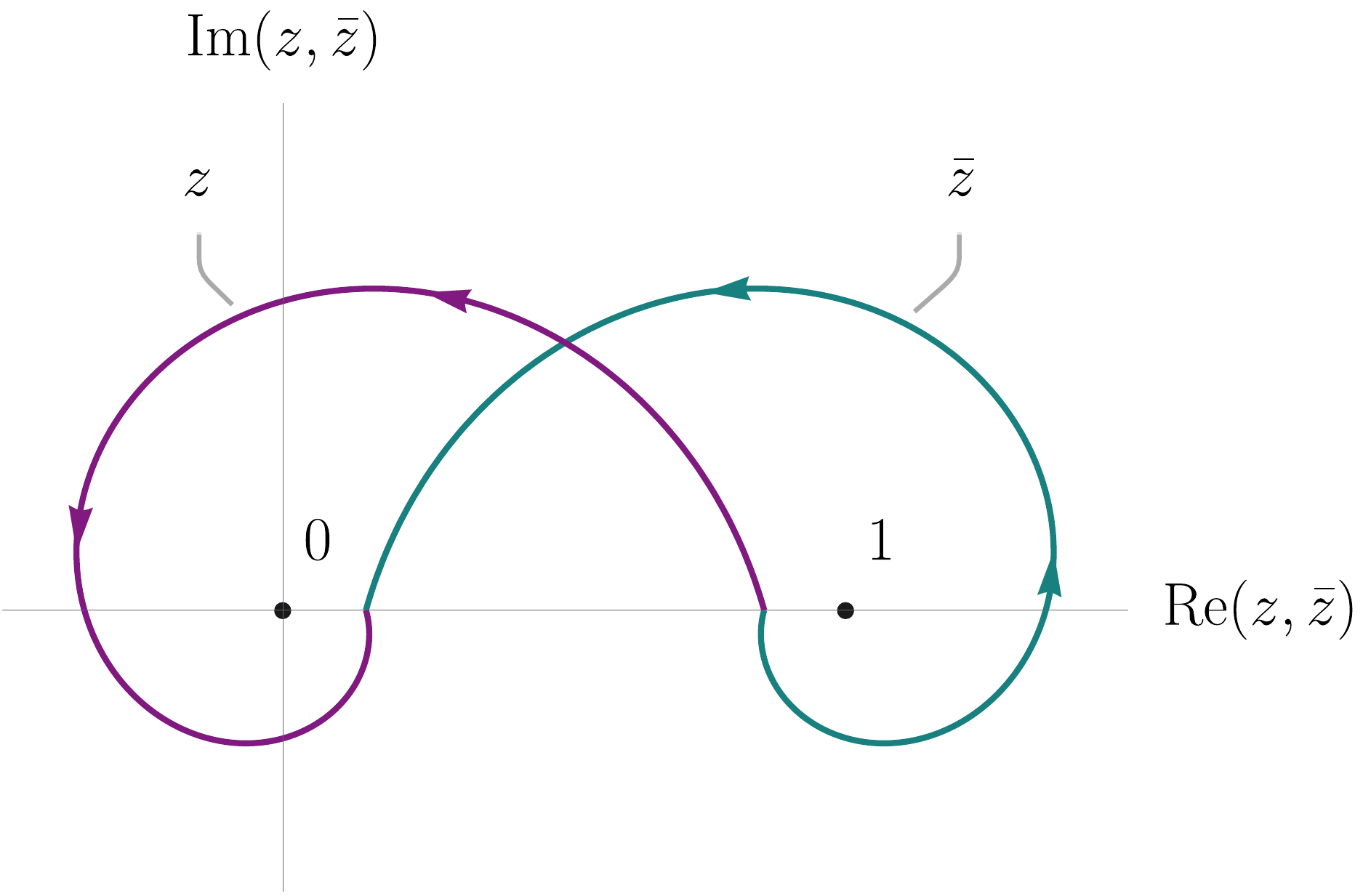}
		\caption{The path traversed in the complex plane by the variables $z$ (purple) and ${\bar z}$ (green) as we lower $\tau$ from $\pi$ to $0$ at fixed $\theta$. The vertical scale in these graphs is greatly exaggerated to make them visible. The actual curves should be thought of as hugging the real axis except in the neighbourhood of the branch points which they circle in the manner shown in this Figure. }
		\label{compz}
	\end{center}
\end{figure}

In summary, as we lower $\tau$ from $\pi$ to zero 
at constant $\theta$ we move along the trajectories in (complex) cross ratio space depicted schematically in Fig \ref{compz} (the vertical scale in these graphs is highly exaggerated; all curves hug the real axis except when ${\bar z}$ is in the neighbourhood of unity or when $z$ is in the neighbourhood of zero). In other words, our path starts on the principal or Causally Euclidean sheet when $\tau=\pi$. As $\tau$ is lowered below 
$\pi -\theta$ our path circles around the branch cut at ${\bar z}=1$ counter-clockwise, bringing us onto the Causally Regge sheet. As $\tau$ is further lowered past $\theta$, the path circles counter-clockwise around the branch cut at $z=0$, taking us to the Causally Scattering sheet (see Fig. 8 of \cite{Maldacena:2015iua} for a very closely related discussion).  

\subsection{Intersections of boundary lightcones} \label{ibl}

When do the lightcones emanating out of the four points $P_1, P_2, P_3$ and $P_4$ have a common intersection point in the bulk? For $\tau \neq 0$ the four vectors $P_1$, $P_2$, $P_3$ and $P_4$ span a four dimensional subspace of $\mathbb{R}^{D,2}$ and this subspace is metrically $\mathbb{R}^{2,2}$. 
It follows from the analysis of Appendix \ref{obl} \footnote{In the language Appendx \ref{obl}\eqref{fpscatnnnn} is a Case 1 configuration of boundary points 
	with ${\hat Q}=4$ .} that the lightcones that emanate 
out of these four points have no common intersection
for $\tau \neq 0$. \footnote{In this generic situation  the subalgebra of the conformal algebra that stabilizes the collection of points $P_1$, $P_2$, $P_3$ and $P_4$ - and so the vector space $\mathbb{R}^{2,2}$ of embedding space vectors spanned by $P_1 \ldots P_4$-  is $SO(D-2)$. }

When $\tau = 0$, on the other hand (i.e. at the edge of the parameter range \eqref{rangepar}, i.e.the limit \eqref{crst} in which $\rho \to 0$) 
\footnote{There are actually two distinct limits 
	of the insertions \eqref{fpscatnnnn} in which $\rho \to 0$. The first of these is the limit $\tau \to 0$
	of interest to this paper. The second is the simpler limit $\tau \to \pi$ in which, once again, $\rho \to 0$. These two limits are physically distinct because they lie on different sheets on the $z$ and ${\bar z}$ complex plane. In the second simpler limit the correlators  lie on the principal or Euclidean sheet 
	(i.e. the Causally Euclidean region of \eqref{causalrel}). Correlation functions at nonzero $\theta$ are manifestly non-singular in this second $\rho \to 0$ limit (see under equation 6.21 of 
	\cite{Maldacena:2015iua}). This non-singular limit is of no interest to this paper and will never again be considered here.} 
the vectors $P_1$, $P_2$, $P_3$ and $P_4$ are linearly dependent. 
In particular, at $\tau=0$
\begin{equation}\label{sumofp} 
P_1+P_2+P_3+P_4=0
\end{equation} 
In this case these vectors span a three dimensional subspace of $\mathbb{R}^{D,2}$. This subspace is metrically $\mathbb{R}^{2,1}$. 
In the language Appendx \ref{obl}, \eqref{fpscatnnnn} is a Case 2 configuration of boundary points 
with ${\hat Q}=3$.  The lightcones that emanate out of these four points intersect on an $\mathbb{H}_{D-2}$ , and the subalgebra of the 
conformal algebra that stabilizes the four point is $SO(D-2,1)$.

In the special case that $\tau=0$ and $\theta=0$ (i.e. the Regge limit \eqref{newlim}), it is easy to check that \eqref{sumofp} breaks up into two equations (see \eqref{reggepoints})
\begin{equation}\label{sumofpbreak}
\begin{split}
&P_1+P_2=0\\
&P_3+P_4=0\\
\end{split}
\end{equation} 
It follows that the subspace of $\mathbb{R}^{D,2}$ spanned by 
the vectors $P_1, P_2, P_3, P_4$ in this limit is 
two dimensional; infact it is the $\mathbb{R}^{1,1}$ spanned by 
$$(a, 0, b, 0, \ldots 0)$$
for arbitrary $a$ and $b$. The space orthogonal to this
$\mathbb{R}^{1,1}$ is $\mathbb{R}^{D-1,1}$ spanned by coordinates of the form 
\begin{equation}\label{parrdmoo}
(0,y_0, 0, y_1,y_i)
\end{equation} 
and the subgroup of the conformal group that stabilizes the collection of points $P_1, P_2, P_3, P_4$ is $SO(D-1, 1)$. 
\footnote{Moving beyond the consideration of the specific boundary configurations studied in this paper, the reader may wonder there exists a general interplay between the  $\mathbb{R}^{p,q}$ classification of boundary points, their possible causal structures and the nature of their cross ratios. The answer to this general question (that has no application to the current paper) is clearly in the affirmative and we  present a preliminary investigation of this point in Appendix \ref{lcr}. For example we demonstrate that whenever the four boundary points are in an $\mathbb{R}^{3,1}$ configuration (this can happen for a number of different causal relations between the points), the $\rho$ cross ratio corresponding to these points is always imaginary. On the other hand $\rho$ could be either real or imaginary when the four points span an $\mathbb{R}^{2,2}$. }

Although this fact will have no implications for the main flow of this paper, we note as an aside
that the neighbourhood of the Regge point 
\eqref{reggepoints} includes points whose span in embedding space is an $\mathbb{R}^{3,1}$ (in addition to the points obtained from small values of $\theta$ and $\tau$ in \eqref{fpscatnnnn} whose span in embedding space make up an $\mathbb{R}^{2,2}$ or an $\mathbb{R}^{2,1}$); this point is explained in some detail in Appendix \ref{reggeblowup}.

\subsection{Scaling limits of the two point function}

In this paper, we will be interested in studying the normalized four-point function of operators inserted at the positions $P_1 \ldots P_4$ defined by 
\begin{equation}\label{normfp} 
G^{\rm norm}= \frac{G}{G_{12} G_{34} }
\end{equation} 
where $G$ is the simple four-point function, and $G_{12}$ and $G_{34}$ are two-point functions. We will be particularly interested in the scaling of the correlator \eqref{normfp} in the Regge and small $\tau$ limits. In the subsequent two sections, we present a study of the four-point function - the numerator of \eqref{normfp} - in these two scaling limits. In this brief subsection, we present the much simpler analysis of the scaling of the denominator of \eqref{normfp} in these limits.

To end this section we study the scaling of the two-point function of spin $J$ operators inserted at the locations $P_1$ and $P_2$ in the Regge and $\tau \to 0$ limits. 

As we have reviewed in Appendix \ref{bbc}, the boundary to boundary two point function for a spin $J$ field is given by 
\begin{equation}\label{fttp2} 
G_{ij}=\mathcal{C}_{\Delta,J}\frac{\left( Z_i.Z_j ~P_i.P_j- {Z_i.P_j~ Z_j.P_i} \right)^J}{(P_i.P_j)^{\Delta+J}}
\end{equation} 
where $Z_i$ is the boundary polarization vector for the $i^{th}$ particle (see under \eqref{equivbound}
for a discussion of boundary tangent vectors in the embedding space formalism) and 
\begin{equation}
\mathcal{C}_{\Delta,J} = \frac{(J+\Delta-1)\Gamma(\Delta)}{2\pi^{d/2}(\Delta-1)\Gamma(\Delta+1-h)}
\end{equation}
Note that on the configurations \eqref{fpscatnnnn}
\begin{equation}\label{pnp}
P_1.P_2=P_3.P_4= \cos \tau -\cos \theta 
\end{equation}

\subsubsection{Regge scaling}

In this limit \eqref{pnp} simplifies to 
\begin{equation}\label{pnpr}
P_1.P_2=P_3.P_4= \frac{\theta^2-\tau^2}{2}=\frac{\theta^2}{2} (1-a^2)
\end{equation}
Also
\begin{equation}\label{pnnnn} \begin{split} 
&Z_1.P_2= Z_1.(P_2+P_1)= \theta
\left( Z_1.(0,a , 0, -1, {\vec 0}) \right)  +{\cal O}(\theta^2) \leq {\cal O}(\theta)\\
& Z_2.P_1= Z_2.(P_2+P_1)= \theta
\left( Z_2.(0,a , 0, -1, {\vec 0}) \right)  +{\cal O}(\theta^2) \leq {\cal O}(\theta)\\
&Z_1.Z_2  \leq {\cal O}(1)
\end{split}
\end{equation} 
where we have used $Z_1.P_1=Z_2.P_2=0$ (see around 
\eqref{equivbound}).
It follows from \eqref{pnnnn} that 
\begin{equation} \label{polc}
\left( Z_1.Z_2 ~P_1.P_2- {Z_1.P_2~ Z_2.P_1} \right)^J \leq {\cal O }(\theta^{2J})
\end{equation} 
Generic polarizations saturate the 
inequality \eqref{polc}. Through this paper and we restrict attention to such correlators; In other words in this paper we restrict attention to generic polarizations - those that obey
\begin{equation} \label{polcdem}
\left( Z_1.Z_2 ~P_1.P_2- {Z_1.P_2~ Z_2.P_1} \right)^J ={\cal O}(\theta^{2J})
\end{equation}
(similar comments apply to $G_{34}$.)
\footnote{	This restriction allows for greater  simplicity of presentation and does not really result in loss of generality. Our assumption is maximally violated when the LHS of \eqref{polcdem} vanishes, in which situation \eqref{normfp}. See section \ref{crgcha} for a discussion of how the omission of such correlators does not result in a lack of generality.}

It follows immediately that in the Regge limit,
the two point function $G_{12}$ scales like 
\begin{equation}\label{itlim}
\begin{split} 
G_{12}&  = \frac{1}{\sigma^{ \Delta_1}} \\
G_{34}&  = \frac{1}{\sigma^{ \Delta_3 }} \\
\end{split} 
\end{equation} 

\subsubsection{Small $\tau$ limit}

In the limit in which $\tau$ is taken to zero

\begin{equation}\label{pnp2}
P_1.P_2=P_3.P_4= 1-\cos \theta
\end{equation} 
If we now also take $\theta$ to be small 
\begin{equation}\label{pnpn}
P_1.P_2=P_3.P_4= \frac{\theta^2}{2}
\end{equation} 
in agreement with \eqref{pnpr} at $a=0$.  

At generic $\theta$ we generically have 
\begin{equation} \label{polcdemt}
\left( Z_1.Z_2 ~P_1.P_2- {Z_1.P_2~ Z_2.P_1} \right)^J ={\cal O}(1)
\end{equation}
When we take $\theta$ small, however,  we  recover the $a \to 0$ limit of \eqref{polcdem}.
While at generic $\theta$ the $G_{12}$ and $G_{34}$ are of order unity, at small $\theta$ 
\eqref{itlim} applies.




\section{The $\sigma \to 0$ Regge limit} \label{sgzl}

Consider a correlation function with four operators 
$O_1, O_2, O_3, O_4$ inserted at the points $P_1$, 
$P_2$, $P_3$, $P_4$ given in \eqref{fpscatnnnn}
In all the concrete formulae presented in this paper we will assume that $O_i$ are primary operators of the conformal group $SO(D,2)$ that have some dimension $\Delta_i=w_i+J_i$ and transform in the traceless symmetric representation with $J_i$ indices 
of $SO(D)$ \footnote{The extension of this discussion to general representations of $SO(D)$ seems conceptually straightforward but is not considered here as it is notationally cumbersome.}.

In this section we study the correlator described above holographically, and  
in the Regge limit \eqref{newlim}, i.e. the limit in which the 
conformal cross ratios are given by \eqref{nun}. 
We emphasize that, in the terminology of this paper, the Regge limit straddles the Causally Regge and Causally Scattering sheets (see the discussion around \eqref{nun}). The discussion in this section is inspired by and has considerable overlap with the analysis of section 5.2 of \cite{Heemskerk:2009pn}. The slight novelty of our presentation lies in two aspects. 
First our emphasis on the fact that the analysis of this section applies both to the Causally Regge as well as the Causally Scattering sheets, and in fact allows for smooth interpolation between the two. Second in our study of the structure of the systematic expansion  (beyond simply leading order) of the correlator in the small $\theta$ limit, an expansion we will return to in Section \ref{ire}.

\subsection{Scaling with $\theta$} \label{swt}

We wish to study a four point function generated by a bulk contact term in the Regge limit. Any such four point function is given by the sum over expressions of the schematic form 
\begin{equation} \label{genformoo}
\int d^{D+1}X \frac{N}{(-2P_1.X+i\epsilon)^{{\tilde a}_1} (-2P_2.X+i\epsilon)^{{\tilde a}_2} (-2P_3.X+i\epsilon)^{{\tilde a}_3} 
	(-2P_4.X+i\epsilon)^{ {\tilde a}_4}}
\end{equation} 
where $N$ is a numerator function; $N=N(Z_i, X)$, 
$Z_i$ are boundary polarization vectors and  
${\tilde a}_i$ are positive numbers (not necessarily integers). Bellow we will have use for the symbol 
\begin{equation}\label{tbt}
2{\tilde B}= \sum_i {\tilde a}_i.
\end{equation} 

As we have explained around \eqref{sumofpbreak}, in the strict Regge limit the four points $P_1 \ldots P_4$ span an $\mathbb{R}^{1,1}$. The space orthogonal to this $\mathbb{R}^{1,1}$ is $\mathbb{R}^{D-1,1}$ spanned by vectors of the form \eqref{parrdmoo}. 
It is thus useful to parameterize the general bulk point in this limit as 
\begin{equation}\label{genbulpoint}
X=(\frac{u+v}{2}, y_0, \frac{v-u}{2}, y_1, y_i)
\end{equation} 
with 
\begin{equation} \label{ygem}
-(y^\mu)^2 +uv=1
\end{equation} 
When $u=v=0$ and in the strict Regge limit the points  \eqref{genbulpoint} are orthogonal to 
all $P_a$,  and so all denominators in \eqref{genformoo} vanish when $u$, $v$, $\theta$, $\tau$ all equal zero. Recall that in the Regge limit $\theta$ and $\tau$ are of the same order of smallness. It will turn out that the leading contribution to the integral in \eqref{genformoo} comes from values of $u$ and $v$ that are also of order $\theta$ (see \cite{Heemskerk:2009pn}). 
It is thus useful to expand the integrand in 
\eqref{genformoo} in a power series expansion in 
the four small variables ($\theta, \tau, u, v$) with all treated as being of the same order in smallness, with all other quantities ($y_i$ and $Z_i$) held fixed \footnote{The $D-1$ variables $y_i$ are the coordinates on the $\mathbb{H}_{D-1}$ below. It is convenient not to view $y_0$ as an independent variable, but instead to use \eqref{ygem} to solve for $y_0$ in a power series expansion in $uv$. It follows that the coefficients of the power series expansion in $\theta, \tau, u, v$ are all functions of $y_i$ and $Z_i$. } Let us suppose that the numerator $N$ is of order $\theta^M$ plus subleading in this expansion. We define
\begin{equation}\label{nunt2}
2B=2{\tilde B}-M.
\end{equation} 
It follows that the integrand of \eqref{genformoo} scales like $\frac{1}{\theta^{2B}}$ in the limit under consideration.

At leading order
\begin{equation}\label{genformPX}
\begin{split} 
&   -2P_1.X = u + \tau ~ {\tilde y}_0 \equiv D_1^0 \\
&-2P_2.X = -u+\theta ~ y_1 \equiv D_2^0\\
& -2P_3.X = v  +\tau ~ {\tilde y}_0 \equiv D_3^0 \\
& -2P_4.X = -v-\theta ~ y_1 \equiv D_4^0\\
&y_0=\sqrt{1+y_i^2-uv} \\
& {\tilde y}_0=\sqrt{1+y_i^2}
\end{split} 
\end{equation}
Note that ${\tilde y}_0$ is simply $y_0$ at leading order in the small $\theta$ expansion - recall the monomial $uv$ is of order $\theta^2$. 

As we have mentioned above, in order to generate the small $\theta$ expansion we proceed to Taylor expand the integrand of \eqref{genformoo} in a Taylor series in small quantities. Notice that the quantities $D_i^0$ are a collection of four linearly independent combinations of these small quantities. As $D_i^0$  will play a distinguished role in our perturbative expansion, we find it useful to use them rather than  (see \eqref{genformPX}) ) as the the basic monomials for our Taylor series expansion. One can easily pass between $D_i^0$ and $u, v, \theta, \tau$ using \eqref{genformPX} and the inverse relations \eqref{ttuv} 
\begin{equation}\label{ttuv}
\begin{split}
u &= \frac{1}{2}  \left(D_0^1-D_0^2-D_0^3-D_0^4\right)\\
v &= -\frac{1}{2} \left(D_0^1+D_0^2-D_0^3+D_0^4\right)\\
\theta &= \frac{1}{2y_1} \left(D_0^1+D_0^2-D_0^3-D_0^4\right)\\
\tau &=\frac{1}{2 \sqrt{1+y_1^2+y_\perp^2}}\left(D_0^1+D_0^2+D_0^3+D_0^4\right)
\end{split}
\end{equation}
In \eqref{ttuv} we have split up the $D-1$ $y_i$ variables into $y_1$ and the remaining $D-2$ variables 
$y_\perp$.

To proceed we now expand both the numerator and each denominator in \eqref{genformoo} in a power series expansion in $D^i_0$.
Let us first consider the numerator
$N$. This quantity is a polynomial in $Z_i$ and $X$ and so can be expanded in a power series in $D^i_0$. The Taylor expansion of the numerator takes the schematic form  
\begin{equation}\label{eachtermnum}
 \sum_{n_i} a_{n_1, n_2, n_3, n_4}(Z_i, y_i) \left(D_0^1\right)^{n_1} \left(D_0^2\right)^{n_2} \left(D_0^3\right)^{n_3} \left(D_0^4\right)^{n_4}
 \end{equation} 
where $M$ is the smallest homogeneity (i.e. smallest value of $n_1+n_2+n_3+n_4$) that appears in this expression)

In a similar manner each term in the  denominator of \eqref{genformoo} can also be expanded. For instance 
\begin{equation}\label{exphut} 
\frac{1}{(-P_1.X+i\epsilon)^{{\tilde a}_1}}
= \sum_{n=0}^\infty E_n
\end{equation} 
where $E_n$ takes the form 
\begin{equation}\label{en}
E_n= \sum Q_{n_1, n_2, n_3, n_4}(y_i)
\frac{(D^2_0)^{n_2} (D^3_0)^{n_3} (D_0^4)^{n_4}}{(D^1_0)^{ {\tilde a}_1 +n_1}} 
\end{equation}
where all $n_i$ are integers, 
$n_2, n_3, n_4$ are all positive, $n_1$ is either positive or negative and 
\begin{equation}\label{condonn}
n_2+n_3+n_4-n_1=n
\end{equation}

Putting it all together, it follows that the integrand in \eqref{genformoo} admits an expansion of the form 
\begin{equation}\label{sumform}
\sum_{a_i}
\frac{1 }{(u + \tau ~ y_0+i\epsilon)^{a_1} (-u+\theta ~ y_1 +i\epsilon)^{a_2} (v + \tau ~ y_0+i\epsilon)^{a_3} 
	(-v-\theta ~ y_1+i\epsilon)^{a_4}}N_{\{a_i\} } (y_i, Z_i)
\end{equation} 
where
\begin{equation}\label{eachai}
\left( \sum_{i=1}^4 a_i \right)   \leq 2B 
\end{equation} 
where, for each $i$, $a_i-{\tilde a}_i$ is a (positive or negative) integer

The summation in \eqref{sumform} includes terms at all orders in the $\theta$ expansion. Terms of leading order in this expansion (those which scale like 
$\frac{1}{\theta^{2B}}$)are those for which $$ \sum_i a_i=2 B$$ On the other hand terms at $n^{th}$ subleading order in the small $\theta$ expansion (those which scale like 
$\frac{1}{\theta^{2B-n}}$) are those for which 
$$ \sum_{i} a_i = 2 B-n$$

. 

\eqref{sumform}, \eqref{eachai} in particular express the fact that the integrand in \eqref{genformoo} is of order $\frac{1}{\theta^{2 {\tilde B}-M}}=\frac{1}{\theta^{2B}}$ in the small $\theta$ limit  (the $M$ arises from the scaling of the numerators) and that there are additional power series corrections to this leading small $\theta$ scaling behaviour.

We now turn to the bulk integral over $u$ and $v$. It is convenient to break up this integral into two regions; the first $R_1$ being disk of radius (say) A in the $u, v$ plane and the second, $R_2$, which is the complement of $R_1$. $A$ is a fixed number, independent of $\theta$, and is chosen to be smaller than the radius of convergence of the power series expansions for $u$ and $v$. \footnote{This expansion has a finite radius of convergence because it arises from the expansion of simple rational functions of $u$ and $v$.}
The point of this split is the following; in order to generate a $\sigma$ expansion of our result, it is useful to use the Taylor expansion \eqref{sumform}. However this expansion is only valid at small enough $u$, $v$. \footnote{In particular if we illegally used the expansion at large $u$ and $v$, terms that appear at high enough orders in this integrand would have divergent integrals over $u$ and $v$.}, We are allowed to use this expansion only in the region $R_1$ but not in $R_2$. 

Indeed, the appropriate Taylor expansion of the integrand in $R_2$ is in a power series in $\theta$ and $\tau$, but with $u$, $v$ being treated of order unity, i.e. $u$ and $v$ dependence being dealt with exactly and not in expansion. Performing this expansion we immediately find that the integral 
\eqref{genformoo} in $R_2$ is of order unity (in $\theta$, $\tau$ smallness or smaller. This will be parametrically smaller than the $\theta$ dependence we will obtain from region $R_1$, and so the integral over $R_2$ will be irrelevant for the determination of the behavior of the integral at leading order in $\theta$. 
The region $R_2$ is also everywhere finitely separated from the `bulk point singularity' $H_{D-2}$ of section \ref{rgzl} (this is the surface $u=v=y_1=0$, see around \eqref{coordinateshup}) and so the integral over this region is also non-singular in the bulk point $\rho \to 0$ $(a \to 0)$ limit. For this reason the integral over $R_2$ will play absolutely no role in the analysis of this paper, and  will be ignored both in the rest of this section and also all through section \ref{ire}.

For the integral over $R_1$ we employ the expansion \eqref{sumform} and perform the integral term by term. Upto corrections of order unity or smaller (that come from the fact that the region of integration in $R_1$ is bounded)\footnote{As mentioned above these corrections are non-singular at small $\rho$ and so play no role - and will be ignored - through the rest of this paper.} we can perform the integral as follows. We make
the variable change $u=\theta U$ and $v=\theta V$ to obtain 
\begin{equation}\label{sumform2}
\sum_i   \theta^{\left(2-\sum_{m} a_m \right) } 
N_{\{ a_i\} }(y_\mu, Z_i) f_{\{a_m \}}(a, y_0, y_1)
\end{equation} 
where 
\begin{equation}\label{foz}  
f_{\{ a^i \}}(a, y_0, y_i)= \int 
\frac{dU d V }{(U+a ~ y_0+i\epsilon)^{a_1} (-U+~ y_1 +i\epsilon)^{a_2} (V +a~ ~ y_0+i\epsilon)^{a_3} 
	(-V- ~ y_1+i\epsilon)^{a_4}}
\end{equation} 
Recall $a=\frac{\tau}{\theta}$. The integration  \eqref{foz} is performed over the full real line for $U$ and $V$; the correction from the finite integration range is of order unity or smaller, as we have explained above. In the integral above $y_0=\sqrt{1+y_i^2}$. The integrals over in \eqref{foz} are easily evaluated using Schwinger parameters; one obtains \cite{Heemskerk:2009pn}
\begin{equation}\label{fozAS2}
\begin{split}   
f_{\{ a^i\}}(a, y_0, y_i)&= \frac{C_{a_1,a_2,a_3,a_4}}{\left(a~y_0+y_1 +i\epsilon\right)^{a_1+a_2-1}\left(a~y_0-y_1 +i\epsilon\right)^{a_3+a_4-1}}\\
&C_{a_1,a_2,a_3,a_4}=\frac{ \Gamma \left(a_1+a_2-1\right) \Gamma \left(a_3+a_4-1\right) }{\Gamma \left(a_1\right) \Gamma \left( a_2 \right)\Gamma\left(a_3\right) \Gamma \left( a_4 \right)}
\end{split}
\end{equation}

In the small $\theta$ limit the dominant term in 
\eqref{sumform2} is of order 
$\frac{1}{\theta^{2B-2}} \sim \frac{1}{\sigma^{B-1}}$. The contribution of this term to the full integral \eqref{genformoo} is
\begin{equation} \label{finans} 
\theta^{-2 B +2} H(a, Z_i) 
\end{equation} 
where 
\begin{equation}\label{hadef} 
H(a, Z) = \int_{\mathbb{H}_{D-1}}  \sum_i N_i(y_i, Z_i) f_i(a, \sqrt{1+y_i^2}, y_1)
\end{equation} 
where the summation in \eqref{hadef} runs only over those terms for which $\sum_i a^i_m=2 B$.

The integral in \eqref{hadef} is taken over the hyperboloid $y_0^2-y_i^2=1$, more precisely on 
its branch in which 
\begin{equation}\label{bhy}
y^0= + \sqrt{1+y_i^2}, ~~~~~ 0 \leq \tau \leq \pi
\end{equation} 
\footnote{In embedding space the curve $y_0^2-y_i^2=1$ has two branches 
	$$y^0= \pm \sqrt{1+y_i^2}$$
	Each of these branches maps into an infinite number of branches in covering space. The branch 
	\eqref{bhy} is the one of relevance to this paper. This is the branch on which the bulk point singularity lies, and is also the branch on which the  $i \epsilon $ assignment in \eqref{foz} is correct.}

\subsection{Analyticity in $a$}\label{anina} 

$H(a, Z_i)$ is a function of $a$ as well as the boundary polarizations $Z_i$. In what follows we allow these polarizations to be a function of $a$, but demand that this function is chosen to be analytic. For any such choice $H(a, Z_i)={\tilde H}(a)$ and our four-point function in the Regge limit takes the form 
\begin{equation} \label{finansn}
\theta^{-2 B+2} {\tilde H}(a)
\end{equation} 
While the function ${\tilde H}(a)$ is in general  complicated, in this paper, we are concerned only with its analytic properties, which are easy to understand. 

${\tilde H}(a)$ has two singularities; the power-law `bulk point' singularity at $a=0$ and a lightcone singularity at $a=1$. 
 In section \ref{ire} we study the bulk point $a \to 0$  singularity in great detail. In this section, we focus our attention on the lightcone singularity at $a=1$.

The main concern we address in this section is the following. Given that ${\tilde H}(a)$ has a singularity at 
$a=1$, one might worry that it is possible for 
${\tilde H}(a) \propto \theta(1-a)$. If ${\tilde H}(a)$ behaved in this manner then the effective value of $B$ in 
\eqref{finans} could be smaller for $a>1$ than it is for $a<1$. In this subsection we explain that this cannot happen. 

Our result follows almost immediately once we analytically continue the function ${\tilde H}(a)$ to complex values of $a$. The fact that such an analytic continuation is clear  from the integral representations 
\eqref{hadef} and \eqref{fozAS2} (see Appendix 
\ref{reggecomp} for a class of worked examples). 
It is also expected on general grounds; the continuation to complex $a$ is effectively a continuation to complex cross ratios (see \eqref{nun}), and conformal correlators are, of course, famously analytic function of cross ratios with branch cuts at lightcone singularities (the branch cut nature of the complex singularity is explicitly displayed in a class of simple examples in Appendix \ref{reggecomp}). Elsewhere the correlators are analytic functions (see Appendix \eqref{singcor} for a review). 

Indeed the $i \epsilon$ prescription effectively tells us that we need to perform an analytic continuation even to work at physical values of parameters. In particular, we see from \eqref{fozAS2} \footnote{And using the fact that $y_0>0$ everywhere on the integration domain in \eqref{hadef}.}, the ${\tilde H}(a)$ is actually a function of the combination of variables

\begin{equation}\label{at} 
{\tilde a}=a+i{\tilde \epsilon}
\end{equation}

for an appropriate definition of ${\tilde \epsilon}>0$. It is useful to view ${\tilde H}$ as a function of ${\tilde a}$. With this convention, the function ${\tilde H}$ has its branch cut singularity exactly at ${\tilde a}=1$.  \footnote{The analytic function ${\tilde H}^{\tilde \epsilon}(a)$ is most conveniently defined to have a branch cut running from $a=1$ down to $a=0$.}. Physics instructs us to work at real values of $a$, and so at slightly complex values of ${\tilde a}$, more particularly to evaluate the function ${\tilde H}({\tilde a})$  on a contour that passes just above the real axis. 

Restated,  both for $a>1$ and $a<1$, ${\tilde H}(a)$ is the restriction of the analytic function ${\tilde H}({\tilde a})$ on the real axis, with the limit to the real axis being taken from above (i.e. from positive values of ${\rm Im}({\tilde a})$). It follows that ${\tilde H}(a)$ for $a<1$ is related to ${\tilde H}(a)$ for $a>1$ via analytic continuation, the continuation being taken in the upper half complex ${\tilde a}$ plane. Despite the singularity at $a=1$ \footnote{When we take $\epsilon \to 0$}, therefore, it follows that 
${\tilde H}(a)$ for $a>1$ and $a<1$ are analytic continuations of each other . In particular if 
${\tilde H}(a)=0$ for all $a>1$ then it follows that ${\tilde H}(a)$ also vanish for all $a<1$, and so in the limit $a \to 0$. Restated, if $H(a)$ does not vanish at small $a$, it cannot vanish identically all $a>1$.

\begin{figure}[H]
	\begin{center}
		\includegraphics[width=10cm]{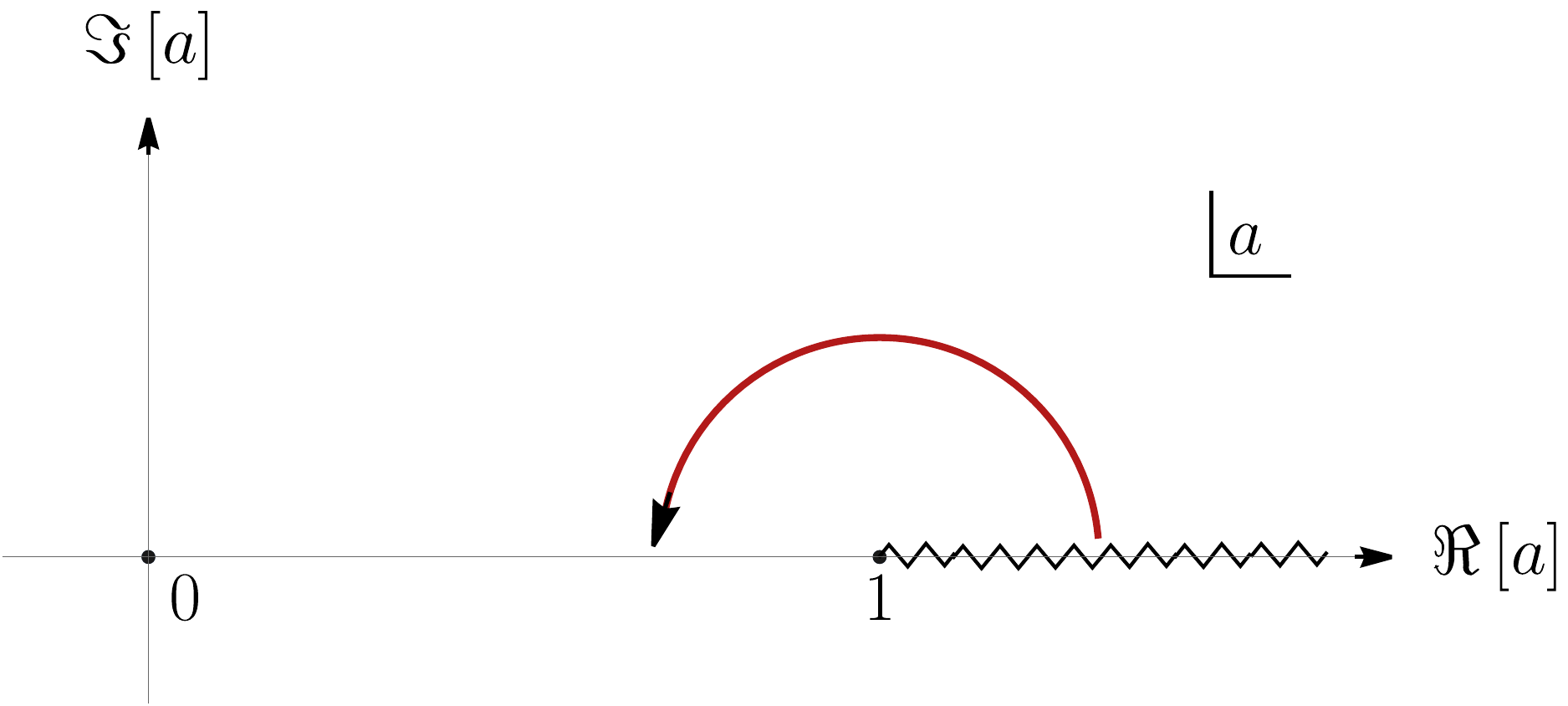}
		\caption{The analytic continuation that takes one from $a>1$ to $a<1$. }
		\label{roundz}
	\end{center}
\end{figure}

\subsection{Analyticity in the cross ratio $e^{2 \rho}$}

The discussion above about the analytic properties of the function $H(a)$ can also be understood in what may be more familiar terms when worded in terms of cross ratio function 
$e^{2 \rho}= \frac{z}{{\bar z}}$ . Recall from 
\eqref{nun} that 
\begin{equation}\label{nuno}
\begin{split} 
& e^{2 \rho} = \left( \frac{1-a-i\epsilon}{1+a +i\epsilon} \right)^2
\end{split} 
\end{equation}

Or in other words 
\begin{equation}\label{nunt}
\begin{split} 
& e^{2 \rho} = \left( \frac{{\tilde a}-1}{1+
	{\tilde a}} \right)^2
\end{split} 
\end{equation}
with ${\tilde a}$ given in \eqref{at}. If we ensure that ${\tilde a}$ has a small positive imaginary part and we take the real part of ${\tilde a}$ from greater than one less than one then the variable $\zeta =e^{ 2\rho}$ effectively circles counter-clockwise round the branch cut at $\zeta=0$ 
\footnote{To see this note that that as we pass from  ${\rm Re}({\tilde a})>1$ to ${\rm Re}({\tilde a}<1)$ with ${\rm Im}(a)=\epsilon>0$, the argument of 
	${\tilde a}-1$ changes from $0$ to $\pi$. It follows, therefore, that when this happens the argument of $({\tilde a}-1)^2$ increase from $0$ to $2 \pi$.   } 
In other words, the passage from $a>1$ to $a<1$ 
(roughly speaking a $\pi$ rotation in the complex $a$ plane) corresponds to winding counter-clockwise around the branch point at zero of the cross ratio variable $e^{2 \rho}$ (roughly speaking performing a $2 \pi$ rotation in the counter clockwise direction in the $e^{2 \rho}$ plane). The analytic continuation that takes us from $a>1$ to $a<1$ is simply the analytic continuation that takes us from just above to just below the branch cut of the correlator when viewed as a function of the  cross ratio variable $e^{2 \rho}$.
\footnote{ Noting that $z={\bar z}
	e^{2 \rho}$ and recalling that ${\bar z}$ is
	fixed and finite at $a=1$ we see that this is basically the same as the fact, established in the previous section, that passing from 
	$\tau >\theta$ to $\tau<\theta$ corresponds to circling counter-clockwise round the branch cut singularity at 	$z=0$.} The impossibility of the function ${\tilde H}(a)$ being nontrivial for $a<1$ but vanishing for $a>1$ now follows as in the previous subsection.

\begin{figure}[H]
	\begin{center}
		\includegraphics[width=10cm]{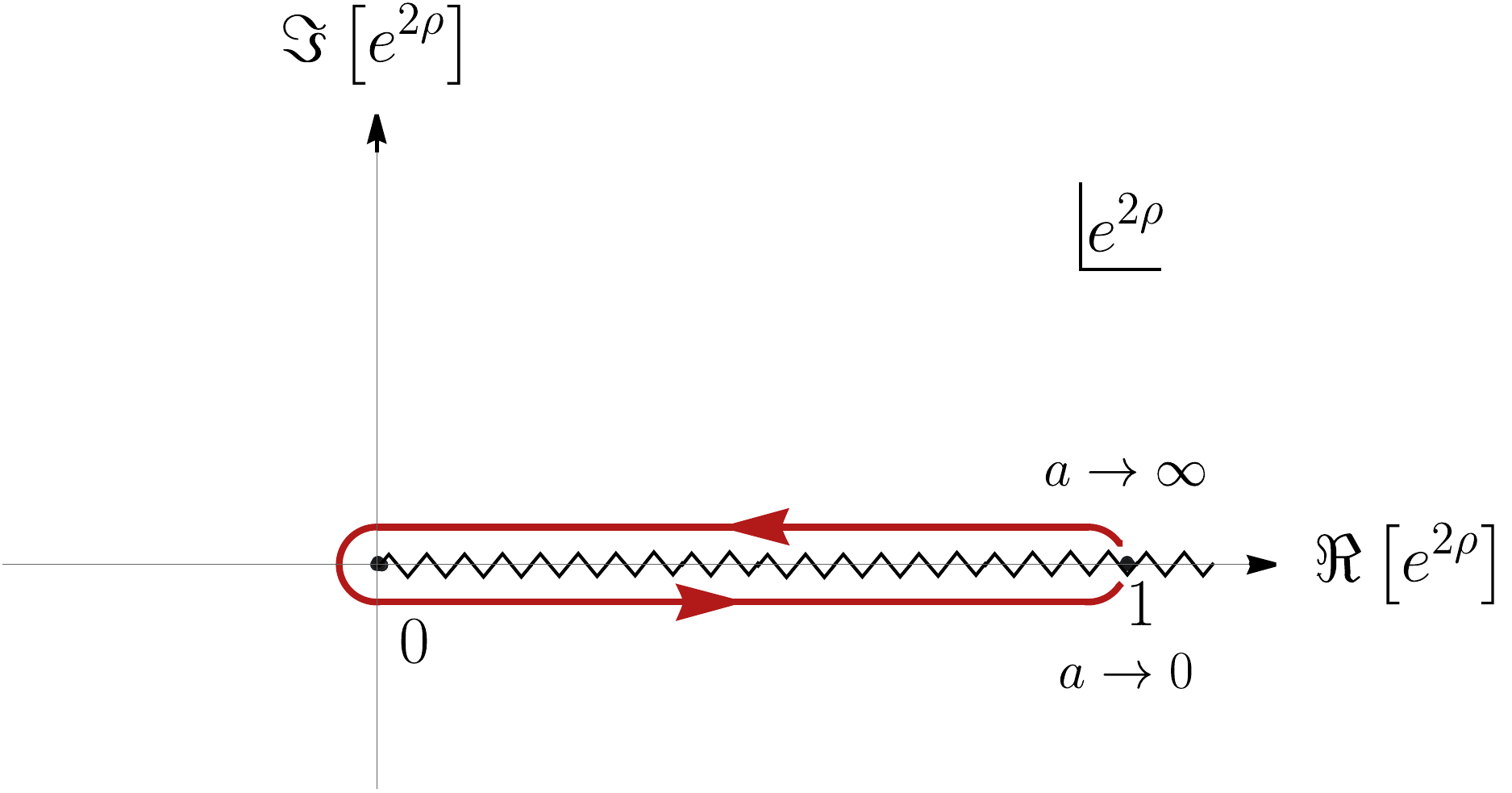}
		\caption{The path traversed in the complex plane by the cross ratio $e^{2 \rho}$ as $a$ moves from greater to one to less than one.  $e^{2 \rho}$ (and so $z$) circles round the branch point at zero in  a counter-clockwise direction. }
		\label{roundz2}
	\end{center}
\end{figure}

All the analytic properties described above are explicitly illustrated in Appendix \ref{singcor}
in the context of a simple example; the correlator generated by a $\phi^4$ interaction for four identical bulk scalars dual to boundary operators of dimension $\Delta$ 
in Appendix \ref{reggecomp}. 

\subsection{Regge scaling of normalized correlators}

In this paper our interest lies mainly in the normalized correlator \eqref{normfp} (because this is the correlator that is constrained by the chaos bound). Recall this normalized correlator is only defined when $\Delta_3=\Delta_1$ and $\Delta_4=\Delta_2$. Defining
\begin{equation}\label{AB}
A'=B-\Delta_1-\Delta_2
\end{equation} 
it follows immediately from \eqref{finansn} and
\eqref{itlim} that the leading behavior of normalized four point function \eqref{normfp} in the Regge limit takes the form 
\begin{equation} \label{normlim}
\frac{H(a)}{\theta^{2A'-2}}
\end{equation} 
The fact that \eqref{normlim} holds uniformly - i.e. with the same value of $A$ both for $a>1$ and $a<1$, i.e. with the same value of $A$ in both the  Causally Scattering and the Causally Regge sheets - is the a central result of this section.




\section{The $\rho \to 0$ ($\tau \to 0$) limit} \label{rgzl}

As we have mentioned in the introduction, it is possible to show on very general grounds that the correlation function studied in this paper is an analytic function of the insertion points away from $\tau=1$ (see the previous section) and $\tau=0$ (see Appendix \ref{singcor} following  \cite{Maldacena:2015iua}). In this section, we will study the singularity at $\tau=0$ more closely, and in particular, determine its precise structure including its coefficient. 
The analysis of this section follows section 6 of \cite{Maldacena:2015iua}, and also overlaps with \cite{Gary:2009ae, Penedones:2010ue, Okuda:2010ym}. 

\subsection{Modulated propagators} 

Recall that every boundary to bulk propagator is a polynomial times 
\begin{equation}\label{pdenom}
\frac{1}{(-2P.X+i\epsilon)^A} 
\end{equation} 
for some value of $A$. In this section we adopt the following strategy. We choose an arbitrary function $f(a)$ with the following properties 
\begin{equation}\label{propf}
\begin{split} 
f_A(a) &= 1 -a^{A}P(a), ~~~ a \ll 1 \\
P(a) &= \sum_{m=0}^\infty c_m a^{2m} \\
\lim_{a \to \infty}& f_A(a) =0\\
\end{split} 
\end{equation} 
In words, the function $f$ vanishes at large values of its argument. It tends to unity at $a=0$, and the deviations from unity take a very particular form. An example of an $f$ function with these properties is 
\begin{equation}\label{exampleoff}
f_A(a)= 1 -\frac{a^{A}}{(1+ a^2)^\frac{A}{2}}
\end{equation} 

Given any such $f_A$ function we next use the trivial identity 
\begin{equation}\label{obx}
\frac{1}{(x+i\epsilon)^A}=
\frac{1-f_A(\frac{x}{L})}{(x+i\epsilon)^A}
+ \frac{f_A(\frac{x}{L})}{(x+i\epsilon)^A}
\end{equation} 
to split every propagator into two pieces. At the moment $L$ is an arbitrary constant. We will, choose its value so that 
\begin{equation}\label{rangeL}
\tau \ll L \ll 1
\end{equation} 
(this is possible in the small $\tau$ limit, the regime of interest to this section). 

A key point is that the first term on the RHS of 
\eqref{obx} is non-singular at $x=0$; we call this the smooth piece.  On the other hand, the second term in \eqref{obx} continues to be singular at $x=0$; indeed this term (which we call the modulated term) is essentially indistinguishable from \eqref{pdenom} for $x \ll L$. 

Now  bulk contact interaction contribution to any correlator is given by the product of four boundaries to bulk propagators (to the bulk point $X$) sewn together by the bulk interaction and integrated over the bulk. Our strategy is to replace each of the four bulk to boundary propagators by the sum of two terms (the regular part of the propagator and the singular part of the propagator) in the integrand above  using the identity \eqref{obx}. The integrand now breaks up into $2^4=16$ terms, depending on whether we retain the regular or modulated part of each propagator. Upon performing the integral over $X$, it follows immediately from the analysis 
of Appendix \ref{singcor} that the integral of any term in which even one of the propagators is replaced by its regular part evaluates to an expression that is non-singular at $\tau=0$. The singularity in our correlator - the object of study of the current section - comes entirely from the integrand with every propagator replaced by the corresponding modulated propagator. 

Let us summarize. We have demonstrated that the singular term in the correlation is unaffected if we make the replacement 
\begin{equation}\label{repstrat}
\frac{1}{(-2P.X+i\epsilon)^A}  \rightarrow
\frac{f_A(\frac{-2P.X}{L})}{(-2P.X+i\epsilon)^A}
\end{equation} 
in every propagator that appears in the integrand of our correlation functions, i.e. if we replace every propagator by a modulated propagator. 

All through this section, we will make the replacement \eqref{repstrat} on every propagator in our integrand.  The utility of this manoeuvre is the following. As the envelope function, $f_A(P.X)$ decays away from the lightcone of the boundary point $P$, once we make the replacement \eqref{repstrat}, it follows immediately that 
the singularity of our correlator receives contributions only from an envelope of width 
$L$ around the common intersections of the lightcones of the boundary points $P_i$, a fact that simplifies our analysis below. 

\subsection{Wave representation for propagators}

The next step in our procedure is to express the modulated propagators as a sum over propagating waves (a sort of Fourier transform). This may be accomplished (\cite{Okuda:2010ym, Maldacena:2015iua}) using the identity 
\begin{equation}\label{ident}
\frac{1}{(-2P.X+i\epsilon)^{A}}= \frac{e^{-\frac{ i \pi  A}{2} }}{2^A\Gamma ( A )}\int_0^\infty d \omega ~\omega^{A-1} e^{i \omega (-P.X+i\epsilon) }
\end{equation} 
It follows that the modulated spin $J$ propagator 
can be written as 
\begin{equation}\label{bbp}
G_J(P, Z, X, W)
=\frac{\mathcal{C}^D_{\Delta,J} e^{-\frac{i\pi (\Delta+J)}{2}}}{2^{\Delta}\Gamma(\Delta+J)} f(\frac{P.X}{L}) \left( -Z.W ~P.X+{Z.X~ W.P} \right)^J
\int d \omega~ \omega^{\Delta + J-1}  e^{i \omega (-P.X+i\epsilon) }
\end{equation} 

where \begin{equation}\label{cdeltaj}
\mathcal{C}^D_{\Delta,J} =\frac{(J+\Delta-1)\Gamma(\Delta)}{2\pi^{\frac{D}{2}}(\Delta -1)\Gamma(\Delta-\frac{D}{2}+1)}
\end{equation}
\eqref{bbp} represents the propagator as a sum over waves of 
momenta $\omega P_M$. These plane waves are multiplied by the envelope function $f$ polynomial dressings. However these factors (as well as the curvature of the underlying AdS background) all vary on scales ranging from $L$ to unity. On the other hand the singularity of our correlator will turn out to have its origin in $\omega$ of order $\frac{1}{\tau} \gg \frac{1}{L} \gg 1$. At distance scales $\tau$ the multiplying factors are effectively constant, and $AdS_{D+1}$ space is effectively flat, so at these scales \eqref{bbp} is literally a Fourier Transform representation of the propagator. 

Following   \cite{Maldacena:2015iua} we will now explain how the wave representation of propagators, 
\eqref{bbp}, can be used to  find an expression for the singularity in $\tau$ of the correlators. We begin by considering very simple situations but building up to more complicated ones.

\subsection{Scalar $\phi^4$ correlator in $AdS_3$} 

Let us first study the correlation function induced by a $\phi_1\phi_2 \phi_3\phi_4$ bulk interaction between 4 scalars of 
dimension $\Delta_1$, $\Delta_2$, $\Delta_3$ and
$\Delta_4$ in $D=2$ (i.e. 2 boundary or 3 bulk dimensions). 

We need to evaluate the integral
\eqref{bbp} is  
\begin{equation} \label{relsimp}
G_{\rm sing}=\left(\prod_a\frac{\mathcal{C}_{\Delta_a} e^{-\frac{i\pi \Delta_a}{2}}}{2^{\Delta_a}\Gamma(\Delta_a)}\right) \int d X \left( \prod_a d\omega_a \omega_a^{\Delta_a -1} f_{\Delta_a}(P_a.X/L) \right) 
e^{-i X. \left( \sum_a \omega_a P_a \right)}
\end{equation}

where $\mathcal{C}_{\Delta_a}$ is short notation for $\mathcal{C}_{\Delta_a,J=0}$.\\

Recall that there are no bulk points $X$ that solve the equation  $P_i.X=0$ for all $i$ except when $\tau=0$. When $\tau=0$, the unique bulk point that obeys this equation in $D=2$ (and also has global time between $0$ and $\pi$) is 

\begin{equation}\label{xnot}
X=X_0=(0, 1, 0, 0).
\end{equation} 
It follows that when $\tau \gg L$ the envelopes around the four lightcones (the lightcones of $P_1, P_2, P_3, P_4$) never intersect and the integral in \eqref{relsimp} is very small. On the other hand when $ \tau \ll L$ the four envelope functions overlap in a spacetime region in $AdS_3$  of `size' $L$ (volume $L^3$) centred around $X_0$. We refer to this region as `the elevator'. 

It is useful to set  
\begin{equation}\label{expofm}
X=X_0 + x
\end{equation} 
As only values of $x$ with $x$ less than or of order $L$ lie in the `elevator' (and so contribute significantly to the integral) and 
as  $L \ll 1$, $x$ effectively a tangent space coordinate about the point $X_0$ in $AdS_3$. It follows that the space parameterized by the coordinate $x$ is effectively flat.  Now 
\begin{equation}\label{projdef} 
{\cal P}_{X_0}^{MN}= \eta^{MN}+X_0^M X_0^N
\end{equation} 
is the projector in $\mathbb{R}^{D,2}$ that projects vectors in this space orthogonal to $X_0$. 
Expanding the equation $(X_0+x)^2=-1$ to first order in $x$, we conclude that 
\begin{equation}\label{xroj}
{\cal P}_{X_0}^{MN} x_M=x_N
\end{equation} 
Now the integral \eqref{relsimp} evaluates to 
\begin{equation}\label{sumgg}
G_{\rm sing}= \left(\prod_a\frac{\mathcal{C}_{\Delta_a} e^{-\frac{i\pi \Delta_a}{2}}}{2^{\Delta_a}\Gamma(\Delta_a)}\right) \int \left( \prod_a d\omega_a \omega_a^{\Delta_a -1} \right) 
e^{-i \sum_a \omega_a P_a. X_0 } (2\pi)^3 {\tilde \delta}^{3}(\sum_a k_a)
\end{equation} 
where 
$$k_a^M = \omega_a ({\cal P}_{X_0})^{M}_N P_a^N$$
${\tilde \delta}$ is a $\delta$ function broadened in a Gaussian manner over scale of order 
$\delta \omega \sim \frac{1}{L}$. 
\footnote{The precise form of this function can be systematically evaluated in a saddle point approximation. For our purposes (i.e. leading order computations in the small $\tau$ limit) , the only thing about about this function that is relevant is that it integrates to unity, and is nonzero only over a range of of order $\frac{1}{\delta L}$ of its arguments.  }

In terms of the vector 
\begin{equation}\label{sumpop}
P= \sum_a \omega_a P_a 
\end{equation} 
\eqref{sumgg} can be recast as
\begin{equation}\label{sumggn}
G_{\rm sing}=(2\pi)^3 \left(\prod_a\frac{\mathcal{C}_{\Delta_a} e^{-\frac{i\pi \Delta_a}{2}}}{2^{\Delta_a}\Gamma(\Delta_a)}\right) \int \left( \prod_a d\omega_a \omega_a^{\Delta_a -1} \right) 
e^{-i P. X_0 }
{\tilde \delta}^{3}\left( ({\cal P}_{X_0})^{M}_N P_N \right) 
\end{equation}

The ${\tilde \delta}$ function in \eqref{sumggn} clicks 
when $({\cal P}_{X_0})^{M}_N P_N$ vanishes. This 
happens on a one-parameter set of $\omega^a$. \footnote{The counting is the following. The space in which the vectors $k_a^M$ varies is the tangent space of the point $X_0$ and so is 3 dimensional. Any four vectors in 3-dimensional space are linearly dependent and so obey an equation of the form $\zeta_a k_a=0$. 
	It follows that the $\delta$ function in 
	\eqref{sumggn} clicks for $\omega_a=\omega \zeta_a$. Note all this is true at generic $\tau$, despite the fact that the four 4 dimensional vectors $P_a$ are linearly independent unless $\tau=0$.} The values of $\omega^a$ at which this happens is easy to deduce. Expanding to first order in  $\tau$ we find 
\begin{equation}\label{sumoverP}
P_1+P_2+P_3+P_4=2 \tau X_0 
\end{equation} 
so that 
\begin{equation}\label{projP}
\sum_a k_a=0
\end{equation} 
It follows that the ${\tilde \delta}$ function is 
nonzero if and only if for every value of $a$  
\begin{equation} \label{tildem}
\omega_a= \omega 
\end{equation} 
\footnote{As the delta function has a spread of order $1/L$, the solution \eqref{tildem} also has the same fuzz. This fuzz will not be important in what follows and we ignore it.}
With these values of $\omega_a$ we find  
\begin{equation}\label{forvec} \begin{split} 
&k_1=\omega (1,0, 1, 0)\\
&k_3=\omega( 1,0,  -1, 0)\\
&k_2=-\omega(1, 0, \cos \theta, \sin \theta)\\
&k_4=-\omega(1, 0, -\cos\theta, -\sin \theta)
\end{split} 
\end{equation} 

Now  it is easy to check that
\begin{equation}\label{intev}
\int d\omega_1 d\omega_2 d \omega_3 d \omega_4~
{\tilde \delta}^{3}\left( ({\cal P}_{X_0})^{M}_N P_N \right) = 
-\frac{1}{2\sin\theta} \int d \omega 
\end{equation} 
(where $\omega$ is defined in \eqref{tildem})
and so \eqref{sumggn} reduces to 
\begin{equation}\label{nnt}
G_{\rm sing}=-\frac{4\pi^3 e^{-\frac{i\pi \Delta}{2}}}{\sin\theta} \left(\prod_a\frac{\mathcal{C}_{\Delta_a} }{2^{\Delta_a}\Gamma(\Delta_a)}\right) \int_0^\infty d\omega  \omega^{\Delta-4} e^{-2i \omega \tau}
\end{equation} 
where, $$\Delta = \sum_a \Delta_a$$

Note that \eqref{nnt} would have diverged at large $\omega$ in the absence of the phase factor $e^{-i \omega \tau}$. The phase factor regulates this divergence at $\omega \sim \frac{1}{\tau}$, converting it into a singularity at $\tau=0$. 
As promised the singularity has its origin in waves with  
$$\omega \sim 1/\tau$$ 
and so is a UV effect at small $\tau$. The precise form of the singularity is 

\begin{equation}\label{finrest_test}
\begin{split}
G_{\rm sing} &=-\frac{4\pi^3 e^{-\frac{i\pi \Delta}{2}}}{\sin\theta} \left(\prod_a\frac{\mathcal{C}_{\Delta_a} }{2^{\Delta_a}\Gamma(\Delta_a)}\right)\left( - i e^{-\frac{i \pi  \Delta}{2}} \Gamma (\Delta -3)\frac{1}{(2\tau) ^{\Delta -3}}\right)\\
&=\frac{4i\pi^3 e^{-i\pi \Delta}}{\sin\theta} \left( \Gamma (\Delta -3)\prod_a\frac{\mathcal{C}_{\Delta_a} }{2^{\Delta_a}\Gamma(\Delta_a)}\right)\frac{1}{(2\tau) ^{\Delta -3}}
\end{split}
\end{equation}
\footnote{The final result \eqref{finrest_test} is correct only at leading order in the limit $\tau \to 0$. The corrections to this answer come from the fact that the ${\tilde \delta}$ function is not a completely sharp but has a width $\delta w$. It follows, for instance, that the integrand in \eqref{nnt} is actually a power series in $\omega$ of the schematic form 
	$$\omega^a \rightarrow \omega^a + \omega^{a-1} \delta \omega + \omega^{a-2} \delta \omega^2 + \ldots $$
	As every factor of $\omega$ is effectively of order 
	$\frac{1}{\tau}$ while $\delta \omega$ is independent of $\tau$, these correction give corrections to the singularity in \eqref{finrest_test} of order ${\cal O}(1/\tau^{\Delta_1+\Delta_2+\Delta_3 +\Delta_4 -m})
	$ where $m$ is an integer.}

\subsection{General scalar contact correlator in $AdS_3$} 

The generalization of the discussion of the previous subsubsection to the study of the singularity of a scalar correlator resulting from a more general bulk contact interaction than $\phi^4$ (i.e. some number of derivatives acting on the $\phi$ 
fields) is completely straightforward.

The analogue of \eqref{sumgg} in this case is 
\begin{equation}\label{sumggn2}
G_{\rm sing}= \left( \prod_a \tilde{\mathcal{C}}_{\Delta_a} \int d\omega_a  \omega_a^{\Delta_a -1} \right) S(k_a) e^{-i \sum_a \omega_a P_a. X_0 } (2\pi)^3 {\tilde \delta}^{3} (\sum_a k_a)
\end{equation}

where,
\begin{equation}
\begin{split}
\tilde{\mathcal{C}}_{\Delta,J} &= \frac{\mathcal{C}_{\Delta,J} e^{-\frac{i\pi (\Delta+J)}{2}}}{2^{\Delta}\Gamma(\Delta+J)}\\
\text{and, } \tilde{\mathcal{C}}_{\Delta} &= \tilde{\mathcal{C}}_{\Delta,0}
\end{split}
\end{equation}

and, $S(k_a)$ is the contact interaction evaluated on the waves $k_a=\omega P_a$. Evaluating the integral over $\omega_a$, as in the previous subsection, gives the analogue of 
\eqref{nnt} 
\begin{equation}\label{nntn}
G_{\rm sing}=-\frac{4\pi^3}{\sin\theta} \left( \prod_a \tilde{\mathcal{C}}_{\Delta_a}\right) \int d\omega  \omega^{\Delta-4} S(\omega) e^{-2i \omega \tau}
\end{equation} 
where $S(\omega)$ is now the interaction contact term evaluated on the waves \eqref{forvecm}, i.e. the invariant transition amplitude of the S matrix  generated in flat space by the contact term in question for the scattering momenta \eqref{forvecm}.

The vectors on which $S(\omega)$ is evaluated are 
\eqref{forvec} where they are displayed as momenta in the embedding space $\mathbb{R}^{2,2}$. All the 
vectors in \eqref{forvec}, however, are orthogonal to $X_0$ and so lie in its tangent space. This tangent space consists of the collection of 
$\mathbb{R}^{2,2}$ vectors with second component set to zero. Effectively these vectors lie in the $\mathbb{R}^{2,1}$ obtained by deleting the second component of all vectors in \eqref{forvec}. With this convention
\begin{equation}\label{forvecm} \begin{split} 
&k_1=\omega (1, 1, 0)\\
&k_3=\omega( 1, -1, 0)\\
&k_2=-\omega(1, \cos \theta, \sin \theta)\\
&k_4=-\omega(1, -\cos\theta, -\sin \theta)
\end{split} 
\end{equation} 
\footnote{For complete clarity we reiterate that the first coordinate of each of the vectors here is timelike, the second and third coordinates are spacelike. The vectors in \eqref{forvecm} are obtained from those in \eqref{forvec} by deleting the zero in the second component of all vectors in \eqref{forvec}.}	

Note that $k_1$ and $k_3$ are 3 momenta with positive energy, and so particles $1$ and $3$ are initial states. On the other hand the momenta $k_2$ and $k_4$ have negative energy, so that particles $2$ and $4$ are the final states. The 3 momenta \eqref{forvecm} are conserved (this is, of course, a consequence of the delta function in \eqref{sumggn2}). And $S(\omega)$ in \eqref{nntn} is simply the S matrix \footnote{Or more precisely the invariant transition amplitude, the S matrix with the momentum conserving delta function deleted.} for the scattering process with the momenta \eqref{forvec}. The Mandlestam variables for this scattering process are given by 
\begin{equation}\label{scatdef} \begin{split} 
&s \equiv -(k_1+k_3)^2 = 4 \omega^2,\\ 
&t \equiv -(k_1+k_2)^2 = -2\omega^2(1-\cos \theta)\\
&u\equiv (k_1+k_4)^2= -2\omega^2(1+ \cos \theta)
\end{split} 
\end{equation}

From \eqref{crst} we see that $\sin \theta = 2 \sqrt{\sigma (1-\sigma)}$ so that \eqref{nntn} can be rewritten more invariantly as 
\begin{equation}\label{nntn2}
G_{\rm sing}=-\frac{2\pi^3}{\sqrt{\sigma(1-\sigma)}} \left( \prod_a \tilde{\mathcal{C}}_{\Delta_a}\right) \int d\omega  \omega^{\Delta-4} S(\omega) e^{-2i \omega \tau}
\end{equation} 

In the case of the scattering of four scalars, the S matrix is a function only of $s$ and $t$. 
Let the invariant amplitude as a function of 
$s$ and $t$ be denoted by $T(s,t)$. It follows that 
\begin{equation}\label{sitt}
{\cal S}(\omega)=i T\left(4 \omega^2, -2\omega^2(1-\cos \theta) \right)
\end{equation} 
It follows that \eqref{nntn} can be rewritten in terms of the invariant transition amplitude $T(s,t)$ generated by the contact interaction as  
\begin{equation}\label{nntre}
\begin{split}
G_{\rm sing} &=-\frac{i 2\pi^3}{\sqrt{\sigma(1-\sigma)}} \left( \prod_a \tilde{\mathcal{C}}_{\Delta_a}\right)  \int d\omega \omega^{\Delta -4}
e^{-2i\omega \tau} T\left(4 \omega^2, -2\omega^2(1-\cos \theta) \right)
\end{split}
\end{equation} 
Let us suppose that the contact term in question is of $r^{th}$ order in spacetime derivatives. Then
\begin{equation}\label{rhthord}
T\left(4 \omega^2, -2\omega^2(1-\cos \theta) \right)=\omega^r T\left(4, -2(1-\cos \theta) \right)
\end{equation}
so that 
\begin{equation}\label{nntre1}
\begin{split}
G_{\rm sing} &=-\frac{2\pi^3}{\sqrt{\sigma(1-\sigma)}} \left( \prod_a \tilde{\mathcal{C}}_{\Delta_a}\right)  \int d\omega ~\omega^{\Delta +r -4}
e^{-2i\omega \tau } (iT\left(4, -2(1-\cos \theta) \right))\\
&=-\frac{2\pi^3 (iT\left(4, -2(1-\cos \theta) \right))}{\sqrt{\sigma(1-\sigma)}} \left( \prod_a \tilde{\mathcal{C}}_{\Delta_a}\right)  \int d\omega ~\omega^{\Delta +r -4}
e^{-2i\omega \tau }\\
&=-\frac{2\pi^3 (iT\left(4, -2(1-\cos \theta) \right))}{\sqrt{\sigma(1-\sigma)}} \left( \prod_a \tilde{\mathcal{C}}_{\Delta_a}\right)   \left( - i e^{-\frac{i \pi  (\Delta+r)}{2}} \Gamma (\Delta+r -3)\frac{1}{(2\tau) ^{\Delta+r -3}}\right)
\end{split}
\end{equation}

which simplifies to,

\begin{equation}\label{finrestn1}
\begin{split}
G_{\rm sing} &= \frac{2i\pi^3 e^{-\frac{i \pi  (\Delta+r)}{2}}}{\sqrt{\sigma(1-\sigma)}} \left( \prod_a \tilde{\mathcal{C}}_{\Delta_a}\right)  \Gamma (\Delta+r -3) \frac{(iT\left(4, -2(1-\cos \theta) \right))}{(2\tau)^{\Delta+r-3}}
\end{split}
\end{equation} 

Using the relationship \eqref{crst}, \eqref{finrestn1} can be rewritten as 

\begin{equation}\label{finrestn_TEST}
\begin{split}
G_{\rm sing} &= 2i\pi^3 \frac{e^{-\frac{i \pi  (\Delta+r)}{2}}}{2^{\Delta+r-3}} \left( \prod_a \tilde{\mathcal{C}}_{\Delta_a}\right)  \Gamma (\Delta+r -3)~~~ \frac{\sqrt{1-\sigma}^{\left( \Delta+r-4 \right)}  (iT\left(4, -4\sigma \right)) }{\sigma^{\frac{\Delta + r-2}{2}} (-\rho)^{\Delta+r-3}}
\end{split}
\end{equation} 	
\eqref{finrestn_TEST} is a completely explicit expression for the leading singularity at small $\rho$ of the four point function generated by a particular contact term in $AdS_{3}$ in terms of the flat space S matrix generated 
by the same contact term. \eqref{gsingnn2} agrees perfectly with the  expression for the S-matrix in terms of the coefficient of singularity of the CFT four point function as derived using a slightly different method (consisting of producing localized bulk waves by smearing boundary insertion points) in \cite{Gary:2009ae} (see eq (3.37)).

As an aside let us highlight an initially puzzling aspect of \eqref{finrestn_TEST}. As $\Delta$ is not necessarily an integer, the correlator \eqref{finrestn_TEST} at fixed $\sigma$ has a branch cut around $\rho=0$ in the complex $\rho$ plane. This
is puzzling, and branch cuts in correlation functions in conformal field theories are usually associated with operator ordering ambiguities across lightcones; however, there is no causal significance to the point $\rho=0$ (the boundary separation between no two points changes from spacelike to timelike across $\rho=0$). We present a brief discussion of this question in Appendix \ref{rous}.

\subsection{General contact scalar interaction in $AdS_{D+1}$} \label{gcsia}

The new element in this story when $D>2$ is that the boundary lightcones intersect over a manifold ($\mathbb{H}_{D-2}$) rather than at a point. 

The integrand in the analogue of the expression \eqref{relsimp} is now nonzero everywhere in an envelope around an $\mathbb{H}_{D-2}$. To describe this hyperboloid it is useful to choose coordinates so that 
\begin{equation}\label{coordinateshup}
X_M=(V^0, Y^0, V^1, V^2, Y^i), ~~~~i=1 \ldots D-2
\end{equation} 
\footnote{ 
	Let us recall that the space spanned by the $P_a$ is $\mathbb{R}^{2,2}$ (preserving the symmetry group $SO(D-2)$) when $\tau \neq 0$, but is 
	$\mathbb{R}^{2,1}$ (preserving the symmetry group $SO(D-2,1)$) when $\tau=0$. The $SO(D-2,1)$ is the subgroup that rotates 
	the $Y_\mu$ coordinates into each other in the usual Minkowskian fashion. The $SO(D-2)$ rotates 
	the $Y^i$ into each other.  }
The hyperboloid over which boundary lightcones intersect is located at $V^\mu=0$. Points on the hyperboloid are parameterized by the coordinates 
$Y^\mu$, subject to the relation
$$Y^2=-1.$$

It is useful to proceed as follows. We break up the $\mathbb{H}_{D-2}$ into little cells of volume $V_{D-2}$. Like the variable $L$ earlier in this section, these cells are chosen to be large compared to $\tau$ but small compared to unity. Restricting ourselves to a particular cell and performing the integral over $x$ we find 
\begin{equation}\label{sumggnn}
G_{\rm sing}=  \left( \prod_a \tilde{\mathcal{C}}_{\Delta_a}\right) \int \left( \prod_a d\omega_a \omega_a^{\Delta_a -1} \right) S(k_a)
e^{i \sum_a \omega_a P_a. X_0 } (2\pi)^{(D+1)}
{\tilde \delta}^{D+1}(\sum_a k_a)
\end{equation} 
Note that \eqref{sumggnn} is the same as  \eqref{sumggn2} except for the replacement
\begin{equation}\label{tdfm}
{\tilde \delta}^{3}\left( ({\cal P}_{X_0})^{M}_N P_N) \right) \rightarrow 
{\tilde \delta}^{D+1}\left( ({\cal P}_{X_0})^{M}_N P_N) \right)
\end{equation}
The manipulations leading up \eqref{nntre} continue 
to work, except that the integral over 3 of the four $\omega_a$ variables (see \eqref{sumgg}) are sufficient 
to use up only 3 of the $D-1$ ${\tilde \delta}$ functions on the RHS of \eqref{tdfm}. The analogue of \eqref{intev} is 
\begin{equation}\label{intevn}
(2\pi)^{(D+1)}\int d\omega_1 d\omega_2 d \omega_3 d \omega_4~
{\tilde \delta}^{D+1}\left( ({\cal P}_{X_0})^{M}_N P_N \right) = 
-\frac{(2\pi)^{(D+1)}\delta^{D-2}(0)}{2\sin\theta}  \int d \omega = -(2 \pi)^3 \frac{V_{D-2} }{2\sin\theta} \int d \omega
\end{equation}

We now need to sum the results \eqref{intevn} over all the cells in $\mathbb{H}_{D-2}$. Note that this summation is weighted by the volume of those cells. As the size of each cell is small compared to the $AdS$ radius (unity in our units) the summation is well approximated by an integral and \eqref{nntre} turns into 
\begin{equation}\label{nntre2}
G_{\rm sing}=-\frac{2\pi^3 \left( \prod_a \tilde{\mathcal{C}}_{\Delta_a}\right)}{ \sqrt{\sigma(1-\sigma)}}
\int_{\mathbb{H}_{D-2}} \sqrt{g_{D-2}} ~d^{D-2} X \int d\omega \omega^{\Delta -4}
e^{i\omega P.X } {\cal S}\left(\omega \right)
\end{equation} 
where $\sqrt{g_{D-2}} d^{D-2}X$ is the volume element on $\mathbb{H}_{D-2}$ where $S(k)$ is the S matrix 
(more precisely invariant transition amplitude) for four scalars with momenta 
\begin{equation}\label{forvecg} \begin{split} 
&k_1=\omega (1,0, 1, 0, {\vec 0})\\
&k_3=\omega( 1,0,  -1, 0, {\vec 0})\\
&k_2=-\omega(1, 0, \cos \theta, \sin \theta, {\vec 0})\\
&k_4=-\omega(1, 0, -\cos\theta, -\sin \theta, {\vec 0})
\end{split} 
\end{equation} 

Note that all the vectors $k_a^M$ have vanishing components in all $Y_M$ directions. It follows that the vectors $k_a$ lie in the tangent space of the $\mathbb{H}_{D-2}$ at every value of $Y^\mu$. 
It follows that \eqref{scatdef} and \eqref{sitt} apply without modification to this $D+1$ dimensional scattering process. Note, in particular, that in this case $S(k)$ is independent of the point $Y_\mu$ on the hyperboloid. For this reason the integral over the hyperboloid in \eqref{nntre2} is easily performed (the same will not be true for scattering of spinning particles, as we will see in the next section). To do this is is useful to parameterize points on $\mathbb{H}_{D-2}$ by 
\begin{equation}\label{yinzeta}
Y_i = \sinh \zeta {\hat n}_i, ~~~ Y_0=\cosh \zeta 
\end{equation}
where ${\hat n}_i$ is a unit vector in $\mathbb{R}^{D-2}$. 
It is easily verified that 
$$ d^{D-2} X= d\zeta ~\sinh^{D-3} \zeta 
~d \omega_{D-3}$$. It follows from \eqref{sumoverP} that 
$$X. P= -2 \tau \cosh \zeta $$
The integral over angles produce the volume of the unit $D-3$ sphere, $\Omega_{D-3}$. It remains to perform the integrals over $\zeta$ and $\omega$. While we can perform these integrals in any order, it is easier to perform the (elementary) integral over $\omega$ first. 
Assuming that it is of $r^{th}$ order in derivatives, we find 
\begin{equation}\label{gsingnn}
G_{\rm sing}=i(2\pi^3)\left( \prod_a \tilde{\mathcal{C}}_{\Delta_a}\right)\Omega_{D-3}e^{-\frac{i(\Delta +r)}{2}}\frac{\Gamma (\Delta +r -3)}{2^{\Delta+r-3}} 
\frac{\sqrt{1-\sigma}^{\left( \Delta+r-4 \right)} ~
	(i{T}\left(4, -4 \sigma \right)) }{\sigma^{\frac{\Delta + r-2}{2}} (-\rho)^{\Delta+r-3}}
\int  d\zeta \frac{\sinh^{D-3} \zeta}{ \cosh \zeta^{\Delta +r-3 }}
\end{equation} 
The integral over $\zeta$ is now simply a finite number. Our final answer is 
\begin{equation}\label{gsingnn2} \begin{split} 
G_{\rm sing}&=\left[i(2\pi^3)N_{D, \Delta} \left( \prod_a \tilde{\mathcal{C}}_{\Delta_a}\right)\Omega_{D-3}e^{-\frac{i(\Delta +r)}{2}}\frac{\Gamma (\Delta +r -3)}{2^{\Delta+r-3}}\right]
\frac{\sqrt{1-\sigma}^{\left( \Delta+r-4 \right)} ~
		(i{T}\left(4, -4 \sigma \right)) }{\sigma^{\frac{\Delta + r-2}{2}} (-\rho)^{\Delta+r-3}},\\
N_{D, \Delta}&=\int_0^\infty  d\zeta \frac{\sinh^{D-3} \zeta}{ \cosh \zeta^{\Delta +r-3 }}=\frac{\Gamma \left(\frac{D-2}{2}\right) \Gamma \left(\frac{1}{2} (\Delta+r-D )\right)}{2 ~\Gamma \left(\frac{1}{2} (\Delta+r -2)\right)},\qquad \Omega_{n}=\frac{2\pi^{\frac{n+1}{2}}}{\Gamma\left(\frac{n+1}{2}\right)} \\
\end{split} 
\end{equation} 
As in the case $D=2$ \eqref{gsingnn2} agrees exactly with the expression for the S-matrix in terms of the coefficient of singularity of the CFT four point function derived in \cite{Gary:2009ae} (see eq (3.37)) using slightly different methods. The  method of derivation employed in this paper will allow for easy generalization to the scattering of massless spinning particles in the rest of this section.

As in the previous subsection, it might seem odd to the reader that \eqref{gsingnn2} displays a branch cut type singularity at $\rho=0$, a point at which no causal relations get changed. We investigate this question in Appendix \ref{rgtt}.

\subsection{Contact interactions involving gauge bosons and gravitons}

The analysis of the previous subsection goes through when studying photons (or non-abelian gauge bosons) and gravitons with a few extra twists. From the point of view of the analysis of this section, the main difference between gauge bosons, gravitons, and scalars lies in their boundary to bulk propagators. 

\subsubsection{The spin one propagator}

Let us denote the propagator of a spin $J$ particle of dimension $\Delta$ by $(G_\Delta)_{A_1 \ldots A_J}^{M_1 \ldots M_J}$ where $A_i$ are the boundary indices and $M_J$ are the bulk indices. Then (see Appendix \ref{bbprop} and \cite{Costa:2014kfa})
\begin{equation}\label{propsj} 
Z^{A_1} \ldots Z^{ A_J} (G_\Delta)_{A_1 \ldots A_J}^{M_1 \ldots M_J}W_{M_1} \ldots W_{M_J} =\frac{\mathcal{C}_{\Delta,J}}{e^{-i \pi J} 2^\Delta}\frac{\left( Z.W ~P.X- {Z.X~ W.P} \right)^J}{(-P.X)^{\Delta + J}}
\end{equation}
where $Z^A$ and $W_M$, respectively, are arbitrary boundary and bulk tangent vectors that obey $Z_A Z^A=W_M $ and the explicit form of $\mathcal{C}_{\Delta,J}$ is given by \eqref{cdeltaj}. When $J=1$, the bulk gauge field at position $X$, sourced by `current' with polarization $Z^M$, is given by 
\begin{equation}\label{polg} 
Z^A (G_\Delta)_{AM}=\frac{-\mathcal{C}_{\Delta,1}(D-1)}{(-2)^{\Delta+1}}\left(\frac{Z_M}{(P.X)^\Delta} - \frac{(Z.X) P_M}{(P.X)^{\Delta +1}}\right) 
\end{equation}
where $Z^A$ is an 

Here $A$ is a bulk index which is understood to be projected orthogonal to $X_A$. It is useful to make the projection more explicit. Let 
\begin{equation}\label{xoproj}
\left( {\cal P}_{X} \right)^M_N \equiv 
\delta^M_N+X^M X_N
\end{equation}  
denote the projector orthogonal to $X_0$. Given any vector `field $A^M$ let us also define 
\begin{equation}\label{xoproj2}
A_M^\perp \equiv \left( {\cal P}_{X} \right)^M_N A^N 
\end{equation} 
It follows that the gauge field at the point $X$ is given by 
\begin{equation}\label{polgn} 
Z^A (G_\Delta)_{AM}=\frac{-\mathcal{C}_{\Delta,1}(D-1)}{(-2)^{\Delta+1}}\left(\frac{Z_M^\perp}{(P.X)^\Delta} - \frac{(Z.X) P^\perp_M}{(P.X)^{\Delta +1}}\right) 
\end{equation} 

This expression can further be manipulated to\footnote{The boundary tangent vector $Z_A$ has $D$ independent components, which is the same as the number of degrees of freedom in a massive vector field propagating in $D+1$ bulk dimensions. \eqref{polgnn} makes clear that  - roughly speaking - the $D-1$ components in $Z^\perp$ parameterize the transverse degrees of freedom while 
	$Z.X$ labels the longitudinal degree of freedom.}	 
\begin{equation}\label{polgnn} 
Z^A (G_\Delta)_{AM}=\frac{-\mathcal{C}_{\Delta,1}(D-1)}{(-2)^{\Delta+1}}\left(\left(1- \frac{1}{\Delta} \right) \frac{Z_M^\perp}{(P.X)^\Delta} 
+  \nabla_M \left( \frac{Z.X}{\Delta (P.X)^\Delta}\right) \right)
\end{equation} 
where the $AdS_{D+1}$ covariant derivative is defined by 
$$ \nabla_M = ({\cal P}_X)^N_M \partial_N.$$

It follows that 	 
\begin{equation}\label{twotermdecomp}
(G_\Delta)_{AM}=(G_\Delta^1)_{AM} + (G_\Delta^2)_{AM}
\end{equation} 
with 
\begin{equation}\label{grad}
\begin{split} 
(G_\Delta^1)_{AM}&=\frac{-\mathcal{C}_{\Delta,1}(D-1)}{(-2)^{\Delta+1}}\left(1- \frac{1}{\Delta} \right) \frac{\eta_{AM}+X_A X_M}{(P.X)^\Delta},\\
 (G_\Delta^2)_{AM}&= \nabla_M (\zeta_\Delta)_A,  ~~~\zeta_\Delta ^A=\frac{-\mathcal{C}_{\Delta,1}(D-1)}{(-2)^{\Delta+1}}\frac{X^A}{\Delta (P.X)^\Delta},
\end{split} 
\end{equation}
It is easily verified (see Appendix \ref{bces}) that 
\begin{equation}\label{boundsig}
\nabla^A (G_\Delta^1)_{AM}=0
\end{equation} 
\footnote{On the other hand \begin{equation}\label{boundsig2}
	\nabla^A \left( G^2_{AM} \right) = \nabla_M\nabla^A\zeta_A= \nabla_M \left( \frac{1}{(P.X)^{\Delta +1}} \right) 
	\end{equation} }
It follows that \eqref{twotermdecomp} decomposes the boundary propagator into two pieces, the first of which is conserved (divergenceless) on the boundary while the second of which is a gradient in the bulk. 			
\footnote{ It follows that the contribution of $G^1_{AM}$ to Witten diagrams yields a term in the boundary correlator that is  identically conserved.  On the other hand the contribution of $G^{2}_{AM}$ to Witten diagrams yields a term in the bulk correlator that is generically non conserved. However this second contribution sometimes  vanishes. This happens, for instance, when  all bulk interactions happen to be `gauge invariant' - i.e. built only out of field strengths of the bulk vector field and their derivatives. For general $\Delta$ there is no reason for bulk interactions to be gauge invariant. The case of  $\Delta=D-1$ is special and will be dealt with in detail below. } 

So far we have worked at arbitrary values of the dimension, $\Delta$ of our spin one operator. In this paper, we will focus our attention on the special case $\Delta=D-1$. In this case, the bulk gauge field enjoys invariance under bulk gauge transformations. In this case, the second terms in \eqref{polgnn} and \eqref{twotermdecomp} are pure gauge and can be dropped 
(as their contribution to any correlator vanishes). 
It follows in this case that, effectively, 
\begin{equation}\label{polgnnon} \begin{split} 
&Z^A (G_{D-1})_{AM}=\frac{\mathcal{C}_{D-1,1}(D-1)}{2^{D}}\left( \frac{D-2}{D-1} \right)  \frac{Z_M^\perp}{(-P.X)^{D-1}}\\
&(G_{D-1})_{AM}=	\frac{\mathcal{C}_{D-1,1}(D-1)}{2^{D}}\left( \frac{D-2}{D-1} \right)  \frac{\eta_{AM}+X_A X_M}{(-P.X)^{D-1}} 
\end{split} 
\end{equation} 

The `plane wave' representation of this propagator is
\begin{equation}\label{pwrop}
\begin{split} 
Z^A (G_{D-1})_{AM}&=\frac{\mathcal{C}_{D-1,1}\left( D-2 \right) }{2^{D}i^{D-1}\Gamma(D-1)} \int_0^\infty d\omega~\omega^{D-2} Z_M^\perp~ e^{-i \omega P.X -\epsilon \omega}
\end{split}
\end{equation}

\subsubsection{The spin two propagator}

In the case of the spin two propagator, similar manipulations to those performed in the previous subsubsection yield the analogue of \eqref{polgn} 
\begin{equation}\label{polgnspint} 
\begin{split}
Z^{A_1} Z^{A_2} (G_\Delta)_{A_1 A_2 M_1 M_2}&=\frac{\mathcal{C}_{\Delta,2}(D^2-1)}{2(-2)^{\Delta}}\left(\frac{Z_{M_1}^\perp Z_{M_2}^\perp}{(P.X)^\Delta} - \frac{(Z.X) (P_{M_1}^\perp Z_{M_2}^\perp+Z_{M_1}^\perp P_{M_2}^\perp)}{(P.X)^{\Delta +1}}\right.\\
&\left.+ \frac{(Z.X)^2}{(P.X)^{\Delta+2}}P_{M_1}^\perp P_{M_2}^\perp\right)
\end{split}
\end{equation} 

and of \eqref{polgnn} 
\begin{equation}\label{poltwo} 
\begin{split}
Z^{A_1} Z^{A_2} (G_\Delta)_{A_1 A_2 M_1 M_2}=&  \frac{\mathcal{C}_{\Delta,2}}{(-2)^{\Delta}}\left(-\frac{\left(D^2-1\right) (X.P)^{-\Delta } \left((X.Z)^2 \left(\eta _{M_1 M_2}+X_{M_1} X_{M_2}\right)-(\Delta -1) Z^\perp_{M_1} Z^\perp_{M_2}\right)}{2 (\Delta +1)}\right.\\
&\left.+\left(\text{D}^2-1\right) \left(\nabla_{M_1} \xi^\perp_{M_2}+\nabla_{M_1} \xi^\perp_{M_2}\right)\right) 
\end{split}
\end{equation} 
 where,
 \begin{equation}
 	\xi^\perp_A = \frac{1}{2(\Delta+1)}\frac{Z.X}{(P.X)^\Delta} \left(-Z^\perp_A +\frac{1}{2} \frac{Z.X}{P.X} P^\perp_A\right)
\end{equation}

Note that when $\Delta=D$, the second term in \eqref{poltwo} is pure gauge. As in the previous subsection it follows that the propagator can be decomposed into two pieces
\begin{eqnarray}
Z^{A_1} Z^{A_2} (G^1_\Delta)_{A_1 A_2 M_1 M_2}&=&\frac{\mathcal{C}_{\Delta,2}}{(-2)^{\Delta}}\left(-\frac{\left(D^2-1\right) (X.P)^{-\Delta } \left((X.Z)^2 \left(\eta _{M_1 M_2}+X_{M_1} X_{M_2}\right)-(\Delta -1) Z^\perp_{M_1} Z^\perp_{M_2}\right)}{2 (\Delta +1)}\right)\nonumber\\
Z^{A_1} Z^{A_2} (G^2_\Delta)_{A_1 A_2 M_1 M_2}&=&\frac{\mathcal{C}_{\Delta,2}}{(-2)^{\Delta}}\left(\left(\text{D}^2-1\right) \left(\nabla_{M_1} \xi^\perp_{M_2}+\nabla_{M_1} \xi^\perp_{M_2}\right)\right)
\end{eqnarray}

It is then easily verified (see Appendix \ref{bces}) that for $\Delta=D$
\begin{equation}\label{boundsig3}
\nabla^{A_1} Z_2^{A_2}(G^1_D)_{A_1A_2M_1M_2}=0
\end{equation} 
We will focus on the special case $\Delta=D$. In this case $G_D^2$ is pure gauge and can be ignored. Effectively  
\begin{equation}\label{polgnnontwo} \begin{split} 
Z^{A_1}Z^{A_2} (G_{D})_{A_1 A_2 M_1 M_2}&=-\frac{\mathcal{C}_{\Delta,2}\left(D-1\right)}{2(-2)^{D}(X.P)^{D}} \left((X.Z)^2 \left(\eta _{M_1 M_2}+X_{M_1} X_{M_2}\right)-(D-1) Z^\perp_{M_1} Z^\perp_{M_2}\right)\\
\end{split} 
\end{equation} 

The `plane wave' representation of this propagator is 
\begin{equation}\label{pgrop}
\begin{split} 
&Z^{A_1}Z^{A_2} (G_{D})_{A_1 A_2 M_1 M_2}\\
&=\frac{\mathcal{C}_{D,2}}{i^{D}\Gamma(D)}\left( \frac{(D-1)^2}{2^{D+1}} \right)  \int_0^\infty d\omega~\omega^{D-1} \left(Z^\perp_{M_1} Z^\perp_{M_2}- \frac{(X.Z)^2 \left(\eta _{M_1 M_2}+X_{M_1} X_{M_2}\right)}{D-1}\right)~ e^{-i \omega P.X -\epsilon \omega}
\end{split}
\end{equation} 

\subsection{Singularity in terms of scattering} 

It follows immediately from \eqref{pwrop} and 
\eqref{pgrop} that the singular part of correlator involving boundary scalars, conserved current and conserved stress tensor operators is given 
by the following generalization of \eqref{nntre2}
\begin{equation}\label{nntrespin}
G_{\rm sing}=-\frac{2\pi^3 \left( \prod_a \tilde{\mathcal{C}}_{\Delta_a, J}\right)}{ \sqrt{\sigma(1-\sigma)}}
\int_{\mathbb{H}_{D-2}} \sqrt{g_{D-2}} ~d^{D-2} X \int d\omega \omega^{\Delta -4}
e^{i\omega P.X } {\cal S}_X\left(\omega \right)
\end{equation} 
with
\begin{eqnarray}
\tilde{\mathcal{C}}_{\Delta,0} = \frac{\mathcal{C}_{\Delta,0} e^{-\frac{i\pi (\Delta)}{2}}}{2^{\Delta}\Gamma(\Delta)},\qquad
\tilde{\mathcal{C}}_{D-1,1}=\frac{\mathcal{C}_{D-1,1}\left( D-2 \right) }{2^{D}i^{D-1}\Gamma(D-1)}, \qquad \tilde{\mathcal{C}}_{D,2}=\frac{\mathcal{C}_{D,2}}{i^{D}\Gamma(D)}\left( \frac{(D-1)^2}{2^{D+1}} \right) \nonumber\\
\end{eqnarray}
for scalars, photons and gravitons respectively and ${\cal S}_X$ is the S matrix for the scattering of the waves 
\begin{equation} \begin{split} \label{scatwaves1}
\phi &= e^{i k.x} ~~~~~~~~~~~~~~~~~~~~~~~~~~~~~~~~~~~~~~~~~~~~~~~~~~~~~~~~~{Scalar}\\
A_M&= Z^\perp_M~ e^{i k.x} ~~~~~~~~~~~~~~~~~~~~~~~~~~~~~~~~~~~~~~~~~~~~~~~~~~~~{Vector}\\
h_{MN}&=  \left(Z^\perp_{M_1} Z^\perp_{M_2}- \frac{(X.Z)^2 \left(\eta _{M_1 M_2}+X_{M_1} X_{M_2}\right)}{D-1}\right)~~~~~~{Graviton}\\
\end{split}
\end{equation} 
Here $Z_M$ is the boundary polarization of the 
current or stress tensor operator, and $k$ is the momentum for the appropriate particle  listed in \eqref{forvecg}).

Notice that while all four scattering momenta, given by \eqref{forvecg}, are independent of the point $X$, the polarizations listed in \eqref{scatwaves1} depend on the point $X$ on the hyperboloid (over which the integral in \eqref{nntrespin} is performed) because 
\begin{equation}\label{polphograv}
Z_M^\perp= ({\cal P}_X)^N_M Z_N
\end{equation} 
(see \eqref{xoproj} for the definition of the projector
$({\cal P}_X)$). It follows that the $S$ matrix ${\cal S}_X(\omega)$ that appears in 
\eqref{nntrespin} depends on $X$, and so we cannot trivially perform the integral over the $S^{D-3}$ in \eqref{nntrespin} (as we were able to do in subsection \eqref{gcsia}).

Performing the integral over $\omega$ in \eqref{nntrespin} we find the analogue of \eqref{gsingnn}:
\begin{equation}\label{gsingnngen}
G_{\rm sing}=i\left( 2\pi^3 \left(\tilde{\mathcal{C}}_{\Delta,J}\right)^4\right)\Gamma (\Delta +r -3)e^{-\frac{i(\Delta +r)}{2}}  ~~~
\frac{\sqrt{1-\sigma}^{\left( \Delta+r-4 \right)} ~
}{\sigma^{\frac{\Delta + r-2}{2}} \rho^{\Delta+r-3}}
\int  d \Omega_{D-3} d\zeta \frac{\sinh^{D-3} \zeta}{ \cosh \zeta^{\Delta +r-3 }} \left( \frac{{\cal S}_X(\omega)}{\omega^r } \right) 
\end{equation} 
where we have assumed, as in the subsection \eqref{gcsia}, that the interaction term is of order $r$ in derivatives so that the S matrix scales with overall energy scale like ${\cal S}(\omega) \sim \omega^r $ so that the quantity 
\begin{equation}\label{smat}
{\tilde {\cal S}}(\omega)=\left( \frac{{\cal S}_X(\omega)}{\omega^r} \right)
\end{equation} 
that appears in \eqref{gsingnngen} is actually independent of $\omega$. 
\footnote{Note that the RHS of \eqref{gsingnngen} involves an integral over the hyperboloid $\mathbb{H}_{D-2}$. $\zeta$ and the angles on the $D-3$ sphere $\Omega_{D-3}$ are coordinates on this hyperboloid. $ d \Omega_{D-3} d\zeta {\sinh^{D-3} \zeta}$ 
	is the usual volume element on the hyperboloid. The factor of $\cosh \zeta^{\Delta +r-3 }$ in the denominator is 
	a result of doing the $\omega$ integral, and is a consequence of the fact that the phase factor $e^{i P.X}$ breaks the $SO(D-2,1)$ isometry of the hyperboloid.}

\subsection{Transforming to standard graviton polarizations}

While the formulae presented in the previous subsection are all accurate, the flat space graviton polarization that appears in \eqref{scatwaves1} is 
presented in an unusual gauge. In this subsection, we will explain this fact and also gauge transform to a more standard gauge. 

A linearized onshell graviton in flat space always obeys the massless onshell condition $k^2=0$. In addition, the linearized Einstein equations also impose the following condition on $h_{MN}$ 
\begin{equation} \label{lineeq}
k^M h_{MN} - \frac{k_N}{2} h=0
\end{equation}
where $h=h^M_M$ is the trace of $h_{MN}$. 

When studying gravitational scattering, it is conventional to work in a Lorentz type  gauge in which 
\begin{equation}\label{ltg}
k^M h_{MN}=0
\end{equation} 
If we impose this condition, it follows immediately from \eqref{lineeq} that 
\begin{equation}\label{traceless}
h=0
\end{equation} 

Let us now turn to the graviton wave presented in \eqref{scatwaves1}. It is easy to verify that $h_{MN}$ that appears in \eqref{scatwaves1} obeys 
\begin{equation} \label{hmnn}
k^M h_{MN}=-\frac{k_N \left(X.Z\right){}^2}{D-1}, ~~~~~ h=-\frac{2 \left(X.Z\right){}^2}{D-1}
\end{equation} 
It follows that $h_{MN}$ listed in \eqref{scatwaves1} obeys \eqref{lineeq}, even though it does not obey the equations \eqref{ltg} or \eqref{traceless} individually.

It is, of course, possible to gauge transform
$h_{MN}$ listed in  \eqref{scatwaves1} to ensure that it obeys \eqref{ltg} (and so, automatically, \eqref{traceless}). Under an infinitesimal  gauge transformation 
\begin{equation}\label{gtongrav}
{\tilde h}_{MN} =h_{MN}+ \zeta_M k_N+\zeta_N k_M
\end{equation} 
It follows from \eqref{hmnn} that the transfrormed 
${\tilde h}_{MN}$ obeys both \eqref{ltg} and \eqref{traceless} provided
\begin{equation}\label{condonzet}
k.\zeta= \frac{ \left(X.Z\right){}^2}{D-1} 
\end{equation} 
Of course \eqref{condonzet} is one condition on 
$D+1$ variables, and so does not completely determine $\zeta$. If we want a concrete particular formula for the transformed $h_{MN}$ we  need to impose $D$ additional conditions; these conditions are arbitrary and can be chosen as per convenience. 
\footnote{The reason for this ambiguity is the following. In $D+1$ spacetime dimensions the number of metric components is $\frac{(D+2)(D+1)}{2}$ while the number of independent graviton polarizations is $\frac{D(D-1)}{2}-1$. 
The difference between these two numbers is $2(D+1)$. If we impose the 
$D+1$ conditions \eqref{ltg}, we get the one additional condition \eqref{traceless} free. 
This still leaves us with $D$ more parameters in $h_{MN}$
than the number of physical gravitons. We need to fix these additional parameters by imposing an additional (arbitrary)
gauge condition; these are the $D$ undetermined parameters in 
$\zeta$.  The Maxwell analogue of the discussion of this footnote is simply the fact that Lorentz gauge does not completely fix photon polarizations; the remaining ambiguity is 
$\epsilon_M \rightarrow \epsilon_M + k_M$.}
A physically natural additional gauge condition  - the one we will choose to adopt -  is to demand that all gravitons polarizations are transverse to the centre of mass momentum $k_1+k_3= -(k_2+k_4)$. Since we have already demanded that $h^a_{MN}$ is transverse to $k_a$, this additional requirement requires us only to impose the conditions 
\begin{equation}\label{arbitconven}
(k_3)^M h^1_{MN} =(k_1)^M h^3_{MN} =(k_4)^M h^2_{MN} =(k_2)^M h^4_{MN} =0
\end{equation} 
Despite first appearances, \eqref{arbitconven} imposes $D$ (rather than $D+1$) conditions on each graviton as one of each of the four groups of $D+1$ equations in \eqref{arbitconven} is automatic from \eqref{ltg}.\footnote{For the graviton inserted at the point $P_1$, for instance, the condition $k_3^M h^1_{MN} k_1^N$ is 
already ensured by \eqref{ltg}.}

If we choose to impose both \eqref{ltg} as well as \eqref{arbitconven} on our scattering gravitons, the scattering graviton waves reported in \eqref{scatwaves1} are modified to 
\begin{equation}\label{newgravwave} \begin{split} 
h^1_{MN}&=\left(Z^{1\perp}_{M} Z^{1\perp}_{N}- \frac{(X.Z^1)^2 \left(\eta _{MN}+X_{M} X_{N}\right)}{D-1}\right)+\frac{(X.Z^1)^2 \left(k^3{}_M k^1_N+k^1_M k^3_N\right)}{(D-1) k^1.k^3}\\
&-\frac{k^3.Z^1 \left(k^1_N Z^{1\perp}_M+k^1_M Z^{1\perp}_N\right)}{k^1.k^3}+\frac{k^1_M k^1_N \left(k^3.Z^1\right)^2}{\left(k^1.k^3\right)^2}\\
h^2_{MN}&=\left(Z^{2\perp}_{M} Z^{2\perp}_{N}- \frac{(X.Z^2)^2 \left(\eta _{MN}+X_{M} X_{N}\right)}{D-1}\right)+\frac{(X.Z^2)^2 \left(k^4_M k^2_N+k^2_M k^4_N\right)}{(D-1) k^2.k^4}\\
&-\frac{k^4.Z^2 \left(k^2_N Z^{2\perp}_M+k^2_M Z^{2\perp}_N\right)}{k^2.k^4}+\frac{k^2_M k^2_N \left(k^4.Z^2\right)^2}{\left(k^2.k^4\right)^2}\\
h^3_{MN}&= \left(Z^{3\perp}_{M} Z^{3\perp}_{N}- \frac{(X.Z^3)^2 \left(\eta _{MN}+X_{M} X_{N}\right)}{D-1}\right)+\frac{(X.Z^3)^2 \left(k^1{}_M k^3_N+k^3_M k^1_N\right)}{(D-1) k^1.k^3}\\
&-\frac{k^1.Z^3 \left(k^3_N Z^{3\perp}_M+k^3_M Z^{3\perp}_N\right)}{k^1.k^3}+\frac{k^3_M k^3_N \left(k^1.Z^3\right)^2}{\left(k^1.k^3\right)^2}\\
h^4_{MN}&=\left(Z^{4\perp}_{M} Z^{4\perp}_{N}- \frac{(X.Z^4)^2 \left(\eta _{MN}+X_{M} X_{N}\right)}{D-1}\right)+\frac{(X.Z^4)^2 \left(k^4_M k^2_N+k^2_M k^4_N\right)}{(D-1) k^2.k^4}\\
&-\frac{k^2.Z^4 \left(k^4_N Z^{4\perp}_M+k^4_M Z^{4\perp}_N\right)}{k^2.k^4}+\frac{k^4_M k^4_N \left(k^2.Z^4\right)^2}{\left(k^2.k^4\right)^2}\\
\end{split}
\end{equation} 

In final summary, the S matrix that appears in \eqref{gsingnngen} is the S matrix for the scalars and vectors reported in \eqref{scatwaves1} and either the gravitons reported in \eqref{scatwaves1} or the gravitons reported in \eqref{newgravwave}. The two sets of gravitons are gauge related and have equal S matrices. The advantage of expressions presented  in \eqref{scatwaves1} is that the polarization for the $i^{th}$ particle does not refer to any other particle. Its disadvantage is that it appears in an unfamiliar gauge. These advantages and disadvantages are reversed in the polarizations 
\eqref{newgravwave}. The reader is free to choose either of these gauges (or any other) according to her convenience.

\subsection{Regge scaling} \label{rsm}

In \eqref{gsingnngen}, the small $\rho$ behavior of the four-point function is expressed in terms of the integral over flat space S matrices over an $\mathbb{H}_{D-2}$. As we have emphasized above, all the S matrices that appear in this formula have the same effective scattering momenta, but the effective scattering polarizations depend on 
the scattering point $X$.

The Regge scaling of the S matrices ${\cal S}_X$ that appear in \eqref{gsingnngen} is not necessarily the same 
for all values of $X$. Recall, however, that the dependence of the S matrix on $X$ is very simple; it arises entirely through the dependence on the scattering polarizations on $X$. The dependence of the S matrix is very simple. In the case of photons, it is a linear function of polarizations (separately in the polarization of every particle) while in the case of gravitons, it is a bilinear function of polarizations. It follows from this fact that if the S matrix displays a certain Regge growth at some point $X$ on the hyperboloid, it must grow at least as fast at all but possibly (measure zero sets of) isolated 
points on the hyperboloid. Ignoring these isolated points that make no significant contribution to the integral,  it follows that all the S matrices that appear in \eqref{gsingnngen} scale in the same manner in the 
Regge limit. 

Let us suppose that in the small angle limit, this generic S matrix behaves like 
\begin{equation}\label{smalltn}
{\hat {\cal S}} \propto \theta^{r-2A}
\end{equation} 
Using the fact that our ${\cal S}$ matrix scales with energy like $\omega^r$, and that in the small angle limit 
$t \sim \theta^2 \omega^2$, it follows from \eqref{smalltn} that the Regge (i.e. fixed $t$) scaling of
${\cal S}$ is  
\begin{equation}\label{smallt}
{\cal S} \sim \omega^{2 A} \sim s^A
\end{equation} 

Plugging this estimate into \eqref{gsingnngen} we find that in the small $\sigma$ limit 
\begin{equation}\label{hom}
G_{\rm sing} \propto 
\frac{1}{\sigma^{\frac{\Delta}{2} }} \times \frac{1}{\sigma^{A-1}} 
\end{equation} 
Note in particular that the scaling \eqref{hom} is independent of $r$. 

\subsection{Regge scaling of normalized correlators} \label{rsmn} 

In applications related to the chaos bound it is useful to study the normalized four point correlator \eqref{normfp}. Recall that in order for the denominator 
of \eqref{normfp} to be nonvanishing, $\Delta_1=\Delta_2$ and $\Delta_3=\Delta_4$. It follows that 
$$ \Delta=2 \Delta_1+2 \Delta_2$$
Combining \eqref{itlim} and \eqref{hom}, it follows that 
\begin{equation}\label{homtt}
G^{\rm norm}_{\rm sing} \propto 
\frac{1}{\sigma^{A-1}} 
\end{equation}

\subsection{Massive higher spin scattering} \label{mhss}

In this subsection, we have restricted our attention to the study of the massless spin one and spin two fields (dual to the correlators of conserved currents and the stress tensor on the boundary). While the  generalization to the study of massive higher spin particles should certainly be possible, such a generalization involves new complications, which will be briefly outlined in this subsection (and discuss in much greater detail in Appendix \ref{mhsp}).

Consider the case of a massive vector field. Because the bulk interactions of such a field are not necessarily `gauge invariant', in this case we are forced to deal with the full propagator \eqref{polgnn} rather than a simplified propagator analogous to \eqref{polgnnon}. As a consequence, the scattering modes include an additional polarization; a  longitudinal polarization (which is pure gauge in the massless theory). In equations, at leading order, the wave representation of his propagator, tells us that the effective scattering wave of the massive spin one particle is given by
\begin{equation}\label{scatwavesprelmain} 
\begin{split} 
A^i_M&=  \epsilon^i_M e^{i {k}_i.x}\\
\epsilon^i_M&= \left( 1- \frac{1}{\Delta_i} \right) 
(Z_i^\perp)_M + \frac{i k^i_M}{\Delta^i}(Z_i.X_0) 
\end{split} 
\end{equation}
($k^i_M$ are as listed in \eqref{forvecm}).
The second term in \eqref{scatwavesprelmain} is the new longitudinal polarization. 

Notice that the coefficient of the longitudinal mode in \eqref{scatwavesprelmain} is of order $\omega$, while that of the transverse mode is of order unity. Recall also that increasing the number of powers of $\omega$ in the S matrix increases the degree of singularity in $\rho$ (in formulae like \eqref{gsingnngen}). It follows that the analog of the formula \eqref{gsingnngen} involves multiple terms on the RHS with different inverse powers of $\rho$. The coefficient of the maximally singular term in $\rho$ is the S matrix involving the maximal number of longitudinal polarizations. The coefficients of subleading singularities in $\rho$ include the S matrices of modes 
with less than the maximal number of longitudinal polarizations - but also receive contributions from subleading effects from, e.g., the scattering of the maximal number of longitudinal modes. These subleading effects - which presumably involve corrections to the simple flat space S matrix (resulting from the fact that the relevant propagators are not precisely plane waves and from the curvature of $AdS_{D+1}$) - complicate the analysis. While we believe that these complications are tractable, and the main result of this paper - namely that scattering amplitudes that violate the CRG conjecture lead to correlators that violate the chaos bound - likely also holds for these massive modes, the proof of this claim needs more care in these cases, and we leave this to future work. See Appendix \ref{mhsp} for more discussion of this case.  

\section{Inter relationship between the small $\theta$ and small $a$ expansion} \label{ire} 

In section \eqref{sgzl} we studied Regge expansion (the small $\theta$ fixed 
$a$) expansion of our correlator. In section \ref{rgzl}, on the other hand,  we have studied the small $a$ fixed $\theta$ expansion of our correlator on the Causally scattering sheet. In this section, we will explore the interrelationship between these two expansions. We will do this by returning to the expansion of section \eqref{sgzl}, first performing the small $\theta$ fixed $a$ expansion, and then examining how each coefficient in this expansion scales with $a$ in the $a \to 0$ limit. \footnote{We developed the material recorded in this section in part to address questions posed to us by A. Gadde and A. Zhiboedov. We would like to thank them for their probing questions which helped us understand that one of the arguments we presented in an earlier version of this paper was incorrect, and spurred us to better understand the structure of the small $\theta$ perturbation theory (see the material in this section) to correctly argue for the same conclusion.}

As we explained in subsection \ref{swt},  the correlator under study in this paper can be expanded in a  systematic power series expansion in $\theta$ in the Regge limit. In particular we demonstrated that the (unnormalized) correlator takes the form 
\begin{equation}\label{expt}
G= \frac{1}{\theta^{2 B-2}} 
\left( \sum_{n=0}^\infty \theta^n H_n(a) \right) 
\end{equation} 
where $B$ was defined in equations \eqref{tbt} and \eqref{nunt2}. 
($H(a)$ of  subsection \ref{anina}  is the same as $H_0(a)$ in \eqref{expt}).
In this section, we return to the expansion of 
subsection \ref{swt} to investigate the small $a$ scaling of the functions $H_n(a)$. 

Recall that the expansion of subsection \ref{swt} was obtained starting with integral \eqref{genformoo} and Taylor expanding the integrand in the variables $u$ $v$, $\theta$ and $\tau$ with $y_i$ and $y_1$ unexpanded (i.e. treated as order unity). As we explained in subsection \ref{swt} it is actually most convenient to use the variables $D_0^i$  as the expansion variables for the power series expansion; the expression for $u$
$v$, $\tau$ and $\theta$ in terms of $D_0^i$ is given in \eqref{ttuv}. In subsection \ref{swt} $H_n(a)$ was obtained by expanding the integrand of \eqref{genformoo} in a power series expansion in the variables $D_0^i$ with $y_1$ and $y_i$ regarded as order unity, and then performing the integrals, first over $u$ and $v$ and then over $y_1$ and $y_i$. 

\subsection{Statement of the $a$ counting rule} \label{socr}

We are interested in the scaling of the functions $H_n(a)$ in the small $a$ limit. The functions $H_n(a)$ are obtained after performing the integration in 
\eqref{genformoo} over $u$, $v$, $y_1$ and $y_i$. 

As in the case of the small $\theta$ expansion, it is possible to find an effective small $a$ scaling assignment for the integrand in \eqref{genformoo}. This rule is devised to obey the following property. If any particular term in the integrand is assigned scaling $a^n$, then the integral of this term will scale like $a^n$. 

In this subsection, we simply state this `small $a$ integrand scaling rule'. The justification for this rule is provided later in this section. 

The `small $a$ integrand scaling rule' goes as follows.

\begin{itemize} 
	\item $\tau$, $u$ $v$ and $y_1$ are each of order $a$, while $\theta$ and $y_i$ are of order unity. 
	\item Working with the variables $D^0_i$ , $y_1$ and $y_i$ we conclude from \eqref{ttuv} that all four $D_0^i$ and $y_1$  are of order $a$ while $y_i$
	are of order unity. 
	\end{itemize}

\subsection{Consequence of the $a$ scaling rule}\label{car}

Assuming the small $a$ integrand scaling rule of the previous subsection, it is easy to bound the small $a$ behaviour of 
each $H_n(a)$. All we need to do is count the $a$ power of every term in the expansion of the integrand of \eqref{genformoo} that contributes to $H_n(a)$. It is easy to convince oneself that every such terms scales in the small $a$ limit like 
\begin{equation}\label{scalingofterm} 
\frac{P(a)}{a^{{\tilde a}_1 + {\tilde a}_2+ {\tilde a}_3+ {\tilde a}_4}} 
\end{equation} 
(see \eqref{genformoo} for the definition of ${\tilde a}_i$) where $P(a)$ is an expression that has a power series expansion in terms of order $a$. It follows that the integral of this quantity over $u$, $v$ and $y_1$ scales like 
\begin{equation}\label{scalingofe}
\frac{P'(a)}{a^{{\tilde B}-3}}= \frac{P'(a)}{a^{2B+M-3}} 
\end{equation} 
(see \eqref{tbt} for the definition of ${\tilde B}$)
where $P'(a)$ is a function that admits a power series expansion in $a$. \eqref{scalingofe} follows from  \eqref{scalingofterm} upon integrating.  \eqref{scalingofterm} follows from the following facts 
\begin{itemize}
	\item{1.} Each $D^i_0$ is of order $a$. 
	\item{2.} If we write true denominators in \eqref{genformoo} as 
	$D^i= D^i_0+ \delta D_i$ (where $\delta D_i$ is the collection of all terms that are subleading to $D_i^0$ in the $\theta$ expansion, i.e. $\delta D_i$ consists of terms that are of $\theta^2$ or smaller) then each $\delta D_i$ is of order $a P(a)$ where $ P(a)$ is a function that starts out at order unity (for $i=2,4$) or order $a^2$ (for $i=1,3$) and has a power series expansion in $a$.
	(see around \eqref{defofdexp}.) 
	\item{3.} While the numerator in \eqref{genformoo} starts at order $\theta^M$ in the small $\theta$ expansion, it starts either at order unity (this is the generic expectation) or higher (possible but non generic) in the small $a$ expansion. In other words the function in \eqref{genformoo} has a power series expansion in terms of order $a$.
\end{itemize} 
Note in particular that the leading possible  singularity in $a$ in \eqref{scalingofterm} is independent of $n$. In other words, the leading small $a$ scaling of distinct coefficients  of the small $\theta$ expansion does not increase with the order to the expansion; it generically also does not decrease with the order of the expansion. 

The reason that coefficients of higher-order in the $\theta$ expansion do not start out at higher orders in the $a$ expansion is the following. While each of $u$, $v$, $\tau$, and $\theta$ are all of the same order of smallness in the Regge expansion,  
only the first three of these terms are treated as small for the small $a$ expansion.  Consequently a term which is  $\theta^n$ \footnote{In this sentence $\theta^n$ means the $n^{th}$ power of the variable $\theta$ rather than just a term that is of $n^{th}$ order in the small $\theta$ expansion.} times the leading order contribution to the expansion would be counted as a contribution to $H_n(a)$ (as it is $n^{th}$ order suppressed in the small $\theta$ expansion) but would be of leading order in the small $a$ expansion (as $\theta$ is of order unity in the small $a$ expansion).

It is also nontrivial that terms at successively higher-order in the $\theta$ are not increasingly singular in the $a$ expansion. Such a situation could have arisen as follows. As we have reviewed above, 
the leading small $\theta$ approximation to the  denominators, $D^0_i$,  are all of order $a$ in the small $a$ expansion. Now as in item 2 above, let us consider the true denominator $D_i$, and let $D_i=D_i^0+ \delta D_i$ 
where $\delta D_i$ is of order $\theta^2$ or higher in the small $\theta$ expansion. If  had turned out that $\delta D_i$ had included a piece that was of order unity in the small $a$ expansion - e.g. a term proportional to $\theta^r$ with no additional small $a$ suppression factors - then higher orders in the small $\theta$ expansion of the denominator (which would have included a power series in $\frac{\delta D_i}{D_i^0}$)  would have included terms with increasingly high singularities  at small $a$. As we have mentioned in item 2, the explicit expressions for $P_i.X$ make it clear that this never happens. Every term in $\delta D_i$ is always of order $a$ or smaller in the small $a$ expansion. 

As the point made in the previous paragraph is a  crucial link in our derivation of the CRG conjecture from the chaos bound, we pause to explain this point in more detail with the aid of equations. Let us use the notation 
\begin{equation}\label{denomnot}
D_i=-2P_i.X
\end{equation}
It is easy to explicitly evaluate $D_i$ in terms of $\theta, \tau, u, v$ and then reexpress the result in terms of $D^0_i$. We find  
\begin{equation}\label{expexp}
\begin{split}
D_1 &= \frac{1}{2} \left(-\left(D_2^0+D_4^0\right) \cos \tau +\left(D_1^0-D_3^0\right)+2 y_0 \sin \tau \right)\\
D_2 &= \frac{1}{2} \left(\left(D_3^0-D_1^0\right) \cos \left(\frac{D_1^0+D_2^0-D_3^0-D_4^0}{2 y_1}\right)+D_2^0+D_4^0\right) + y_1 \sin \left(\frac{D_1^0+D_2^0-D_3^0-D_4^0}{2 y_1}\right)\\
D_3 &= \frac{1}{2} \left(-\left(D_2^0+D_4^0\right) \cos \tau -D_1^0+D_3^0+2 y_0 \sin \tau \right)\\
D_4 &= \frac{1}{2} \left(\left(D_1^0-D_3^0\right) \cos \left(\frac{D_1^0+D_2^0-D_3^0-D_4^0}{2 y_1}\right)+D_2^0+D_4^0\right)- y_1 \sin \left(\frac{D_1^0+D_2^0-D_3^0-D_4^0}{2 y_1}\right)
\end{split}
\end{equation}
where, $$\tau(D_1^0,D_2^0,D_3^0,D_4^0) = \frac{D_1^0+D_2^0+D_3^0+D_4^0}{\sqrt{\left(D_1^0-D_2^0-D_3^0-D_4^0\right) \left(D_1^0+D_2^0-D_3^0+D_4^0\right)+4 \left(1+y_1^2+y_{\perp}^2\right)}}$$
and
$$y_0(D_1^0,D_2^0,D_3^0,D_4^0) = \sqrt{1+y_1^2+y_{\perp}^2+\frac{1}{4} \left(D_1^0-D_2^0-D_3^0-D_4^0\right) \left(D_1^0+D_2^0-D_3^0+D_4^0\right)}$$
Using the explicit relations \eqref{expexp} it is easy to verify that the small $\theta$ expansion of each of the denominator factors $D_i$ takes the form 
\begin{equation}\label{defofdexp}  
D_i= \sum_{r=0}^\infty D_i^r
\end{equation} 
where $D_i^r$ is of order $2r+1$ in the small $\theta$ expansion. Again using these explicit expressions it is easy to convince oneself that the small $a$ expansion of $D_1^r$ and $D_3^r$ starts at order $a^{2r+1}$ 
\footnote{This is a consequence of the fact that $\tau$ and $uv$ are simultaneously of order $\theta^2$ and $a^2$, while $y_1^2$ is of order $a^2$ but order unity in the $\theta$ expansion.}. On the other hand $D_2^r$ and $D_4^r$ respectively each start at order $a$ in the small $a$ expansion. \footnote{This is a consequence of the fact that the argument of the cosine and sine in these expressions is of order $\theta$ in the Regge expansion but of order unity in the small $a$ expansion. Note its crucial that the pre factor of both the cosine and sine in these expressions is of order $a$ in the small $a$ expansion.} These facts are illustrated by the explicit expressions for the first subleading corrections
\begin{equation}\label{dexpea}
\begin{split} 
&D_1^1= -\frac{(D^0_1-2 D^0_2+D^0_3-2 D^0_4) (D^0_1+D^0_2+D^0_3+D^0_4)^2}{48 \left(1+y_\perp^2+y_1^2\right)}\\
&D_2^1= \frac{(2D^0_1-D^0_2-2D^0_3+D^0_4)(D^0_1+D^0_2-D^0_3-D^0_4)^2 }{48 ~y_1^2}\\
&D_3^1= -\frac{(D^0_1-2 D^0_2+D^0_3-2 D^0_4) (D^0_1+D^0_2+D^0_3+D^0_4)^2}{48 \left(1+y_\perp^2+y_1^2\right)}\\
&D_4^1=-\frac{(2D^0_1-D^0_2-2D^0_3+D^0_4)(D^0_1+D^0_2-D^0_3-D^0_4)^2 }{48~y_1^2}\\
\end{split}
\end{equation}
Note that $D_1^1$ and $D_3^1$ each start at order $a^3$
while $D_2^1$ and $D_4^1$ each start at order $a$, in agreement with the general rule spelt out above.

\subsection{Derivation of the $a$ counting rule} 
\label{dacr}

In this subsection, we explain why the counting rule of subsubsection \ref{socr} is correct. 

Recall from subsection 
\eqref{swt} that the final expression for $H_n(a)$ is given by 
\begin{equation} \label{schematicform}
H_n(a) = \sum_{\{a_i\}} \int d y_i d y_1 
\frac{N_{\{a_i\}}(y_1, y_i, a)}{\left(a~y_0+y_1 +i\epsilon\right)^{a_1+a_2-1}\left(a~y_0-y_1 +i\epsilon\right)^{a_3+a_4-1}} 
\end{equation} 
\footnote{If we had expanded the integrand in \eqref{genformoo} in a power series expansion in $\tau$, $\theta$, $u$ and $v$ then the coefficients of this expansion would have been independent of $a$. However we instead chose to expand the integrand in a power series in $D_0^i$. The linear transformation to the new variables \eqref{ttuv} involves $a$. 
This is why the function $N_{\{\alpha_i\}}$ depends on $a$.}
where the sum is taken over terms with 
\begin{equation}\label{alphasumb}
a_1+a_2+a_3+a_4=2 { B} -n
\end{equation} 
(${\tilde B}$ was defined in \eqref{tbt}) and all factors of $C_{a_1, a_2, a_3, a_4}$
(see \eqref{fozAS2}) have been absorbed into the function $N_{\{a_i\}}(y_1, y_i, a) )$.

Now the small $a$ singularity in \eqref{schematicform}
is a consequence of the pinch in the integral over $y_1$ in the denominators in the integrand of that expression. It is easy to convince oneself (see Appendix \ref{reggecomp}, especially \ref{isa} for some more detail) that if the numerator 
function scales like
\begin{equation}\label{Nscal}
N_{\{a_i\}}(\lambda y_1, y_i, \lambda a) \sim  \lambda^{Y_N}
\end{equation}  
at small $\lambda$ then 
\begin{equation}\label{HNa}
H_n(a) \sim a^{Y_N +3 - a_1 -a_2 -a_3 -a_4}
=a^{Y_N +3 -2B+n}
\end{equation} 
where in the last equality we have used \eqref{alphasumb}. \eqref{HNa} follows from the observation that $y_1$ is of order $a$ in the two poles that participate in the pinch in \eqref{schematicform}; as a consequence the singularity of the integral is determined by small $y_1$ expansion of the numerator in \eqref{schematicform}. The result 
\eqref{HNa} then follows from the change of variables $y_1= ay_1'$. 

The counting rule described in the previous paragraph follows immediately from the observation that  each  additional factor of $D^i_0$ and $y_1$ in the integrand of \eqref{schematicform} gives an additional factor of $a$ in \eqref{HNa}. The fact that $\theta$ is of order unity then follows from \eqref{ttuv}.
\footnote{In particular, the fact that $y_1$ is of order $a$ follows simply from the fact that the pole values of $y_1$ are each of order $a$ in the integrand in \eqref{schematicform}. }

To end this subsection we briefly outline how the RHS of \eqref{HNa} explicitly reproduces \eqref{scalingofe}. Infact \eqref{scalingofe} follows from \eqref{HNa} from the observation that $Y_N$ is bounded from below according to 
\begin{equation}\label{YNineq} 	
	Y_N \geq -n-M
	\end{equation} 
Inserting \eqref{YNineq} in \eqref{HNa} gives \eqref{scalingofe}.

\eqref{YNineq} follows because terms in the  integrand of \eqref{genformoo} which contribute to $H_n(a)$ are of order $\theta^{M+n}$ compared to the leading contribution to the denominator in \eqref{genformoo}. It is possible for some - or all- of these additional small factors to consist of powers of $\theta$ (rather than $\tau$ or $u$ or $v$). As $\theta$ is of order unity rather than $a$ in the small $a$ limit, every insertion of $\theta$ leads to a factor of $D_i^0$ or $\tau$ (already accounted for above) divided by a factor of $a$ or $y_1$ (see \eqref{ttuv}) which have to be accounted for additionally, leading to \eqref{YNineq}.
	 
\subsection{Strength of the $a$ singularity} 

In order to make quantitative contact with the small $\rho$ analysis presented in section \ref{rgzl}, we now specialize to the case of a correlators generated by the interaction of four bulk fields dual to operators of dimension $\Delta_i$ ($i=1 \ldots 4$) interacting via a bulk interaction vertex that is of $r^{th}$ order in derivatives. As in the earlier parts of this section we restrict attention to the case that the bulk particles are either gravitons dual to the stress tensor of dimension $D$, photons dual to a conserved current of dimension $D-1$ or massive scalars dual to scalar operators of arbitrary mass. In this context the largest possible value of 
the ${\tilde a}_1+{\tilde a}_2+ {\tilde a}_3+ {\tilde a}_4$ in \eqref{scalingofterm} is 
\begin{equation}\label{largest} 
{\tilde a}_1+{\tilde a}_2+ {\tilde a}_3+ {\tilde a}_4=\Delta_1 +\Delta_2+ \Delta_3+\Delta_4 +r
\end{equation} 
(this maximum is obtained when all bulk derivatives are allowed to hit the denominators of propagators in the bulk). It follows from \eqref{scalingofe} that the most singular possible small $a$ behaviour from such a bulk interaction is 
\begin{equation}\label{mostsing}
\frac{1}{a^{\Delta_1 +\Delta_2+ \Delta_3+\Delta_4 +r-3}}
\end{equation} 
in perfect agreement with the power of the $\rho$ singularity in \eqref{gsingnngen}.

\subsection{Order of limits}\label{ool}

In this section, we have established that the correlator under study has the following simple analytic structure in the limit that $\theta$ and $a$ are both small. 
\begin{equation}\label{genstructu}
G=\frac{1}{\theta^{2B-2} a^{\Delta_1+\Delta_2+\Delta_3+\Delta_4+r-3}} H(\theta, a)
\end{equation} 
where the function $H(\theta, a)$ admits a double power series expansion in $a$ and $\theta$.  We also know from \eqref{gsingnngen} that the function $H(\theta, a)$ is nonvanishing in the limit $a\to 0$ at generic 
$\theta$. It follows that in the expansion of $H(a, \theta)$, the coefficients of the monomials $\theta^n a^0$ must be nonzero for atleast one value of $n$. Let the smallest value of $n$ for which this is the case be denoted by $n_{BP}$. On the other hand let the smallest power of $\theta$ that appears in the expansion of $H(a, \theta)$ (accompanied by any value of $a^q$, i.e. $q$ not necessarily zero) be $n_{Regge}$. Clearly 
\begin{equation}\label{abt}
n_{Regge} \leq  n_{BP}.
\end{equation} 
As $A'$ measures the $\theta$ scaling of the correlator 
in the small $\theta$ limit while $A$ controls the small $\theta$ scaling  of the coefficient of $a \to 0$ limit, it follows immediately that 
\begin{equation}\label{apia}
A'=A+ n_{BP} -n_{Regge}  \geq A
\end{equation}  
The last inequality in \eqref{apia} follows from 
\eqref{abt}. 

Generically we expect that $n_{BB}=n_R=0$ and $A'=A$. 
However, even if this generic expectation is not met, the inequality \eqref{apia} is always true. As we will see in more detail in the next section, \eqref{apia} is sufficient to establish the main result of this paper, namely that the CRG conjecture follows from the 
chaos bound, in the context studied in this paper.




\section{CRG conjecture from the chaos bound} \label{crgcha}

\subsection{$A' \geq A$}

So far in this paper we have established three key facts about the small $\sigma$ scaling of the normalized correlation function 
$G^{\rm norm}$ (see \eqref{normfp}). 
\begin{itemize} 
	\item In the $\sigma$ fixed $\rho$ limit our 
\begin{equation}\label{parsearg}
\frac{g_{CS}(e^{2\rho})}{\sigma^{A'-1}} ~~~~~{\rm and}~~~~~~
\frac{g_{CR}(e^{2\rho})}{\sigma^{A'-1}}
\end{equation}
on the Causally Scattering and the Causally Regge sheets respectively. $g_{CS}(e^{2\rho})$ and $g_{CR}(e^{2\rho})$ are both nontrivial (neither of them vanishes identically). Crucially, the scaling exponent $A'$ is independent of $\rho$ and is the same 
on both sheets. 
\item 
Upon first taking the small $\rho$ limit (on the Causally Scattering sheet) and then taking the small $\sigma$ limit, on the other hand, we demonstrated that the normalized correlator scales like  
\begin{equation}\label{scalarg}
\frac{1}{\sigma^{A-1}}
\end{equation} 
(see \eqref{hom}) where $A$ characterizes the leading order large 
$s$ fixed $t$ scaling of the flat space S matrix of the corresponding contact term (see \eqref{smallt}). 
\item By examining the small $a$ behaviour of coefficients at all orders in the small $\sigma$ limit, we demonstrated that normalized correlator has a simple analytic structure in the double  $\sigma$ small $\rho$ limit (see around \eqref{genstructu}) and used this result to deduce $A' \geq A$. 
\end{itemize} 

Putting these three items of information together, it follows that $A'$, the coefficient that characterises the small $\sigma$ scaling of the normalised correlator, is greater than or equal to $A$, the coefficient that characterises the Regge growth of flat space scattering governed by the corresponding bulk contact term.

It is of crucial importance to this paper that while the results obtained so far in this paper allow for $A'>A$, they rule out the possibility that $A'<A$. Although we have already established this fact in the previous section, we will pause to explain it again from a slightly different point of view. 

The following function 
\begin{equation}\label{exampfunct}
\frac{1}{\sigma^A \rho^a} + \frac{1}{\sigma^{A'} \rho^{a'}} 
\end{equation}
with $A'>A$ but $a> a'$ gives a very simple 
example of a function with $A' >A$. 
At leading order in the small $\rho$ limit this function reduces to $\frac{1}{\sigma^A \rho^a}$ and so scales like $\frac{1}{\sigma^A}$ as required,  whereas at leading order in the small $\sigma$ limit this function reduces to $\frac{1}{\sigma^{A'} \rho^{a'}}$ and so scales like $\frac{1}{\sigma^{A'}}$ as also required. Thus the example listed in \eqref{exampfunct} obeys all the properties of the normalized correlator 
established in this paper. 

The fact that $A'$ can be larger than $A$ in very simple examples can be understood intuitively in the following terms.
 $A'$ captures the leading $\sigma$ singularity at general values of $\rho$. It is, of course, possible that the coefficient of this leading singularity vanishes at any particular value of $\rho$; if this happens the function at that value of $\rho$ will have a milder small $\sigma$ scaling than at generic $\rho$. \eqref{exampfunct} may be viewed as an implementation of this scenario for the particular value $\rho=0$.

Now let us consider a second (and analytically more complicated) toy model \footnote{The material in the rest of this subsection was developed in discussion with A. Gadde and A. Zhibeodov.} 
\begin{equation}\label{newexamplefunct}
\sqrt{\frac{1}{\sigma^2} + \frac{1}{\rho^2} } - \frac{1}{\rho}
\end{equation} 
In the small $\sigma$ limit this function tends to $\frac{1}{\sigma}$ and so has $A'=1$, but in the small $\rho$ limit it tends to 
$\frac{\rho}{2\sigma^2}$ and so 
$A=2$. Note $A'<A$. 

While the toy model \eqref{newexamplefunct} is a perfectly well behaved analytic function with $A'<A$, 
it (and other such functions) cannot capture the behaviour of the correlators studied in this paper because the coefficients of successive orders in the small $\sigma$ expansion of this function are increasingly singular in the small $\rho$ limit 
(the small $\sigma$ expansion is really an expansion in $\sigma/\rho$), a behaviour that we have ruled out in section \ref{ire}. 

The reader may at first find it intuitively puzzling that it is possible at all for $A'$ to be less than $A$, even putting aside the considerations of section 
\ref{ire}. If $A'$ captures the most singular $\sigma$ scaling of our function at any value of $\rho$, how can the function possibly scale more singularly than $\frac{1}{\sigma^{A'}}$ `at small $\rho$'. The answer to this false puzzle is, of course, the following. Viewed as a function of $\sigma$ 
	\eqref{newexamplefunct} has two distinct domains. The first of these is $\sigma \ll \rho$ which is accessed by the small $\sigma$ limit. 
The second of these is $\sigma \gg \rho$ which is accessed by the small $\rho$ limit. The function changes its character dramatically at $\sigma$ of order $\rho$. This is what allows the $\sigma$ scaling in the small $\sigma$ limit to be slower than the $\sigma$ scaling in the small $\rho$ limit. 

In the example \eqref{newexamplefunct} (and indeed in any example of this nature) the small $\sigma$ expansion becomes singular in the limit $\rho \to 0$. Logically this follows because the scale in $\sigma$ at which the function dramatically changes the character and goes to $0$ in this limit. The results of section \ref{ire} can be thought of as forbidding this possibility; those results disallow the sigma scaling of the normalized correlator to dramatically change character at a value of $\sigma$ that goes to zero as $\rho \to 0$.

In summary, while the results of the Regge expansion and the bulk point expansion by themselves would have allowed for $A'<A$,  our detailed understanding of the interplay between these two expansions, presented in section \ref{ire} rules out the possibility. We thus reiterate that the correlators studied in this paper obey
\begin{equation}\label{Aapp}
A' \geq A
\end{equation} 

As we have mentioned in subsection \ref{ool} we suspect it may be possible to prove that the inequality in \eqref{Aapp} is saturated, so the true scenario is possibility 1. However, we do not need this to establish the central result of our paper and do not pursue this question further.

\subsection{$A' \leq 2$}

In their celebrated chaos bound paper, the authors of \cite{Maldacena:2015waa} demonstrated that out of time order thermal four-point function in a large $N$ theory cannot  grow faster with time than $e^{2 \pi T t}$ where $T$ is the temperature of the ensemble. In Appendix A of the same paper, the authors explained that, in the special case of conformal large N field theories, their bound also constrains the growth of ordinary time-ordered correlators in the Regge limit on the Causally Regge sheet. We present a brief summary of this connection. 

Consider the large N CFT first in Euclidean space, and consider the insertion of four operators in a particular plane. Now consider this theory in `angular quantization', i.e. with the angular coordinate $\theta$ of the plane being regarded as Euclidean time and the radial coordinate $r$ thought of as space. As the angular coordinate $\theta$ is periodic with periodicity $2 \pi$, the theory in this quantization is effectively thermal with $T=\frac{1}{2 \pi}$. 

\begin{figure}[H]
	\begin{center}
		\includegraphics[width=10cm]{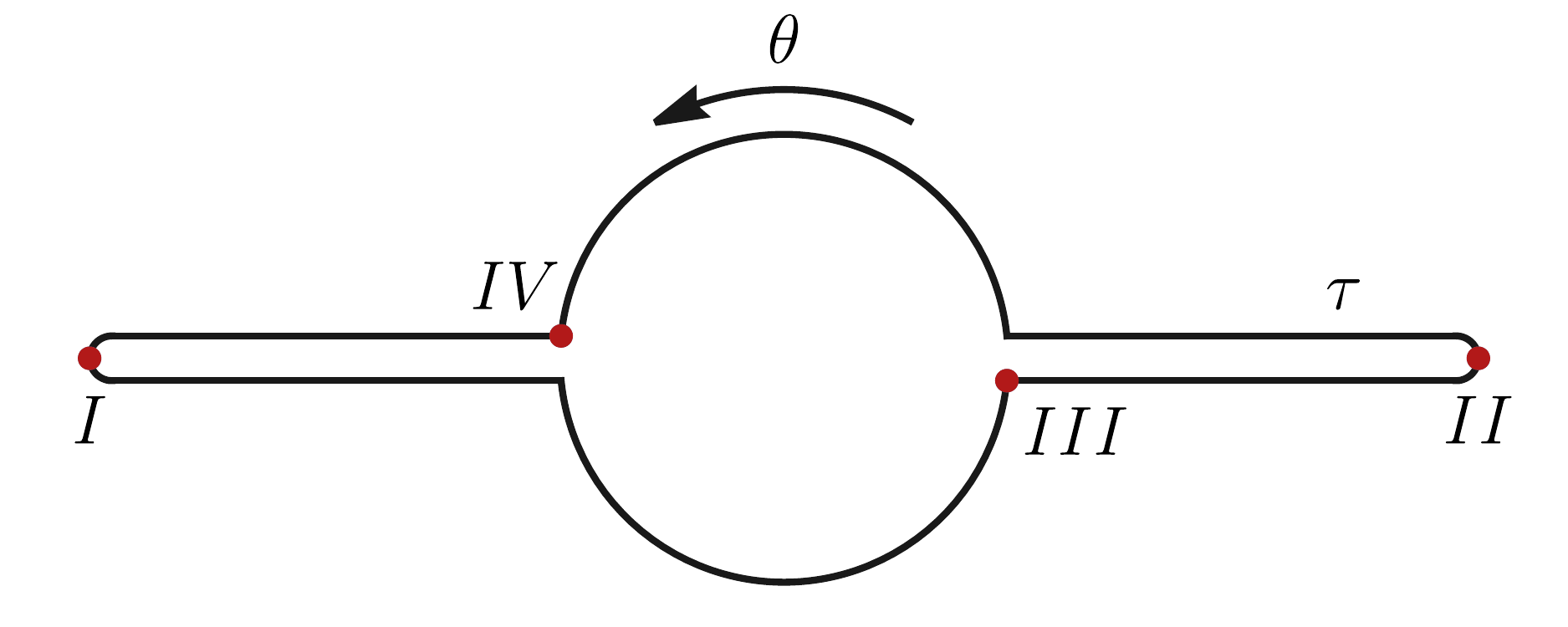}
		\caption{As explained in \cite{Maldacena:2015waa} the Regge correlator can also be viewed as an OTOC in angular quantization}
		\label{otmink}
	\end{center}
\end{figure}

With this choice of quantization we study the out of time ordered correlator depicted in Fig \ref{otmink}. In this correlator Operator $III$ is inserted at $\theta=0$ and $r=1$. Operator $II$ is inserted at $r=x$ ($x<1$) and $\theta=\epsilon+i \tau$. Operator $IV$ is inserted at $\theta=\pi$ and $r=1$ and  operator $I$ at $r=1$ and $\theta=\pi+\epsilon +i \tau$. 
After dividing by the appropriate normalization factor, the  path integral depicted in Fig. \ref{otmink} computes the following operator expectation value (in the Hilbert space obtained from angular quantization)
\begin{equation}\label{normcorchh}
\frac{\langle O_{1} O_{4} O_2 O_{3}\rangle}{\langle O_{2} O_1\rangle \langle O_{4} O_{3}\rangle}
\end{equation}
\footnote{The ordering of operators in \eqref{normcorch} is simply the order in which the operators are inserted along the contour in Fig \ref{otmink}. Note also that this ordering of operators also follows directly from the time assignments of operators in this paragraph, as operators inserted at different Euclidean times have only one consistent ordering - that in which Euclidean time increases moving from right to left. In other words, the contour depicted in \ref{otmink} is the only one consistent with the time assignments of this paragraph, as path integral contours are only allowed to move forward in Euclidean time.}

The chaos bound theorem of \cite{Maldacena:2015waa} asserts that the correlator \eqref{normcorchh} grows no faster with $\tau$ than $e^{\tau}$.    

As explained in Appendix A of \cite{Maldacena:2015waa}, the correlators  in \eqref{normcorchh} have a simple representation in the quantization of the same theory in usual Minkowski time (in the plane $\mathbb{R}^{1,1}$ obtained by starting with the plane $\mathbb{R}^{2}$ and performing the usual analytic continuation to go to $\mathbb{R}^{1,1}$ ). In the Hilbert space obtained by the usual quantization of Minkowski space, the normalized version of the path integral of Fig \ref{otmink} (or equivalently the operator expression \eqref{normcorchh}) has the following representation.
\begin{equation}\label{normcorch}
\frac{\left<  B(\tau) O_2(x, x) B^{-1}(\tau) ~O_{4}(-1,-1) ~O_3(1,1)  ~B(\tau) O_1(-x,-x) B^{-1}(\tau) \right>}{\left<B(\tau) O_{2}(x,x)B^{-1}(\tau)~ B(\tau)O_1(-x,-x)B^{-1}(\tau) \right> ~\left<O_{4}(-1,-1) O_{3}(1,1)\right>}
\end{equation}
 The locations of operator insertions in \eqref{normcorch} is given by $O(w, {\bar w})$ where $w$ and ${\bar w}$ are the lightcone coordinates defined in \eqref{defzbarz}.$B(\tau)$ is the boost operator by rapidity $\tau$. Note that the operators in \eqref{normcorch} are time ordered in Minkowski time (i.e. the earliest insertions are furthest to the right).

The equality of \eqref{normcorch} and \eqref{normcorchh} is a consequence of the fact that time evolution in analytically continued angular time
(i.e. Rindler time) is simply a boost in Minkowski space. The change in ordering of operators between \eqref{normcorchh} and \eqref{normcorch} is a consequence of the fact that angular time and regular time run in opposite directions in the left half of the Minkowskian plane. This has the following implication. Operators $O_2$ and $O_1$ which  are both inserted at  Rindler time $\tau -i \epsilon$ are respectively inserted at Minkowski time 
\begin{equation}\label{minktimeinsert}
t= \pm x \sinh(\tau -i \epsilon)
\approx \pm x \left(  \sinh\tau - i \epsilon \cosh \tau \right)
\end{equation} 
In other words, while $O_2$ is inserted at Minkowski time $x \sinh \tau -i {\tilde \epsilon}$ and so at a positive value of Euclidean time,  $O_1$ is inserted at Minkowski time $-x \sinh \tau +i {\tilde \epsilon}$ 
and so at a negative value of Euclidean time. The fact that operator insertions must always be ordered so that Euclidean time  increases from right to left then determines the ordering of operator insertions in \eqref{normcorch}. 

We can simplify \eqref{normcorch} as follows. Let us assume that the operator $O_m$ has weight 
$\lambda_m$ under boosts. It follows that 
\begin{equation}\label{boostweight}
\begin{split} 
&B(\tau) O_2(x, x) B^{-1}(\tau)=e^{\lambda_2 \tau} O_2(xe^{-\tau}, x e^{\tau})\\
&B(\tau) O_1(-x, -x) B^{-1}(\tau)=e^{\lambda_1 \tau} O_1(-xe^{-\tau}, -x e^{\tau})\\
\end{split}
\end{equation} 
Inserting \eqref{boostweight} into \eqref{normcorch} and asserting the chaos bound, we find that the expression in that equation simplifies to 
\begin{equation}\label{normcorchnew}
\frac{\left< O_2(e^{-\tau}x, e^{\tau} x) ~O_{4}(-1,-1) ~O_3(1,1)   O_1(-e^{-\tau}x,-e^{\tau} x)  \right>}{\left<O_{2}(e^{-\tau} x,e^{\tau} x) O_1(-e^{-\tau} x,-e^{\tau} x) \right> ~\left<O_{4}(-1,-1) O_{3}(1,1)\right>}
\end{equation} 
Note that the factors of $e^{\lambda_1 \tau}$ and 
$e^{\lambda_2 \tau}$ have cancelled between the numerator and denominator. 

When $\tau$ is large enough operator insertions \eqref{normcorchnew} lie on the Causally Regge sheet.
The conformal cross ratios associated with the insertions \eqref{normcorchnew} may be computed using 
\eqref{labn}; in the large $\tau$ limit we find 
\begin{equation}\label{forex}
z= 4 x e^{-\tau}, ~~~~{\bar z}= \frac{4}{x}  e^{-\tau}, 
~~~\sigma = 4 e^{-\tau},~~~~ e^{2 \rho}=x^2 
\end{equation}
The chaos bound theorem, which asserts that 
\eqref{normcorchnew} grows no faster with $\tau$ than
$e^{\tau}$, tells us that the correlator \eqref{normcorchnew} can grow no faster in the small $\sigma$ limit than $\frac{1}{\sigma}$ at any fixed $\rho$. In other words the Chaos bound asserts that $A' \leq 2$.  

Although the chaos bound holds only on the Causally Regge sheet, the fact that $A'$ is the same on the Causally Regge and Causally Scattering sheets tells us that $A' \leq 2$ even on the Causally Scattering sheet in the situation studied in this paper (i.e. correlators induced by a local bulk contact term).

\subsection{The CRG bound from the chaos bound}

Putting together the results of the previous two subsections it follows that  
\begin{equation}\label{aineq2} 
A \leq A' \leq 2, 
\end{equation}

Now using \eqref{gsingnngen} it is easy to convince oneself that there is always a choice of boundary polarizations for which any given bulk polarization appears on the scattering $\mathbb{H}_{D-2}$. 
It follows that \eqref{aineq} must hold for every choice of bulk polarizations. In other words, the chaos bound implies the CRG conjecture.

\subsection{Cross correlators}

The argument presented earlier in this section made important use of the normalized four-point function. As this object is undefined if either $G_{12}$ or $G_{34}$ vanish, it may at first seem that the argument of this paper is restricted to the scattering of particles that are `equal in pairs'. This is not the case. Suppose we are given four particles that are created by the Hermitian boundary operators $O_1$, $O_2$, $O_3$ and $O_4$ such that $\langle O_1 O_2\rangle=\langle O_3 O_4\rangle=0$. The argument presented in this paper applies unmodified to the scattering created by the four correlators $\langle O_1 O_1 O_3 O_3\rangle$, $\langle O_1 O_1 O_4 O_4\rangle$, $\langle O_1 O_1 O_3 O_3\rangle$, $\langle O_2 O_2 O_3 O_3\rangle$ and $\langle O_2 O_2 O_4 O_4\rangle$. This tells us that the scattering processes 
$$ 13 \rightarrow 13, ~~~14 \rightarrow 14, ~~~
23 \rightarrow 23, ~~~24 \rightarrow 24$$ all obey the CRG conjecture. We can now consider the argument of this paper to the correlator 
$\langle(O_1+\alpha O_2) (O_1+\alpha O_2) O_3 O_3\rangle$. The argument of this paper tells us that a linear combination of
the scattering processes
$$ 13 \rightarrow 13, ~~~23 \rightarrow 23, ~~~13  \rightarrow 23 ~~~$$  
obey the CRG conjecture. But as we already know the first two processes obey this conjecture, this tells us that the same is also true for 
$$ 13 \rightarrow 23$$
Similar arguments establish that all scattering processes in which only one of the product particles is different from one of the reactant particles obey the CRG conjecture. Finally, we  study the correlator $\langle(O_1+\alpha O_2) (O_1+\alpha O_2) (O_3+\beta O_4)  (O_3+ \beta O_4)\rangle$. The argument of this paper tells us that a linear combination of several scattering amplitudes obeys the CRG conjecture. But we have already established that 
all of these amplitudes except 
$$13 \rightarrow 24$$
obey CRG scaling. It thus follows that this last scattering amplitude also obeys the CRG conjecture \footnote{Similar arguments have been used in \cite{Turiaci:2019nwa} to constrain the growth of OTOC of mixed correlators using chaos bound.}. 

\section{Discussion and conclusions}\label{disc}

In this paper, we have demonstrated the following. Consider a conformal field theory that has a local bulk AdS/CFT dual. If the contact interaction terms in this bulk dual lead to a flat space S matrix that grows faster than $s^2$ in the Regge limit, then the corresponding four-point functions of the boundary conformal field theory violate the chaos bound. 

We have established the result of the previous paragraph by studying a two-parameter family of CFT correlators with insertions at the points \eqref{fpscatnnnn} that interpolate between the Causally Regge sheet and the Causally Scattering sheet (see \eqref{causalrel}). Our argument proceeds by first demonstrating that on each of these sheets and in the limit that the conformal cross ratio $\sigma \to 0$, our correlators respectively take the form 
\begin{equation}\label{parseconc}
\frac{g_{CS}(e^{2\rho})}{\sigma^{A'-1}} ~~~~~{\rm and}~~~~~~
\frac{g_{CR}(e^{2\rho})}{\sigma^{A'-1}}
\end{equation}
(here $\rho$ is the second cross ratio). The key point here is that the exponent $A'$ is the same on the Causally Regge and Causally Scattering sheets. We then used the fact that the correlator has a singularity - the so-called bulk point singularity - at $\rho=0$ on the Causally scattering sheet. The coefficient of this singularity is the flat space S matrix of the corresponding bulk modes which we assume to scale like 
$s^{A}$ at fixed $t$. This connection allows us to demonstrate that $A' \geq A$. However, the chaos bound applied to the Causally Regge sheet tells us that $A'\leq 2$. Putting these results demonstrates that 
$A \leq 2$, i.e. implies the CRG conjecture.

The fact the bulk duals to `good' conformal field theories - i.e. CFTs that obey the chaos bound - always obey the CRG conjecture (see the introduction for terminology) seems to us to be very strong evidence for the correctness of the CRG conjecture, and therefore of the recent results of \cite{Chowdhury:2019kaq}.

We emphasize that the correlators studied in this paper (i.e. those generated by local bulk contact terms using holography) are analytically very simple. In particular, a key feature of the small $\sigma$ scaling of these correlators is that the scaling exponent $A'$ is a constant independent of $\rho$. This is the key feature that allows for a simple interpolation between the causally Regge and Causally scattering sheets.

While a link between the chaos bound and the allowed Regge scaling of S matrices has been suspected for some time (see e.g. \cite{Maldacena:2015waa, Caron-Huot:2017vep, Turiaci:2018dht, Chowdhury:2019kaq}, and also Appendix \ref{cb}), to the best of our knowledge this connection has never previously been made precise, especially in the context of the scattering of particles with spin (like the photons and gravitons studied in this paper).

At the technical level, one of the accomplishments of this paper is the generalization of the results of 
\cite{Gary:2009ae}  - which determined the coefficient of the bulk point singularity of a four-point function of scalar operators in terms of the coefficient of the S matrix of the corresponding bulk scalar waves - to a similar result for conserved vectors and the conserved stress tensor (the corresponding S matrices are those of scalars, photons, and gravitons). While the techniques employed in this paper can also be used to study four-point functions of  non-conserved spinning operators corresponding to massive spinning bulk particles, the argument linking the singularity in these correlators to the Regge scaling of flat space S matrices is complicated by the enhanced high energy behavior of the scattering of longitudinal polarizations in the bulk (see subsection \ref{mhss} and Appendix \ref{mhsp}). It would be interesting (and should not prove too difficult) to work through these complications and generalize the tight connection between the CRG conjecture and the chaos bound to particles of arbitrary mass and spin.  

As a slight aside from the main flow of this paper, we highlight an issue that we do not understand. To set the context for this question, let us first recall that the weakened version of the 
bulk point singularity \eqref{zepaimp} was conjectured by the authors of \cite{Maldacena:2015iua} to occur in suitable correlators of scalar operators in all non-interacting theories, even at finite $N$. One explanation for this conjecture, presented in \cite{Maldacena:2015iua}, is  the observation that the boundary configurations that have a bulk point singularity are stabilized by  a non-compact $SO(D-2,1)$ subgroup of the conformal group (see Appendix \ref{obl}). \footnote{If we imagine computing a boundary correlator in conformal perturbation theory, the integration of the  `interaction points' over the orbit of this symmetry would thus appear to give an infinite answer, leading to the bulk point singularity.} Now while boundary configurations that have a bulk point singularity are co-dimension $1$ in cross ratio space and so are special, there exist a less special collection of boundary insertion points for scalar operators that span an $\mathbb{R}^{3,1}$ subspace of embedding space (see Appendix \ref{obl}). These points are co-dimension zero or generic in cross ratio space. \footnote{While the two parameter set of insertions  \eqref{fpscatnnnn} that we have focussed on in this paper are never of this form - indeed this is part of the reason  we chose to study the special set of insertions \eqref{fpscatnnnn} -  points with this property occur in the `neighbourhood' of the Regge point 
$\tau=\theta=0$ of \eqref{fpscatnnnn} (see Appendix 
\ref{reggeblowup}). Indeed in Appendix \ref{reggeblowup} it is the  set of `bulk point singular' $\mathbb{R}^{2,1}$ configurations that separate those configurations that are $\mathbb{R}^{3,1}$ from the `good' $\mathbb{R}^{2,2}$ configurations like 
all of \eqref{fpscatnnnn} at $\tau \neq 0$.}  
The subgroup of the conformal symmetry that is preserved by these  $\mathbb{R}^{3,1}$  configurations,  $SO(D-3, 1)$, is also non-compact when $D \geq 4$, suggesting that these correlators are also ill defined (infinite) unless the coefficient of this divergence vanishes for some unknown reason. 
\footnote{In the context of a holographic computation for a classical bulk dual, such configurations do not give rise to a pinch singularity of the form that we 
get for $\mathbb{R}^{2,1}$ configurations, see around \eqref{sumofpapp}. The origin of this apparent divergence is more elementary; it is simply the infinite volume of the orbits of the $H_{D-3}$ foliations of $AdS_{D+1}$ that is generated by  the preserved symmetry subgroup $SO(D-3,1)$. We thank J. Penedones for discussions on this point. }
This conclusion seems to us to be unphysical. It is presumably possible to define the 
correlator in such `$\mathbb{R}^{3,1}$ configurations' by analytically continuing the 
answer from the better behaved `$\mathbb{R}^{2,2}$ configurations'
\footnote{The fact that conformal blocks are well defined in `$\mathbb{R}^{3,1}$ configurations' (though they diverge in `$\mathbb{R}^{2,1}$ configurations' as pointed out in \cite{Maldacena:2015iua}) suggests this should be possible.}
but we do not understand how such correlators can be computed directly, without resorting to an analytic continuation,  even in the simple context of a holographic computation at leading order in large  $N$ in a theory with a local bulk dual.  As we have mentioned above, in this context the integral of the integration point over foliation hyperboloids $\mathbb{H}_{D-3}$ appears to give an infinite result. While this issue has no bearing on the current paper, it would be nice to clear it up.

In this paper, we have focussed our attention on correlators generated by contact diagrams in the AdS bulk. While we have not thought the issue through very carefully, we believe that the generalization of our paper to the study of bulk exchange diagrams is likely to be straightforward, and the final result of this paper is likely to apply without modification to this case. More ambitiously, in this paper, we have focussed our attention on correlation functions generated by classical dynamics in the bulk. However, the logical flow of the argument presented in this paper would appear, at least at first sight, to go through even accounting for quantum effects in the bulk. As the boundary dual to a quantum bulk theory is a CFT at finite $N$, 
it may be possible to use the `finite $N$ chaos bound' i.e. the finite $N$ Cauchy Schwarz inequality (which forms the starting point of the large $N$ analysis in \cite{Maldacena:2015waa}) to obtain a stronger bound for the Regge growth of bulk quantum S matrices than the $s^2$ bound we have derived for classical S matrices in our paper. We think this is a very interesting direction for future work. \footnote{We thank A. Gadde and S. Caron Huot for related discussion.}

While the results of this paper may be taken to be strong evidence for the correctness of the CRG conjecture at least in the context studied in this paper, the argument presented here is very indirect. It relies on the AdS CFT, the flat space limit of $AdS$, and a theorem (the chaos bound) on quantum field theories to constrain the growth of classical bulk scattering amplitudes. 
It should be possible to give a simple general - possibly classical - argument for the CRG conjecture that does not rely on all these bells and whistles. Given the results of this paper, such an argument could also allow one to `derive' the chaos bound directly from the bulk. We think this is a very interesting problem for the future. \footnote{A. Zhiboedov already has some very interesting results on this question. We thank him for discussions on this point.}

Finally, just as the chaos bound only constrains four-point functions in a conformal field theory, the CRG conjecture only constrains the growth of $2 \rightarrow 2$ scattering amplitudes. It would be very interesting to find nontrivial generalizations of both the chaos bound and the CRG conjecture to high point correlators and multi-particle S matrices. We leave this to the future. 




\acknowledgments
We would like to thank Soumangsu Chakraborty for the initial collaboration and A. Gadde, I. Haldar, S. Kundu, R. Loganayagam, M. Mezei, D. Stanford and A. Zhiboedov for very useful discussions. We would also like to thank S. Caron-Huot, T. Hartman, S. Kundu, J. Penedones, G. Mandal,  A. Sinha and D. Stanford for comments on a preliminary version of this manuscript. Part of this work was presented in `Recent Developments in S-matrix theory' program (code: ICTS/rdst2020/07) in ICTS, Bengaluru, and also in the  YITP workshop YITP-W-20-03 on ``Strings and Fields 2020''. The work of all authors was supported by the Infosys Endowment for the study of the Quantum Structure of Spacetime and by the J C Bose Fellowship JCB/2019/000052.  We would all also like to acknowledge our debt to the people of India for their steady support to the study of the basic sciences.

\appendix

\section{Discussion of remarks in \cite{Maldacena:2015waa} relevant to the CRG conjecture} \label{cb} 

The main theorem of the remarkable paper \cite{Maldacena:2015waa} is the starting point of the analysis of the current paper.

In a discussion of their results, the authors of \cite{Maldacena:2015waa} also anticipated the connection between the chaos bound and the CRG conjecture (see the last paragraph of Section 3 of \cite{Maldacena:2015waa}). Somewhat confusingly, however, in a separate discussion, (the second paragraph of Section 3), the authors of 
\cite{Maldacena:2015waa} assert that no finite set of higher derivative corrections to Einstein's equations affect the fact that Einstein gravity saturates the chaos bound.  As individual higher derivative corrections to Einstein's equations certainly violate the CRG conjecture	(see \cite{Chowdhury:2019kaq}) this assertion apparently contradicts the connection between the chaos bound and the CRG conjecture.  

We believe that the resolution to this apparent contradiction is that the  claim that higher derivative corrections do not modify the chaos scaling of Einstein gravity is incorrect. 
The authors of \cite{Maldacena:2015waa} appears to have based their claim on the expectation that scattering amplitudes that involve only spin two particles always scale like $s^2$ in the Regge limit. As explained around 1.5.2 in \cite{Chowdhury:2019kaq}, while this is correct for pole exchange diagrams in the  $t$ channel, it is not correct either for pole exchange diagrams in the $s$ and $u$ channels or for contact diagrams.

All diagrams that violate the `spin two implies $s^2$' intuition (diagrams whose contributions to S matrices potentially grow faster than $s^2$ in the Regge limit) are polynomials in $t$. It follows that these diagrams contribute to bulk scattering only at zero impact parameter. 
\footnote{Recall that the transformation from $t$ to impact parameters is, roughly, a Fourier transformation.} Despite this fact, these diagrams contribute to correlators in the Regge limit at generic values of the cross ratio $\rho$ and not only at very special values of $\rho$. This follows formally from the fact that correlators are analytic functions of $\rho$, and more physically from the spreading of waves between boundary and bulk. 

In summary, we agree with the expectation (expressed in the last paragraph of Section 3 of  \cite{Maldacena:2015waa}) for a tight connection between the chaos bound and CRG scaling. Indeed this paper may be thought of as an attempt to establish this connection more clearly. We believe, however, that the atleast naively contradictory claim of universality of the saturation of the chaos bound in a class of higher derivative gravitational theories is incorrect. 

Happily, the remarks about universality were made only in a motivational context in \cite{Maldacena:2015waa}; their validity or otherwise does not affect any of the actual conclusions of that remarkable paper.

We thank M. Mezei and D. Stanford for discussion related to this Appendix.



\section{Review of the embedding space formalism} \label{ncbm}

\subsection{Definition of $AdS_{D+1}$} \label{de} 

Through this paper we work with $AdS_{D+1}$ using the so called embedding space formalism, within 
which $AdS_{D+1}$ is thought of as the `unfolding' or universal cover of the sub-manifold 
\begin{equation}\label{embed2}
Y. Y \equiv \eta_{MN} Y^M Y^N=-1
\end{equation} 
in the space $\mathbb{R}^{D,2}$ with line element 
\begin{equation}\label{embdelinel2}
ds^2= \eta_{MN} dY^M dY^N ~~~~
\end{equation}
($\eta_{MN}$ has eigenvalues $(-1,-1, 1 \ldots 1))$.

\subsubsection{Global coordinates} 

For some purposes it is  convenient to choose an arbitrary decomposition of $\mathbb{R}^{D,2}$ as $\mathbb{R}^{D,2}=\mathbb{R}^{0,2} \otimes \mathbb{R}^{D,0}$
\footnote{The inequivalent ways of making this decomposition are labelled by elements of the 
	$2 D$ parameter coset 
	\begin{equation}\label{cogs}
	\frac{SO(D,2)}{SO(D) \times SO(2)}.
	\end{equation} .}
Let $Y_{-1}$ and $Y_{0}$ be Cartesian coordinates on $\mathbb{R}^{0,2}$ and $Y_a$ ($a=1, \ldots, D$) be Cartesian coordinates on $\mathbb{R}^{D,0}$ so the line element on $\mathbb{R}^{D,2}$ is 
\begin{equation}\label{embdelinel}
ds^2= -(dY_{-1})^2 -(dY_{0})^2+ dY_a^2
\end{equation}
and the AdS manifold is given by the equation 
\begin{equation}\label{embed}
-Y_{-1}^2 -Y_{0}^2+Y_M^2=-1
\end{equation} 
The very natural `global AdS' coordinate system 
associated with any such split parametrizes points on 
$AdS_{D+1}$ according to the formulae
\begin{equation}\label{globalpar} \begin{split} 
&Y_{-1} = \cosh \zeta \cos \tau \\
&Y_{0}= \cosh \zeta \sin \tau\\
&Y_a= \sinh \zeta~ {\vec n}^a\\
\end{split} 
\end{equation}
where ${\vec n^a}$ is a unit vector on a unit $D-1$ sphere
The metric in these coordinates is given by
\begin{equation}\label{gcmet}
ds^2=d \zeta^2 -\cosh^2 \zeta d \tau^2 + \sinh^2 \zeta 
d \Omega_{D-1}^2
\end{equation}

\subsubsection{Poincare coordinates} 

The construction of the Poincare Patch begins with the choice of an $\mathbb{R}^{1,1}$ hyperplane (of $\mathbb{R}^{D,2}$) that passes through the origin. \footnote{ Clearly the set of such hyperplanes are parameterized by the $2D$ dimensional coset  
	\begin{equation}\label{cosetp} 
	SO(D,2)/ \left( SO(D-,1) \times SO(1,1) \right) 
	\end{equation} 
	It follows that we have $2D$ inequivalent Poincare patches. }
We label the two lightlike directions of $\mathbb{R}^{1,1}$ as 
$Y_+$ and $Y_-$. These are chosen so that the metric on $\mathbb{R}^{1,1}$ is 
$$ds^2 =-dY_+ dY_-.$$ 
These conditions fix $Y_+$ and $Y_-$: upto the 
$Z_2$ ambiguity of interchanging $Y_+$ and $Y_-$. 

One choice for the $\mathbb{R}^{1,1}$ is the space spanned by $Y_{-1}$ and $Y_D$; once we have made this choice one of the (two possible) choices for the coordinates $Y_+$ and $Y_-$ are 
\begin{equation}\label{ypm} 
Y_+=Y_{-1} +Y_{D}, ~~~~Y_+=Y_{-1}+Y_{D}.
\end{equation} 
With these choices the equation \eqref{embed} and the metric in the embedding space 
can be rewritten as 
\begin{equation}\label{surf} 
-Y_+ Y_- + Y_\mu Y^\mu=-1, ~~~~~~~~~~~~ds^2=- d Y_+ d Y_- + d X_\mu dX^\mu 
\end{equation}
where $\mu$ is an index in $D$ dimensional 
Minkowski space with a mostly positive metric. 

We can obtain an explicit parameterization of the half of space with $Y_->0$   by using the first of \eqref{surf} 
to solve for $Y_+$ and plugging the solution back into \eqref{surf}. In this process it is useful to perform the  
redefinition  
$$ Y^\mu = Y_- ~x_\mu . $$
Renaming $Y_-$ as $u$ we now have 
\begin{equation}\label{ymm}
(Y_+, Y_-, Y^\mu)= (u x^2 + \frac{1}{u}, u, ux^\mu )
\end{equation} 

We find \footnote{
	Using 
	$$dY_+= - \frac{ dY_{-}}{Y_{-}^2} + 
	2(Y_-)x^\mu dx_\mu  +  d Y_- x_\mu x^\mu $$. 
	so that
	$$ - d Y_+ d Y_- + d X_\mu dX^\mu =
	+\frac{ dY_{-}^2}{Y_{-}^2} + 
	- 2(Y_-)x^\mu dx_\mu dY_-  +  (d Y_-)^2 x_\mu x^\mu
	+ d(Y_- x^\mu) d (Y_- x^\mu)  $$}
\begin{equation}\label{ppmet} \begin{split} 
&ds^2=
\frac{ d u}{u^2} + u^2 dx_\mu  dx^\mu
\end{split} 
\end{equation}
$u$ is the usual Maldacena (energy scale) coordinate
for the Poincare patch metric. Note that this metric 
in this coordinate becomes singular when $u=Y_-=0$.

 The 
region of negative $Y_-$ is a second Poincare Patch. 
This two Poincare patches together cover all of the manifold \eqref{embed}.

It is not difficult to visualize the half of $AdS_{D+1}$ that has $Y_->0$ and so is contained in a single Poincare patch. We work in global coordinates, \eqref{globalpar} and choose coordinates on the sphere so that $Y_D=\sinh \zeta \cos \phi$. $\phi$ is the `angle with the $Y_D$ axis'. $\phi=0$ is the `north pole' of the sphere, while $\phi=\pi$ is the `south pole' of the sphere. Note $0 \leq \phi \leq \pi$. Let us first characterize the intersection of the Poincare patch with the boundary of $AdS_{D+1}$. To do this we take limit $\zeta \to \infty$ where
$$ Y_+ \propto (\cos \tau +\cos \phi), ~~~
Y_- \propto (\cos \tau -\cos \phi) $$
The intersection of the Poincare patch with the boundary consists of points with  $\phi > |\tau|$. \footnote{And so includes all of the boundary sphere at $\tau=0$, progressively less of this sphere (a region around the north pole is excluded) as $|\tau|$ increases, and only a neighbourhood around the south pole at $|\tau|\to \pi$.}

The region above may be characterized invariantly as follows. Consider the north pole at $\tau=0$. 
\footnote{In the labelling of boundary points \eqref{nullboundary} and \eqref{equivbound}), this is the point $Y_+=0$}. The intersection of the Poincare patch and the boundary is the complement of the past and future boundary lightcones of this point. 

The description of the previous paragraph applies with very little modification in the bulk as well. 
The full Poincare patch is the complement of the (past and future) bulk lightcones of this distinguished boundary point.

\subsection{General analysis in embedding space} \label{es}

While the `global' coordinate system \eqref{globalpar} is familiar and useful for many purposes, and the Poincare coordinates above are convenient for other purposes, each of them suffers from defects. For example, any choice of global coordinates involves an arbitrary splitting of $\mathbb{R}^{D,2}$ into $\mathbb{R}^D \otimes \mathbb{R}^2$ and so  obscures the $SO(D,2)$ invariance of our space. Poincare coordinates share a similar (and more severe) defect of this sort. They also 
cover only half of $AdS$ space. It is often convenient not to tie ourselves to any particular coordinate system, but to employ a more intrinsically geometrical view, regarding \eqref{embed} and \eqref{embdelinel} as the fundamental coordinate independent definitions of our space. In this subsection, we describe some details of this approach.

The geodesic distance, $d(U, V)$  between two points $U$ and $V$ on \eqref{embdelinel} is given by 
\begin{equation}\label{geodist} 
\cos \left( d(U, V) \right) = U.V 
\end{equation}
where $U.V$ is the standard dot product in the embedding space $\mathbb{R}^{D,2}$. 
\footnote{This formula is an analytic continuation of a familiar fact of Euclidean geometry. Recall that the geodesic distance between two points on the Euclidean unit sphere is the angle between them. It follows that the cosine of the geodesic distance is the dot product of the unit vectors from the center of the sphere to the two points.}
In particular, $U$ and $V$ are null related (i.e. lie on the same null geodesic) if and only if $U.V=0$. 

Within the embedding space formalism the  boundary of 
$AdS_{D+1}$ is the collection of null rays in $\mathbb{R}^{D,2}$, i.e. by points $P$ in $\mathbb{R}^{D,2}$ such that 
\begin{equation} \label{nullboundary}
P^2=0
\end{equation}
subject to the equivalence relationship 
\begin{equation} \label{equivbound}
P \sim \lambda(P) P, ~~~~\lambda(P) >0
\end{equation} 
It follows that the tangent space to the boundary point $P$ given by the set of vectors $\delta P$ orthogonal to $P$ (i.e. $P.\delta P=0$) subject to the equivalence relation $ \delta P \sim \delta P+ a P$ where $a$ is any real number. 

If we wish to parameterize boundary points by particular null vectors $P$ rather than equivalence 
classes \eqref{equivbound} of such vectors we need to 
choose a `gauge' - say of the schematic form 
\begin{equation} \label{gcc}
\chi(P)=0, ~~~~~~~~~\chi(P) \sim \chi(P) + P^2 \chi'(P)
\end{equation}
to fix the ambiguity \eqref{equivbound} (the equivalence relationship in \eqref{gcc} follows because $\chi$ should be evaluated only at the boundary). The (gauge dependent) one-form field 
\begin{equation}\label{defn}
n= d\chi, ~~~~~~~~(n \rightarrow n+ 2\chi' P)
\end{equation} 
\footnote{The bracketed equation in \eqref{defn} displays how $n$ transforms under the `gauge' transformations (second o \eqref{gcc}).}
allows us to define the tangent space of the boundary 
in the gauge slicing \eqref{gcc}; the allowed class of $\delta P$ are those vectors that are orthogonal 
to the plane generated by  $n$ and $P$. This two plane is typically an $\mathbb{R}^{1,1}$ - and so $\delta P$  is constrained to lie in the orthogonal $\mathbb{R}^{D-1,1}$. 
Note that unlike $n$ itself this two-plane is gauge invariant under $\chi'$ shifts \eqref{gcc}.

Any such choice of gauge leads to a metric on the boundary given by 
\begin{equation}
ds^2= \left( \delta P \right)^2
\end{equation} 
Under the shift \eqref{equivbound} 
induces the shift 
$$ds^2 \rightarrow \lambda(P)^2 \left(\delta P \right)^2 $$
(upto terms of third order in infinitesimals which we ignore). It follows that $\lambda$ in \eqref{equivbound} is a Weyl factor for an effective Weyl transformation. 

One example of a  gauge choice which fixes the choice of $\lambda$ and hence of Weyl frame is 
\begin{equation}\label{globgc}
\chi(P)= P_{-1}^2+P_{0}^2-1=0.
\end{equation}  
With this choice the boundary is parametrized, in \eqref{globalpar},  by the points 
$$(\cos \tau, \sin \tau, {\hat n})$$
(where ${\hat n}$ is a unit vector on $\mathbb{R}^D$). 
The boundary metric with this choice and with these coordinates is 
$$ds^2 -d \tau^2 + d \omega_{D-1}^2,$$
i.e the metric on a unit sphere times time. Note that with this choice of coordinates 
\begin{equation}\label{tcc}
-2P_1.P_2= 2\cos(\tau_1-\tau_2) -2{\hat n}_1. {\hat n}_2
\end{equation} 

Another choice of Weyl frame is 
\begin{equation}\label{globg}
\chi(P)= (Y_{-}-1)=0
\end{equation} 
With this choice the boundary is parameterized, in the coordinates of \eqref{ymm} by 
$$(x^2, 1, x^\mu)$$ 
The boundary metric with this choice of Weyl frame and these coordinates is 
$$ dx_\mu dx^\mu$$
the usual metric on Minkowski space. With this choice of coordinates
\begin{equation}\label{tccp}
-2P_1.P_2= (x_1-x_2)^2
\end{equation} 

\subsection{Light-cones emanating out of a boundary point in embedding space} \label{leb} 

Consider the boundary point $P$. The light sheet that emanates out of the boundary point $P$  is 
given by the set of points $X.P=0$ and $X^2=-1$. This is a $D$ dimensional null sub-manifold of 
$AdS_{D+1}$; the normal one form to this manifold is $P$. The non-degenerate sections of this manifold - parameterized by the equivalence classes $X \sim X+P$ - are labeled by  the distinct null geodesics that generate this manifold. The tangent along every point along  each of these geodesics is given by the vector $P$; note that this embedding space vector also lies in the tangent space of $AdS_{D+1}$ because 
$P.X=0$.

The set of all equivalence classes of  vectors in $\mathbb{R}^{D,2}$ that are orthogonal to a null vector $P$ and are subject to the identification 
$V \sim V+ a P$ (for any $a$) is an $\mathbb{R}^{D-1,1}$. 
The space of geodesics (the space of equivalence classes of the previous paragraph) is the set of points in this $\mathbb{R}^{D-1,1}$ that obey the $AdS_{D+1}$ equation $X^2=-1$. It follows that the set of geodesics is the space $\mathbb{H}_{D-1}$.  

The space of polarizations along a geodesic parameterized by the point $X$ in $\mathbb{H}_{D-1}$ is given by the set of vectors $\epsilon$ such that 
$\epsilon. P=\epsilon.X=0$, modulo the shifts 
$\epsilon \sim \epsilon + P$ \footnote{The first condition is the Lorentz gauge condition $\partial.A=0$. The second condition asserts that our polarization lies in the $AdS_{D+1}$ manifold. The last condition is the residual gauge invariance on onshell configurations after we have imposed the Lorentz gauge; $\epsilon \sim 
	\epsilon + \alpha k$. }. This space of polarizations spans an $\mathbb{R}^{D-1}$. This $\mathbb{R}^{D-1}$ can be thought of as the tangent space of $\mathbb{H}_{D-1}$. Equivalently it is  the subspace of the tangent space of $P$ (i.e. vectors that are orthogonal to $P$ modulo shifts of $P$) that is orthogonal to the point $X$.

The full $\mathbb{H}_{D-1}$ family of distinct geodesics `meet' at the boundary point $P$. It follows that the set of allowed polarizations at the boundary point $P$  - the vector space of associated spanned by the full set  of the polarizations 
above over all allowed values of $X$ -  is simply the tangent space of $P$. 

\subsection{Overlap of boundary lightcones of $Q'$ points in embedding space} \label{obl}

Consider a set of $Q'$ boundary points $P^M_a$ $(a=1 \ldots Q')$. The overlap of the lightcones of these $Q'$ points is given by the set of simultaneous solutions to the equations 
\begin{equation}\label{pam}
P_a^M X_M=0
\end{equation} 
and the $AdS_{D+1}$ condition $$X^2=-1.$$ 
Let us suppose ${\tilde Q}$ of these equations are linearly independent, i.e. the set of points 
$P^M_a$ span a ${\tilde Q}$ dimensional subspace.
Clearly ${\tilde Q} \leq D+2$ (the dimensionality of embedding space).  If ${\tilde Q} =D+2$ these lightcones never intersect. 
\footnote{We see this as follows. In this situation the only solution of \eqref{pam} is 
	$X_M=0$ for all $M$. But this solution does not obey $X^2=-1$.}. For the rest of this subsection we assume that ${\tilde Q} \leq D+1$. The space spanned by the  vectors $P^M_a$ is either 
\begin{enumerate}
	\item $\mathbb{R}^{{\tilde Q}-2,2}$
	\item $\mathbb{R}^{{\tilde Q}-1,1}$
	\item Null with non-degenerate sections $\mathbb{R}^{{\tilde Q}-2,1}$
	\item Null with non-degenerate sections 
	$\mathbb{R}^{{\tilde Q}-1,0}$
	\item Doubly null with non-degenerate sections $\mathbb{R}^{{\tilde Q}-3,0}$. 
\end{enumerate}

For future convenience we define 
$$Q=D-{\tilde Q}+1$$
Note $Q \geq 0$. 

In case (1) above the  the isometry group that 
stabilizes all $P_a^M$ is $SO(Q+1)$. 
The space of solutions to \eqref{pam} is an $\mathbb{R}^{Q+1,0}$. This solution set has no intersection with the $AdS$ equation $X^2=-1$. It follows that the lightcones of corresponding four points never intersect. 

In case (2) above the isometry subgroup that stabilizes the points $P^M_a$ is 
$SO(Q,1)$. The space of solutions of \eqref{pam} is $\mathbb{R}^{Q,1}$. The intersection of this space with $X^2=-1$ is $\mathbb{H}_{Q}$, a $Q$ dimensional  (Euclidean) hyperboloid.

In case (3) above the space of solutions of 
\eqref{pam} is a null manifold with non-degenerate sections $\mathbb{R}^{Q,0}$. None of the 
solutions obey $X^2=-1$ and so the light cones do not intersect.

In case (4) above the space of solutions of \eqref{pam} is a null manifold with spatial sections  
$\mathbb{R}^{Q-1,1}$. In this case, the intersection 
of lightcones is a null sub-manifold, whose non-degenerate sections are an $\mathbb{H}_{Q-1}$ 
(the null combination of $P^M_a$ is the normal to this sub-manifold).

In case (5) above the space of solutions of 
\eqref{pam} is a doubly null sub-manifold with 
non-degenerate sections are $\mathbb{R}^{Q-2,0}$. None of these solutions obey $X^2=-1$ so the light-cones never intersect in this case.  

Note that in cases (2) and (4), where the intersections of light-cones is nontrivial, the intersection manifold is a homogeneous space;  any point on the intersection manifold can be mapped to any other by the action of the isometry subgroup that preserves all $P_a^M$. 

In this subsection, we have presented a complete classification of all possible subspaces preserved
spanned by a collection of $P_a$. In the special case of interest to this paper - namely the insertions \eqref{fpscatnnnn}, we only encounter 
Case 1 with  ${\tilde Q}=4$ (when $\tau \neq 0$)
or Case 1 with ${\tilde Q}=4$ (when $\tau \neq 0$). In our exploration of the neighbourhood of 
the Regge point we also encounter Case 2 with 
${\tilde Q}=4$ (this happens inside the ellipse depicted in Fig. \ref{R31R22}). Nowhere in this paper do we ever encounter Cases 3, 4, or 5.

\subsection{Conformal cross ratios and intersection lightcones for four boundary points} \label{ccra}

In this subsection we study the insertion of four boundary points on $AdS_{D+1}$. 

Any $SO(D,2)$ invariant expression in the  four points $P_a$ that is separate of homogeneity zero in each of the four points  (so that the expression is `gauge' invariant under the gauge scalings $P_a \rightarrow \lambda(P) P_a$) 
is a conformal cross ratio. In this section 
we briefly recall the definitions of commonly used cross ratios. 

\subsubsection{$u$ and $v$}

The conformal cross ratios $u$ and $v$ are defined by 
\begin{equation}\label{uvdef} 
\begin{split} 
&u = \frac{(P_2.P_1) (P_3.P_4)}{(P_2.P_4) (P_3.P_1)}\\
&v=\frac{(P_2.P_3)(P_4.P_1)}{(P_4.P_2)(P_3.P_1)}
\end{split} 
\end{equation} 

\subsubsection{$\sigma$ and $\rho$}

The conformal cross ratios $\sigma$ and $\rho$ are defined by 
\begin{equation}\label{sigrhodef2} 
\begin{split} 
&\sigma^2=u\\
& 1-2 \sigma \cosh \rho + \sigma^2= v\\
\end{split} 
\end{equation} 

\subsubsection{$z$ and $z$ bar }

The conformal cross ratios $z$ and ${\bar z}$ are defined by the relations 
\begin{equation}\label{zbzdef} 
\begin{split} 
&u = z {\bar z}\\
&v= (1-z)(1-{\bar z})\\
\end{split} 
\end{equation} 

\subsubsection{$z$ and $z$ bar in terms of $\sigma$ and $\rho$}

It follows that $z$ and ${\bar z}$ are the two solutions to the quadratic equation 
\begin{equation}\label{qeq}
x^2-2x \sigma \cosh \rho + \sigma^2=0
\end{equation} 
The two solutions to this equation are 
\begin{equation}\label{soleq}
x= \sigma \left( \cosh \rho  \pm \sinh \rho 
\right) 
\end{equation} 
In particular our conventions are 
\begin{equation}\label{zrs}
z=\sigma e^{\rho}, ~~~~{\bar z}= \sigma e^{-\rho}
\end{equation} 

Note that $z$ and ${\bar z}$ are real and independent when $\rho$ is real (i.e. when 
$\cosh \rho >1$) but are complex conjugates of each other when $\rho$ is imaginary (i.e. when 
$\cosh \rho<1$). 

In the second case it is convenient to set 
$\rho =i \phi$ in terms of which 
\begin{equation}\label{soleq2}
z= \sigma e^{  i \phi }, ~~~~{\bar z}= \sigma e^{-i \phi}
\end{equation}

\subsubsection{More symmetric expression for $\sigma$ and $\rho$}

Let us adopt the notation 
\begin{equation}\label{pij}
P_i.P_j= p_{ij}
\end{equation} 
Using
\begin{equation} \label{sigfho} \begin{split} 
\sigma^2 =& \frac{p_{12} p_{34}}{p_{13} p_{24}}, \\
1+\sigma^2 -2 \sigma \cosh \rho=& \frac{p_{23} p_{14}}{p_{13} p_{24}}, \\
\end{split}
\end{equation} 
it follows that 
\begin{equation} \label{sigfhon} \begin{split} 
\sigma^2 =& \frac{p_{12} p_{34}}{p_{13} p_{24}}, \\
\cosh^2 \rho=& \frac{ \left( p_{23} p_{14} -p_{13} p_{24} - p_{12}p_{34} \right)^2}{4 p_{12} p_{34} p_{13} p_{24}}, \\
\end{split}
\end{equation} 
or with a slight rearrangement
\begin{equation} \label{sigfhonn2} \begin{split} 
\sigma^2 =& \frac{p_{12} p_{34}}{p_{13} p_{24}}, \\
\sinh^2 \rho& = \frac{ p_{12}^2 p_{34}^2+ p_{13}^2 p_{24}^2+ p_{14}^2 p_{23}^2
	-2 \left( p_{12}p_{23} p_{34} p_{41} + p_{13}p_{32}p_{24}p_{41} + p_{13}p_{34} p_{42}p_{21} \right) }{4 p_{12} p_{34} p_{13} p_{24}}
\end{split}
\end{equation}
\footnote{The algebra in the last step is 
	\begin{equation} \begin{split} 
	\cosh^2 \rho-1&=\frac{ \left( p_{23} p_{14} -p_{13} p_{24} - p_{12}p_{34} \right)^2 -4 p_{12} p_{34} p_{13} p_{24}}{4 p_{12} p_{34} p_{13} p_{24}}\\ & =
	\frac{ p_{12}^2 p_{34}^2+ p_{13}^2 p_{24}^2+ p_{14}^2 p_{23}^2
		-2 \left( p_{12}p_{23} p_{34} p_{41} + p_{13}p_{32}p_{24}p_{41} + p_{13}p_{34} p_{42}p_{21} \right) }{4 p_{12} p_{34} p_{13} p_{24}}.
	\end{split} 
	\end{equation}
	The equality of the first and second lines can be seen by 
	expanding the square in the first line.}
For any $4 \times 4$ matrix $p_{ij}$  
\begin{equation}\label{detp}
{\rm Det}(p_{ij}) = 
p_{12}^2 p_{34}^2+ p_{13}^2 p_{24}^2+ p_{14}^2 p_{23}^2
-2 \left( p_{12}p_{23} p_{34} p_{41} + p_{13}p_{32}p_{24}p_{41} + p_{13}p_{34} p_{42}p_{21} \right) 
\end{equation} 
and so \eqref{sigfhon} can be rewritten as (see eq. (3.20) of \cite{Gary:2009ae})
\begin{equation} \label{sigfhonn} \begin{split} 
\sigma^2 =& \frac{p_{12} p_{34}}{p_{13} p_{24}}, \\
\sinh^2 \rho& = \frac{ {\rm Det}(p_{ij})}{4 p_{12} p_{34} p_{13} p_{24}}
\end{split}
\end{equation}

For the special case of the insertions \eqref{fpscatnnnn} we have
\begin{equation}\label{dotprod} \begin{split}
&-P_1.P_3=-P_2.P_4=2\\
& -P_1.P_2=-P_3.P_4= \cos\theta -\cos \tau \\
& -P_1.P_4=-P_2.P_3= -\cos \tau -\cos \theta \\
\end{split}
\end{equation} 
\footnote{As a quick qualitative check of \eqref{dotprod}, note that they (together with the fact that $P_a$ and $P_b$ are space/time like separated in embedding space if $-P_a.P_b$ is 
positive/negative see around \eqref{geodist}) are consistent with the causal relations \eqref{causalrel}.} 
Inserting \eqref{dotprod} into 
\eqref{sigfhonn} reproduces \eqref{crrnnnr}.

\subsection{Classes of cross ratios} \label{cocr} 

As we have seen above, the conformal cross ratios fall into two broad classes. Cross ratios of Type I are those for which $z$ and ${\bar z}$ are complex conjugates of each other. Cross ratios of Type II are those for which $z$ and ${\bar z}$ are real and independent of each other. 

Cross ratios of Type I take the form \eqref{soleq2} with $\sigma$ and $\phi$ both real. For such cross ratios the conformal cross ratio $\sigma$ is real while $\rho=i \phi$ is imaginary. In this case 
\begin{equation} \label{rhocc}
\sinh \rho =i \sin \phi, ~~~~\sinh^2 \rho= - \sin^2 \phi \leq 1
\end{equation} 

cross ratios of Type II are of four sorts. Type IIa are those for which $z$ and ${\bar z}$ are both positive. 
Type IIb are those for which $z$ and ${\bar z}$ are both negative. Type IIc are those for $z$ is positive and  ${\bar z}$ is negative. Type IId is the reverse; those for which $z$ is negative and ${\bar z}$ is positive.

In the case of Type IIa configurations, $\sigma>0$ and 
$\rho$ is real. In this case $\sinh^2 \rho$ is real and positive. The configurations \eqref{fpscatnnnn} are all of Type IIa.  In the case of Type IIb configurations 
$\sigma<0$ and $\rho$ is real. Once again, in this case $\sinh^2 \rho$ is real and positive.  configurations of Type IIc are those for $\sigma= -i \alpha$ with $\alpha>0$ and $\rho= \zeta + i \frac{\pi}{2}$ with $\zeta$ real. In this case 
$\sigma^2= -\alpha^2$ and $\sinh^2 \rho = -\cosh^2 \zeta$. Finally configurations of Type IId are those for which $\sigma= i \alpha$ with $\alpha>0$ and $\rho= \zeta + i \frac{\pi}{2}$. Once again in this case $\sigma^2= -\alpha^2$ and $\sinh^2 \rho = - \cosh^2 \zeta$. \footnote{In the case of configurations of type IIc we could also have set 
$\sigma= i \alpha$ with $\alpha>0$ and $\rho= \zeta + -i \frac{\pi}{2}$. In the case of configurations of type IId we could also have set 
$\sigma= -i \alpha$ with $\alpha>0$ and $\rho= \zeta + i \frac{\pi}{2}$. Which choice we make is unimportant; in particular $\sigma^2$ and $\sinh^2 \rho$ are left unaffected.}

Note that $\sinh^2 \rho$ is positive (and varies in the range $(0,\infty)$ for configurations of type IIa and IIb). On the other hand, $\sinh^2 \rho$ and has a modulus less than or equal to unity in the case of configurations of Type I. Finally, for configurations o Type IIc and IId,  $\sinh^2\rho$ is negative and of modulus greater than unity.

\subsection{Lightcones and cross ratios} \label{lcr}

In this subsection, we will initiate an investigation into the relationship between the causal properties of the four boundary points, the reality (or otherwise) of $\rho$ and the $\mathbb{R}^{p,q}$  classification of these points (see section \ref{obl}).

All through this subsection, we work on the manifold \eqref{embed2}, not on its universal cover. By `causal relations' we mean only the following: given a pair of points, is the separation between them  spacelike or timelike on the manifold \eqref{embed2} (recall this manifold has a compact time circle)? \footnote{The relationship between this spacelike/ timelike dichotomy and causality in the covering space is the following. If two points are timelike separated in embedding space manifold \eqref{embed2} then their pre images are necessarily timelike separated in the covering space. However, points that are spacelike separated in the embedding space manifold \eqref{embed2}  may be either spacelike or  timelike separated in covering space.}

\subsubsection{$\mathbb{R}^{3,1}$} \label{rto}

Let us first assume that the vectors $P_i$ span an $\mathbb{R}^{3,1}$. In this case ${\rm Det}(p_{ij})$ - the determinant of the metric in a coordinate system oriented towards $P_a$ - is negative. Recall the vectors $P_a$ are all null; each of these vectors is either past or future directed. Let us define the variable $\epsilon_a=1$ when $P_a$ is future directed but $\epsilon_a=-1$ when $P_a$ is past directed. Then $p_{ab}$ has the sign of  $-\epsilon_a \epsilon_b$. As each $P_a$ appears twice in the expression $p_{12} p_{34} p_{13} p_{24}$, it follows that this expression is positive. We conclude from \eqref{sigfhonn} that in this case $\sinh^2 \rho$ is negative and so $\rho$ is imaginary.

What are the boundary `causal' relations between points in this situation?   With this clarification, two points $P_a$ and $P_b$  are spacelike separated in embedding space if $-2P_a. P_b=\epsilon_a \epsilon_b$ is positive, and are timelike separated if this quantity is negative.

With this terminology, an $\mathbb{R}^{3,1}$ situation is consistent with three distinct `causal' configurations (upto permutations of particle labels). 
\begin{itemize} 
	\item When all $\epsilon_a$ have the same sign all points spacelike separated from each other. \item When - say - $\epsilon_1$ has a different sign from $\epsilon_i$ ($i=2,3,4$) is a configuration in which $P_2$, $P_3$ and $P_4$ are mutually spacelike, but are timelike separated from $P_1$. 
	\item When - say - $\epsilon_1$ and $\epsilon_2$ have the same sign, but this sign is different from the sign of $\epsilon_3$ and $\epsilon_4$, $P_1$ and $P_2$ are mutually spacelike, $P_3$ and $P_4$ are mutually spacelike, but the pair $(P_1, P_2)$ are timelike separated from $(P_3,P_4)$ 
	\footnote{This is a `causal' relation similar to the scattering configuration we study in this paper, even though the $P_a$ in the scattering configuration span an $\mathbb{R}^{2,2}$ not an $\mathbb{R}^{3,1}$.}
\end{itemize} 	
	
	 We reiterate that as long as the $P_a$ span an $\mathbb{R}^{3,1}$, the cross ratio $\rho$ is imaginary in each of these causal configurations.
	 As $\rho$ is imaginary, the cross ratios must, in the classification of the previous subsection,  be either of Type 1 ($z$ and ${\bar z}$ complex conjugates of each other) or of Type IIc or Type IId ($z$ and ${\bar z}$ both real, but one positive and the other negative).

\subsubsection{$\mathbb{R}^{2,2}$} \label{brtt}

When the four points $P_a$ span an $\mathbb{R}^{2,2}$ ${\rm Det}(p_{ij})$ is positive, so it follows from 
\eqref{sigfhonn} that $\rho$ is real if an even number of the pairs of the tuples 
$(12), ~~(13)~~(24)~~(34)$ are timelike separated, but $\rho$ is imaginary if an odd number of these pairs are timelike separated. In other words the reality or otherwise of $\rho$ is determined by the causal relations between points. We now turn to 
examining the possibilities for these causal relations.

Note that, in this case, the vectors $P_a$ can each (by scaling) be put in the form 
\begin{equation} \label{Pant}
P_a=( \cos \tau_a, \sin \tau_a, \cos \theta_a, \sin \theta_a)
\end{equation} 
in a Cartesian coordinate system with signature $\mathbb{R}^{2,2}$ ($\tau_a$ and $\theta_a$ are both angle valued). It follows that 
$$-p_{ab}=\cos(\tau_a-\tau_b)-\cos(\theta_a-\theta_b)$$
Now suppose that all $\tau_a$ are concentrated around a given point while the four $\theta_a$ are widely spread on the circle. 
Then all $-p_{ab}$ are positive, so all four points are spacelike separated from each other. In this case it follows from \eqref{sigfhonn} that $\rho$ is real. In the opposite configuration - when all $\theta_a$ are near to each other and all $\tau_a$ widely separated - then all points are mutually timelike separated, and $\rho$ is still real. 

There are many other possible configurations. For instance let the times of $P_1, P_2, P_3$ be clumped around a given point, the time of $P_4$ be near the opposite end of the time circle , and all angle separated at distances large compared to the first time differences but small compared to $\pi$. Then 
$P_4$ is timelike separated from all other points - while the rest are mutually spacelike separated. This configuration also has real $\rho$. (A similar configuration with the role of $\tau$ and $\theta$ flipped - with $P_1$, $P_2$, $P_3$ mutually timelike separated but all spacelike separated from $P_4$- also have real 
$\rho$).

Similarly if the times of $P_1$ and $P_2$ are near to each other the times of $P_3$ and $P_4$ are also near to each other but the times of each pair are separated by a larger amount, and the angular separation between all particles is large compared to the separation between time separations within a pair, but small compared to the time separation between pairs then $(P_1, P_2)$ are mutually spacelike separated, and $(P_3,P_4)$ are mutually spacelike separated, but the two  pairs are mutually timelike separated. Once again $\rho$ is 
real in this configuration. 

Now consider yet another configuration - one in which $\tau_1=\tau_2=\tau_3=0$, $\tau_4=\frac{3 a}{2}$, $\theta_1=0$, 
$\theta_2=a$, $\theta_3=2a$ and $\theta_4=3a$. In this configuration $-p_{34}$ is negative (so $P_3$ and $P_4$ are spacelike separated) while all other points are timelike separated. In this configuration $\rho$ is imaginary. 

Let us summarize. While we have not attempted a complete careful classification of the allowed causal configurations for $\mathbb{R}^{2,2}$, it appears at first sight that  these configurations allow for virtually any causal combinations; some of these causal configurations have real $\rho$ while other have imaginary $\rho$, as spelled out by the rule enunciated at the beginning of this subsubsection. 

As explained in the previous subsection, configurations with real $\rho$ are either of 
Type IIa or IIb, ($z$ and ${\bar z}$ both real and either both positive or both negative). In the case that $\rho$ is imaginary cross ratios must be either of Type 1 ($z$ and ${\bar z}$ complex conjugates of each other) or of Type IIc or Type IId ($z$ and ${\bar z}$ both real, but one positive and the other negative).

\subsubsection{Null subspaces} 

Finally if the $P_i$ span a three or lower dimensional subspace of $\mathbb{R}^{D,2}$ or if the four dimensional manifold spanned by these vectors is null, ${\rm Det }(p_{ij})$, and so $\rho$ vanishes. We have not carefully investigated the question of which causal configurations are consistent with these null configurations. We leave it to the interested reader to fill this gap.

\subsection{Weyl weights and scaling dimensions} 

Recall that the combination of a conformal diffeomorphism and 
a compensating Weyl transformation leaves the metric invariant. 
A CFT is a theory that is Weyl covariant. In such a theory the correlators of a primary operator $O_{A_1 \ldots A_n}$ on a space with metric $g_{AB}$ are the same as the correlators of 
a primary operator $e^{w \phi} O_{A_1 \ldots A_n}$ on the space 
with metric $e^{2 \phi} g_{AB}$. In a schematic equation 
\begin{equation}\label{scho}
(O_{A_1 \ldots A_n}, g_{AB})= (e^{w \phi} O_{A_1 \ldots A_n}, 
e^{2 \phi} g_{AB})
\end{equation} 
\footnote{This definition is chosen to ensure that a scalar operator of Weyl weight $w$ has scaling dimension $w$. We can see this as follows. Let $e^\phi=\lambda$ be a constant. Consider the two point function of a scalar operator $O$. If $\lambda$ is small then the action of scaling on the metric reduces the proper distance between the insertions by a factor of $\lambda$, and so increases the 2 point function by a factor of 
	$\frac{1}{\lambda^{2 \Delta}}$ where $\Delta$ is the scaling dimension of the operator. The equality \eqref{scho} thus tells us that $\lambda^{w-\Delta}=1$ so that $w=\Delta$. We will figure out the connection between the Weyl weight and the scaling dimension for tensor operators below.}
We call the real number $w$ the Weyl weight of the primary operator $O$. Note that if the operator $O_A$ has weight $w$ by 
this definition then 
\begin{equation} \label{oaba} \begin{split} 
&(O_{A}, g_{CD})= (e^{w \phi} O_{A}, 
e^{2 \phi} g_{CD})\\
\implies &(g^{BA}O_{A}, g_{CD})= (e^{w \phi} g^{BA}O_{A}, 
e^{2 \phi} g_{CD})\\
\implies &(g^{BA}O_{A}, g_{CD})= (e^{(w +2)\phi} ~(e^{-2 \phi} g^{BA})~O_{A}, 
e^{2 \phi} g_{CD})\\
\implies &(O^B, g_{CD})= (e^{(w +2)\phi} O^B, 
e^{2 \phi} g_{CD})\\
\end{split}
\end{equation} 
and so it follows that the operator $O^A$ has weight $w+2$. The general rule is that raising the index of an operator increases its Weyl weight by two. 

Let us now specialize this discussion to the space $\mathbb{R}^{D}$ (or $\mathbb{R}^{D-1,1}$). Let $X^A$ represent Cartesian coordinates in this space. 
and let the metric in $X^A$ coordinates be $\eta_{AB}$. We first perform a Weyl transformation on this space with a constant $\phi$ (and use the variable $e^\phi=\lambda$) and then perform the coordinate transformation
\begin{equation}\label{ct}
X^A = \frac{{\tilde X}^A}{\lambda}.
\end{equation} 
After the Weyl transformation, the metric on our space is $\lambda^2
\eta_{AB}$. It follows that after this Weyl transformation 
$$ds^2=\lambda^2 \eta_{AB} dX^A dX^B= \eta_{AB} d {\tilde X}^A d{\tilde X}^B$$
\footnote{This is just a complicated way of saying that $g_{AB}= e^{2 \phi} {\tilde g}_{AB}$.}
It follows that 
\begin{equation}\label{scht}
(O_{A_1 \ldots A_n}^{B_1 \ldots B_m}(X), \eta_{AB})= (\lambda^w O_{A_1 \ldots A_n}^{B_1 \ldots B_m}(X), 
\lambda^2 \eta_{AB})= (\lambda^{w+n -m} O_{A_1 \ldots A_n}^{B_1 \ldots B_m}(\lambda X),  \eta_{AB})
\end{equation} 
The $=$ in this equation means 
`has the same correlation functions as'. 
In the first equality in \eqref{scht} we have performed a Weyl transformation with the constant Weyl factor $\lambda$. In the second equality, we have made the variable change \eqref{ct}. Note that if an operator is inserted at the point $X$ in the $X$ coordinate system, it is inserted at the point 
${\tilde X}= \lambda X$ in the ${\tilde X}$ coordinate system. This is why the argument of insertions in the third bracket in \eqref{scht} is $\lambda X$. \footnote{We emphasize that in any of the brackets above, the value of the argument is simply the location - in the coordinate system relevant to that bracket - of the operator insertion}.

Comparing the first and the third brackets, setting and suppressing the metric (as it is 
$\eta_{AB}$ on both sides) we conclude in summary that
\begin{equation}\label{schth} 
O_{A_1 \ldots A_n}^{B_1 \ldots B_M}(\lambda X)= \frac{O_{A_1 \ldots A_n}^{B_1 \ldots B_m}(X)}{\lambda^{w+n-m}}
\end{equation} 
(equality means has the same correlators as)
so that the field $O_{A_1 \ldots A_n}$ is of scaling weight 
\begin{equation}\label{deltadef}
\Delta =w+n-m.
\end{equation}
Note, in particular, that raising and lowering indices leaves $\Delta$ invariant (because $w$, $n$ and $m$
all change in a coordinated manner to leave $\Delta$ invariant).

In summary the scaling dimension of a primary operator with $n$ lower indices and $m$ upper indices is its Weyl weight plus $n$ -$m$.  \footnote{We can understand this intuitively as follows. Consider the two point function 
	$$<O^{\mu_1 \ldots \mu_m}_{\nu_1 \ldots 
		\nu_n} O^{\alpha_1 \ldots \alpha_m}_{\beta_1 \ldots 
		\beta_n} >.$$
	If the operator has Dimension $\Delta$, one possible term in the two point function of this operator is 
	$$ \frac{\left(\eta_{\mu_1 \alpha_1} \ldots \eta_{\mu_n \alpha_n} \right) \left(   \eta^{\nu_1 \beta_1}  \ldots  \eta^{\nu_m \beta_m} \right)}{\left(\eta_{\theta \phi} x^{\theta} x^{\phi}
		\right)^\Delta }$$.
	Under a Weyl transformation $O \rightarrow \lambda^w O$ ,  $ \eta_{ab} \rightarrow \lambda^2 \eta_{ab}$ and $\eta^{ab} \rightarrow 
	\frac{\eta^{ab}}{\lambda^2}$ so it follows that the expression above picks up a factor of $\lambda^{2 w -2 \Delta -2m +2n}$. It follows $\Delta=w+n-m$ as deduced above. } 
\footnote{One example of this rule is provided by the world sheet theory of the bosonic string. In this theory $b=b^z$ and $c=c_{zz}$ both have Weyl weight. This is why their scaling dimensions are 
	$-1$ and $2$ respectively.} In particular, 
we know that the scaling dimension of a conserved current is $D-1$. It follows that the Weyl weight of the corresponding operator with an upper index is $D$, while the Weyl weight of 
the corresponding operator with a lower index is $D-2$. Similarly 
the Weyl weight of the stress tensor (scaling dimension $D$) with both lower indices is $D-2$, with one upper and one lower index $D$, and with both upper indices $D+2$.

Now let's say we parameterize boundary points by the real projective coordinates $P_A$, and compute the correlators of insertions with polarizations $Z^A$.
The final expression for the correlator will be a function of the $P_A$ and $Z_A$ coordinates of all of the operators. We wish to find a rule that connects the scaling and Weyl dimensions of all operators with the homogeneity of the final correlators in its arguments. The appropriate rules are the following: 
\begin{itemize}
	\item If an expression scales like $\lambda^{-\Delta}$ when we replace $P_i$ by $\lambda P_i$, the expression has scaling dimension $\Delta$. 
	\item If the RHS scales like $\lambda^{-w}$ 
	when we make the replacements $ P^i_M \rightarrow \lambda P^i_M$, $Z_i^M \rightarrow \lambda Z_i^M$, then the $i^{th}$ operator (viewed as an object with lower indices) has Weyl weight $w$.
\end{itemize} 

To see how these rules work it is useful to  consider an example. 
Consider, for instance, a one-form vector field $A_M$ of definite scaling dimension and Weyl weight. The two point function of such a field could, for instance, have terms of the form
\begin{equation}\label{rhsncont}
Z_1.A_M(P_1)  Z_2. A_N(P_2) = 
a \frac{ Z_1.Z_2}{(-2P_1.P_2)^{w+1}}
+ b \frac{  Z_2.P_2 Z_1. P_2}{(-2P_1.P_2)^{w+2}}
\end{equation} 
which means
\begin{equation}\label{rhsn}
A_M(P_1) A_N(P_2) = 
a \frac{ \eta_{MN}}{(-2P_1.P_2)^{w+1}}
+ b \frac{  (P_1)_M (P_2)_N}{(-2P_1.P_2)^{w+2}}
\end{equation} 
where 
$$-2P_1.P_2= (P_1-P_2)^2 =(P_1-P_2)^P  (P_1-P_2)^Q \eta_{PQ}$$
and $\eta_{PQ}$ is the flat metric in embedding space.

With the choice of Weyl frame (and coordinate system)
\eqref{ymm}, this two point function reduces to 
\begin{equation}\label{bbbun}
A_{\mu}(x_1) A_\nu(x_2) = a \frac{ \eta_{\mu\nu}}{x_{12}^{2(w+1)}}
+ b \frac{x_\mu x_\nu} {x_{12}^{2(w+2)}}
\end{equation} 
We see immediately from \eqref{bbbun} that the operators in question have scaling dimension $w+1$, in agreement with the first rule (see the expressions \eqref{rhsn} and \eqref{rhsncont}). 
We now turn to the Weyl weights. 
 Taking into account the $x_{12}^2= (x_1-x_2)^\mu (x_1-x_2)^\nu \eta_{\mu\nu}$, and that $x_\mu=
\eta_{\mu \nu} x^\nu $ we see that the Weyl scaling of the RHS of \eqref{bbbun} is 
$e^{-w \phi(x_1) -w \phi(x_2)}$, so that 
(in order that the RHS be Weyl invariant), the two operators must each  have Weyl weight $w$, in agreement with the second rule (see the expression 
\eqref{rhsncont}).

\subsection{Bulk to boundary propagators} \label{bbprop}

Consider the bulk field 
\begin{equation}\label{Tdef}
T(X,W)=T_{M_1 \ldots M_J}W^{M_1} \ldots W^{M_J}
\end{equation} 
where $W^M$ is a constant vector field. This field corresponds to a boundary operator $O_{M_1 \ldots M_J}$ of scaling dimension $\Delta$.  
The Weyl weight of $O_{M_1 \ldots M_J}$ (note we have taken it to have all lower indices) is, then 
\begin{equation}\label{Weylwt}
w=\Delta -J
\end{equation}

In brief subsection we determine the bulk to boundary propagator of this field - up to an overall constant - from general considerations. 

The bulk to boundary propagator is a function of a boundary vector $Z^M$ ($Z^A$ has upper indices), a boundary point $P^M$, a bulk point $X^M$ and a 
bulk polarization vector $W^M$.  The propagator is a homogeneous polynomial of degree $J$ separately in $Z$ and $W$. Moreover we have seen in the previous subsection that it is of homogeneity $\Delta$ in $P$. It must also be invariant under the shift 
$Z \rightarrow Z + \alpha P$ for any $\alpha$ (see the discussion around \eqref{defn}). Finally, it must be invariant under $SO(D,2)$ transformations, and so must be made up of dot products of the four vectors that form the data of this correlator. 

It follows from the discussion of the last paragraph that the propagator takes the general form 
$$ \frac{F}{(-2P.X)^{\Delta}}$$
where $F$ is a polynomial in $Z.X$, $Z.W$ and $\frac{W.P}{(P.X)}$. The fact that $F$ is of degree $J$ separately in $Z$ and $W$ tells us that it is a polynomial of degree 
$J$ in the two variables $(Z.W)$ and 
$\left( \frac{Z.X~ W.P}{P.X} \right)^{J-m}$.
However this expression must also be  invariant under the shift $\delta Z = P$. 
It is easy to check that the unique combination of these two monomials that is 
invariant under this shift is 
$$( Z.W-\frac{Z.X~ W.P}{P.X})$$
It follows that the spin $J$ bulk to boundary propagator is proportional to \cite{Costa:2014kfa}
\begin{equation}\label{BBprop}
\frac{\left( Z.W-\frac{Z.X~ W.P}{P.X} \right)^J}{(-2 P.X)^\Delta}
= \frac{\left( Z.W ~P.X- {Z.X~ W.P} \right)^J}{e^{-i \pi J} 2^\Delta (-P.X)^{\Delta+J}}
\end{equation} 

Even though we did not explicitly put in this requirement, 
note that our propagator is invariant under the shift 
$\delta W=X$. It is satisfying, as the bulk gauge field is 
$A_M W^M$. The invariance tells us that the gauge field read off in this fashion is automatically orthogonal to the 
$AdS_{D+1}$ sub-manifold. 

Note that the propagator is singular precisely on the light front of the point $P$, i.e. on the sub-manifold spanned by light rays emanating out of $P$. As explained above, the geodesics that make up this light front each have tangent 
vectors proportional to $P^M$; i.e. the various light rays each move in the direction $P^M$.

\subsection{Boundary to boundary correlators} \label{bbc}

The boundary to boundary correlator is given by the expression \eqref{BBprop} upon replacing $X$ by $P'$ and $W$ by $Z'$. So, in particular, the field theory two point function for spin $J$ operators located at $P_i, Z_i$ and $P_j, Z_j$ 
is equal to \cite{Costa:2011mg}
\begin{equation}\label{fttp} 
G_{ij}=\mathcal{C}_{\Delta,J}\frac{\left( -2 Z_i.Z_j ~P_i.P_j+2 {Z_i.P_j~ Z_j.P_i} \right)^J}{(-2P_i.P_j)^{\Delta+J}}
\end{equation} 

where,
\begin{equation}
\mathcal{C}_{\Delta,J} = \frac{(J+\Delta-1)\Gamma(\Delta)}{2\pi^{d/2}(\Delta-1)\Gamma(\Delta+1-h)}
\end{equation}

\subsection{Boundary calculus in embedding space}\label{bces} 

In the $D$ dimensional boundary CFT, a generic traceless symmetric polynomial tensor is encoded in the embedding space by a $(D+2)$ dimensional polynomial. It is a (polynomial) function of the position $P$ and polarisation vector $Z$ subject to the condition $P^2=Z^2=P\cdot Z=0$. More specifically, we can encode a traceless  symmetric tensor of spin-$l$, in the following way \cite{Costa:2011mg},
\begin{eqnarray}
T(P,Z)&=& T_{A_1A_2A_3\cdots A_l}Z^{A_1}Z^{A_2}\cdots Z^{A_l}
\end{eqnarray}    
Conservation condition of this spin-$l$ tensor implies \cite{Costa:2011mg, Costa:2011dw},
\begin{eqnarray}
(\partial \cdot D_Z) T(P,Z)&=& 0,\nonumber\\
\partial \cdot D_Z &=& \frac{\partial}{\partial P_M} \left( \left( \frac{D}{2}-1+Z\cdot \frac{\partial}{\partial Z}\right)\frac{\partial}{\partial Z^M}-\frac{1}{2}Z_M\frac{\partial^2}{\partial Z\cdot \partial Z} \right) 
\end{eqnarray} 

\section{Exploration of the neighbourhood of the Regge point} \label{reggeblowup} 

Consider the neighbourhood of the Regge configuration \eqref{reggepoints}
\begin{equation}\label{fpregge} \begin{split}
&P_1=\left(\left(1+\frac{a_1^2}{4}\right) , a_1^0, \left(1-\frac{a_1^2}{4}\right), a_1^i \right)\\
&P_3=\left( \left(1+\frac{a_3^2}{4}\right) , a_3^0 ,-\left(1-\frac{a_3^2}{4} \right) , a^i_3\right)\\
&P_2=\left(-\left(1+\frac{a_2^2}{4}\right), a_2^0, -\left(1-\frac{a_2^2}{4} \right), a^i_2 \right)\\
&P_4=\left(-\left(1+\frac{a_4^2}{4}\right), a_4^0, \left(1-\frac{a_4^2}{4} \right), a_4^i\right)\\
\end{split}
\end{equation}
where $a^i_\mu=(a_i^0, a_i^j)$ are vectors in $\mathbb{R}^{D-1,1}$ ($j=1 \ldots D-1$). In the parameterization \eqref{fpregge} we have fixed the scale symmetry of each $P_i$ in a convenient manner (for instance we have fixed the scale symmetry of $P_1$ by the requirement that the sum of the first and third components of $P_1$ equals $2$).

It follows from \eqref{fpregge} that 
\begin{equation}\label{aroundrege}
-2P_1.P_2= -(a_1+a_2)^2, ~~~~
-2P_3.P_4= -(a_3+a_4)^2
\end{equation} 
It follows that if $\frac{(a_1+a_2)^2}{2} >0$ then $P_2$ and $P_1$ are timelike related on the manifold \eqref{embed} and so also that $P_2>P_1$ in the full covering space AdS. If $\frac{(a_1+a_2)^2}{2} <0$, on the other hand, the two points are spacelike separated on the manifold \eqref{embed}. They may be either spacelike or timelike separated on the covering space AdS - we will discover the correct rule below. Similar comments hold for $P_3$ and $P_4$.

\subsection{Relationship to global coordinates}

Let $\delta \tau_i$ represent the deviation in global coordinates \eqref{globalpar} from the $\tau$ value of the point $P_i$ from the Regge point. Similarly let $\delta \theta_i$ represent the deviation in global coordinates from the $\theta$ value of the point $P_i$. It follows that the embedding space coordinates of the boundary points in the gauge 
\eqref{globgc} - working to second order in smallness - are 
\begin{equation}\label{globalcoord} \begin{split} 
& P_1=\left(1 -\frac{\delta\tau_1^2}{2}, \delta \tau_1, 1-\frac{ (b_1^i)^2 + \delta \theta_1^2}{2}, \delta \theta_1, b^i_1 \right)\\
& P_3=\left(1 -\frac{\delta \tau_3^2}{2}, \delta \tau_3, -\left(1-\frac{ (b_3^i)^2 + \delta \theta_3^2}{2}\right) , -\delta \theta_3, b_3^i \right)\\	
& P_2=\left( - \left(1 -\frac{\delta\tau_2^2}{2} \right), -\delta\tau_2, -\left(1-\frac{ (b_2^i)^2 + \delta \theta_2^2}{2}\right), -\delta \theta_2, b^i_2 \right)\\
& P_4=\left( - \left(1 -\frac{\delta\tau_4^2}{2} \right), -\delta\tau_4, \left(1-\frac{ (b_4^i)^2 + \delta \theta_4^2}{2}\right), \delta \theta_4, b^i_2 \right)\\
\end{split}
\end{equation}

It follows that 
\begin{equation}\label{identa} \begin{split} 
&\left( a_1^0, a_1^1, a_1^j \right) 
=\left(\delta \tau_1, \delta \theta_1, b_1^j \right)\\
&\left( a_3^0, a_3^1, a_3^j \right) 
=\left(\delta \tau_3, -\delta \theta_3, b_3^j \right)\\
&\left( a_2^0, a_2^1, a_2^j \right) 
=\left(-\delta \tau_2, -\delta \theta_2, b_2^j \right)\\
&\left( a_4^0, a_4^1, a_4^j \right) 
=\left(-\delta \tau_4, \delta \theta_4, b_4^j \right)\\
\end{split}
\end{equation}
(The points $\eqref{fpregge} $ with \eqref{identa} 
agree with \eqref{globalcoord} upto a scaling for each $P_M$)

Below we will find it useful to define 
\begin{equation}\label{ppo}
\alpha_1=P_1+P_2=a_1+a_2, ~~~~\alpha_2 =P_3+P_4=a_3+a_4
\end{equation} 

Finally note that in the special kinematical configuration \eqref{fpscatnnnn}
\begin{equation}\label{special} \begin{split}
&(\alpha_1^0, \alpha_1^1, \alpha_1^j)=(\delta \tau, -\delta \theta, 0)\\
&(\alpha_2^0, \alpha_2^1, \alpha_2^j)=(\delta \tau, \delta \theta, 0)\\
\end{split}
\end{equation}

\subsection{Causal relations} 

The global time difference (see \eqref{globalpar}) between points 2 and 1 is $\pi+\delta \tau_2-\delta\tau_1$. The spatial angular difference is computed from $P_1.P_2$ in 
\eqref{globalcoord} and is given by  
$$\pi - \sqrt{ (\delta\theta_1-\delta \theta_2)^2+(b_1+b_2)^2}$$
As two points are timelike/ spacelike separated depending on whether their separation in global
time is larger or smaller than their angular difference, it follows that  
\begin{equation}\label{causalrelnn} \begin{split}
&P_2> P_1 ~~~{\rm iff} ~~~\delta \tau_1-\delta \tau_2<\sqrt{ (\delta\theta_1-\delta \theta_2)^2+(b_1+b_2)^2} ~~~{\rm i.e.}\\
&P_4> P_3 ~~~{\rm iff} ~~~\delta \tau_3-\delta \tau_4<\sqrt{ (\delta\theta_3-\delta \theta_4)^2+(b_3+b_4)^2} ~~~{\rm i.e.}\\
\end{split}
\end{equation} 
In other words
\begin{equation}\label{causalrelsim}
\begin{split}
&P_2> P_1 ~~~{\rm iff} ~~~\alpha_1^0<\sqrt{ (\alpha_1^i)^2} ~~~\\
&P_4> P_3 ~~~{\rm iff} ~~~\alpha_2^0<\sqrt{ (\alpha_2^i)^2} ~~~\\
\end{split}
\end{equation} 

As a consistency check on this answer we see from \eqref{causalrelnn} that wheneever $P_1$ and $P_2$ are timelike separated on the manifold \eqref{embed2}  (i.e. when  $-2P_1.P_2<0$ i.e.  when $(\alpha_1^i)^2> (\alpha_1^0)^2)$) then $P_2>P_1$ in the covering AdS space as expected on general grounds.  On the other hand when 
$P_1$ and $P_2$ are spacelike separated on the manifold \eqref{embed2} then they are either spacelike or timelike separated in global $AdS$, depending on whether $\alpha_1^0$ is positive or negative.  Identical remarks hold for the points $3$ and $4$. 

On the special configuration \eqref{fpscatnnnn}, 
both conditions \eqref{causalrelsim} hold whenever  
$$\delta \tau < |\delta \theta|$$ 
and neither apply if this relation does not hold.
This is as we have seen before in the main text.

\subsection{$\mathbb{R}^{p,q}$} 

If all the $a_i$ are small (as we assume) it follows that the subspace generated by the four points $P_i$ is the $\mathbb{R}^{1,1}$ plus the subspace generated by $\alpha_1$ and $\alpha_2$. The signature of this space is determined by the sign of the determinant 
\begin{equation}\label{determinant}
D=\alpha_1^2 \alpha_2^2 - (\alpha_1.\alpha_2)^2
\end{equation} 
The space is $\mathbb{R}^{1,1}$ if this determinant is negative and $\mathbb{R}^{2,0}$ if the determinant is positive. 

Note, in particular, that if one or both of $\alpha_i$ are timelike then the combination in \eqref{determinant} is necessarily negative 
\footnote{When both vectors are timelike, this follows because the dot product of two timelike vectors - in a space with only one timelike direction - is necessarily larger in magnitude than the product of norms of vectors.}

On the special configuration \eqref{special}, the quantity \eqref{determinant} evaluates to 
$$- 4 \delta \tau^2 \delta \theta^2 $$
and so is always negative, consistent with the fact that the configuration \eqref{fpscatnnnn} is always $\mathbb{R}^{2,2}$.

\subsection{Cross ratios}

The conformal cross ratios for these points, at leading order in smallness of the $a^i$ is given by

\begin{equation}\label{confcrossrat}\begin{split} 
z=& \frac{1}{4} \left(-\alpha _1.\alpha _2-\sqrt{\left(\alpha _1.\alpha _2\right){}^2-\alpha _1^2 \alpha _2^2}\right)\\
{\bar z}=& \frac{1}{4} \left(-\alpha _1.\alpha _2+\sqrt{\left(\alpha _1.\alpha _2\right){}^2-\alpha _1^2 \alpha _2^2}\right)\\
\sinh^2 \rho=& \frac{\left(\alpha _1.\alpha _2\right){}^2-\alpha _1^2 \alpha _2^2}{\alpha_1^2 \alpha_2^2}\\
\sigma^2 =& \frac{1}{16} \alpha _1^2 \alpha _2^2\\
\end{split}
\end{equation}

For the special configuration \eqref{special} , using \eqref{special}, the above cross ratios evaluates to,

\begin{equation}\label{confcrossrat_spcl}\begin{split} 
z=& \frac{1}{4} \left(\delta\tau-\delta\theta\right)^2\\
{\bar z}=& \frac{1}{4} \left(\delta\tau+\delta\theta\right)^2\\
e^{2 \rho}=& \left( \frac{\delta \theta - \delta \tau}{\delta \theta +\delta \tau } \right)^2\\
\sigma^2 =& \frac{1}{16} \left(\delta\tau^2-\delta\theta^2\right)^2\\
\end{split}
\end{equation} 
in perfect agreement with \eqref{nun}.

\subsection{A diagram of the neighbourhood of the Regge point}

In order to understand the neighbourhood of the 
Regge point it is useful to vary $\alpha_1^0$ and 
$\alpha_2^0$ keeping $\alpha_1^i$ and 
$\alpha_2^i$ fixed. Let us define 
\begin{equation}\label{definitionsn}
x= \frac{\alpha_1^0}{|{\vec \alpha}_1|}, ~~~
y= \frac{\alpha_2^0}{|{\vec \alpha}_2|}, ~~~
\theta=\frac{{\vec \alpha}_1. {\vec \alpha}_2}{{|{\vec \alpha}_1||{\vec \alpha}_2|}}, ~~~w=\frac{y}{x}
\end{equation} 
Note that 
\begin{equation}\label{thetaineq}
|\theta|\leq 1
\end{equation} 
and also that $\theta=-1$ (and so \eqref{thetaineq} is saturated) on the special configuration \eqref{special}.

Using \eqref{confcrossrat} we find 
\begin{equation}\label{zbzar} \begin{split} 
&z=\frac{{|{\vec \alpha}_1||{\vec \alpha}_2|}}{4}\left( \left( xy -\theta \right) -\sqrt{
	\left(\theta -xy\right)^2-(x^2-1)(y^2-1) } \right) \\
&{\bar z}=\frac{{|{\vec \alpha}_1||{\vec \alpha}_2|}}{4}\left( \left( xy-\theta \right) 
+\sqrt{
	\left(\theta -xy\right)^2-(x^2-1)(y^2-1) } \right) \\
&\frac{z}{{\bar z}}= \frac{ \left( xy -\theta \right) -\sqrt{
		\left(\theta -xy\right)^2-(x^2-1)(y^2-1) }} 
	{\left(xy-\theta  \right) +\sqrt{
			\left(\theta -xy\right)^2-(x^2-1)(y^2-1) } }\\
\end{split}
\end{equation} 

$D$ in \eqref{determinant} is given by 
$|{\vec \alpha}_1|^2|{\vec \alpha}_2|^2$ times 
\begin{equation}\label{determpo} \begin{split}
&-x^2 + 2 x y \theta -y^2 +1 -\theta^2\\
&=-x^2 \left(w^2-2 \theta w +1 \right) + \left(1-\theta^2\right) 
\end{split}  
\end{equation}
(this is the negative of the quantity in the square root in \eqref{zbzar}). 
It follows that $D$ 
vanishes whenever
\begin{equation}\label{xmb} 
x^2=\frac{1-\theta^2}{\left(w^2-2 \theta w +1 \right)}
\end{equation} 
Note that 
\begin{equation}\label{huin} 
\left(w^2-2 \theta w +1 \right) = 
\left( w-\theta \right)^2 + \left( 1-\theta^2 \right) 
\end{equation} 
It follows from\eqref{thetaineq} that the RHS of
\eqref{xmb} is 
positive for every value of $w$. This tells us that \eqref{xmb} has a real solution for $x$ solution for every value of $w$. It follow the curve $\rho=0$ is a closed curve that surrounds the origin (infact it is an ellipse in the $x$, $y$ plane).

\begin{figure}[H]
	\begin{center}
		\includegraphics[width=10cm]{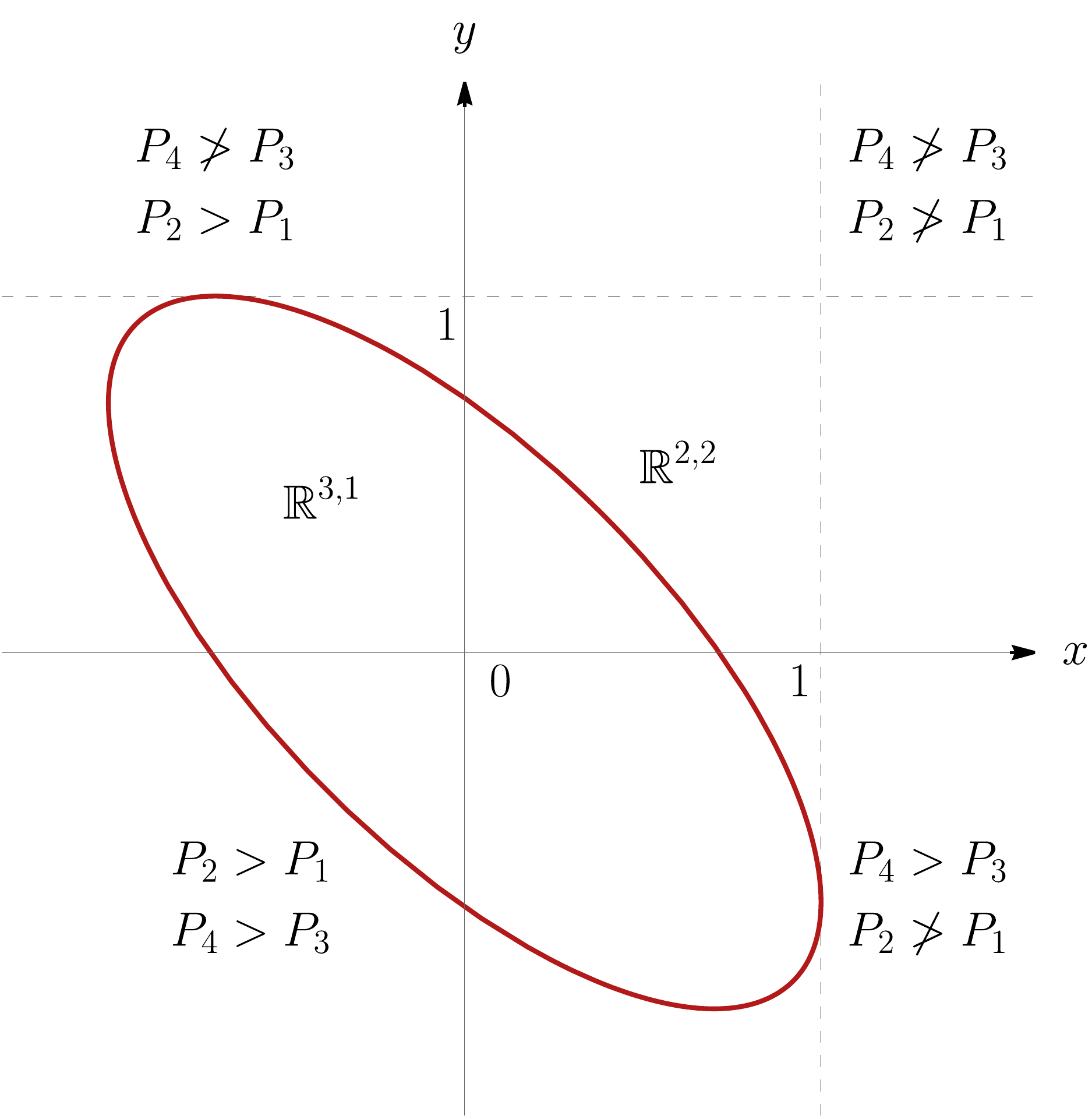}
		\caption{A plot of the causal and $\mathbb{R}^{p, q}$ structure of the neighbourhood of the Regge point. The red line in this plot represents the boundary between $\mathbb{R}^{3,1}$ and $\mathbb{R}^{2,2}$ for a typical value of $\theta$. $\mathbb{R}^{3,1}$ region is represented by the deep blue shaded region. $P_i> P_j$, $P_i \ngtr P_j$ means  that $P_i$ is in the causal future of $P_j$ or not respectively. The two dashed lines divide the plane into four distinct causal configurations. }
		\label{R31R22}
	\end{center}
\end{figure}

\begin{figure}[H]
	\begin{center}
		\includegraphics[width=10cm]{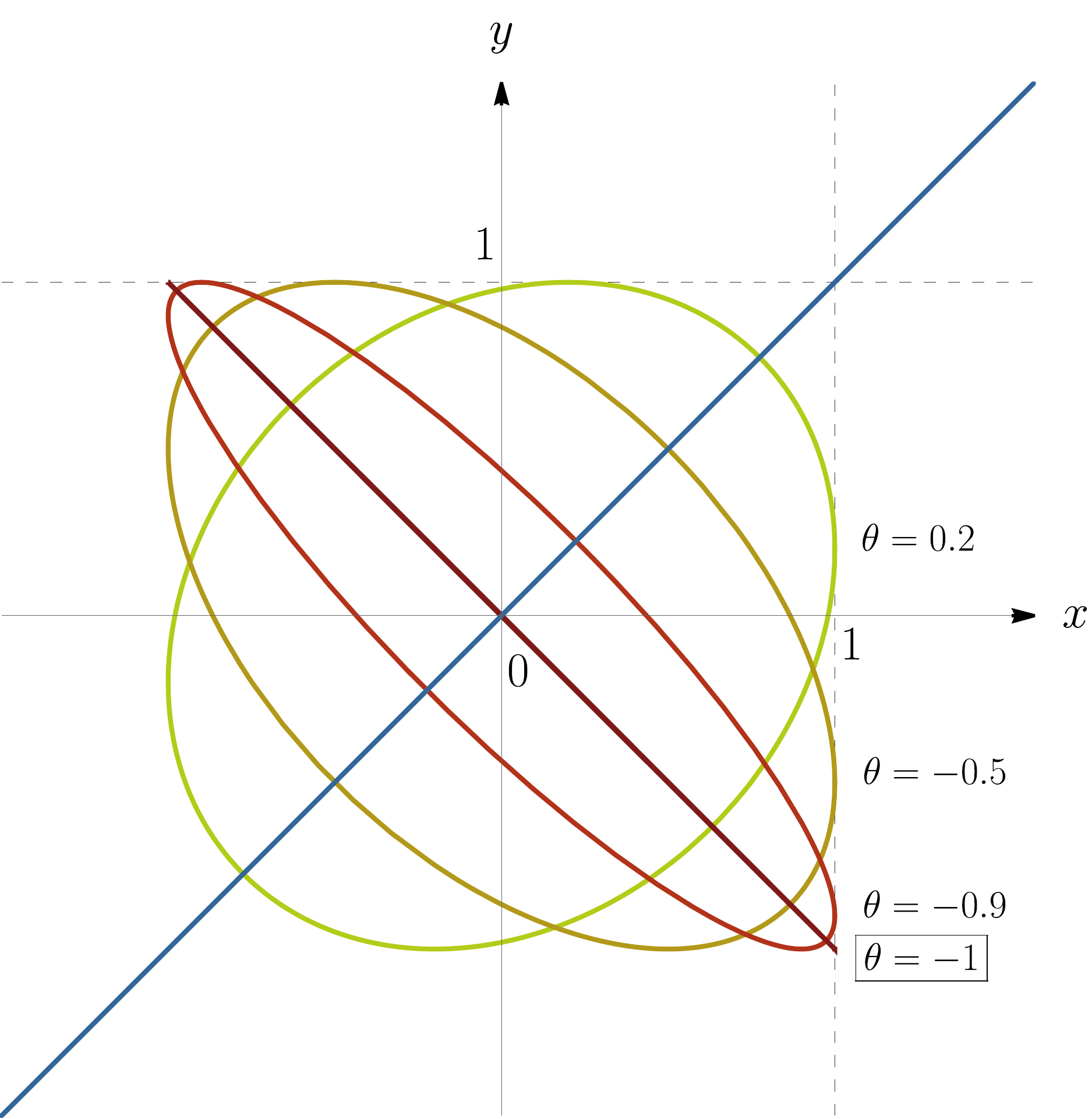}
		\caption{Various colored ellipses (Brown for $\theta=-1$, Red for $\theta=-0.9$, Yellow for $\theta=-0.5$ and Green for $\theta=0.2$) represents the contour for $\rho^2=0$ for various $\theta$ including a special case $\theta=0$. Blue line represents parameter space for configuration \eqref{fpscatnnnn}.}
		\label{R31R22_1}
	\end{center}
\end{figure}

Note that the precise shape of the ellipse depicted in Fig. \ref{R31R22} depends on the value of $\theta$. The slope of the major axis of the ellipse has the same sign as $\theta$ (Fig. \ref{R31R22} is plotted with $\theta=-.7$). At $\theta=0$ the ellipse becomes a circle, and at positive $\theta$ it begins to `tilt to the right' (see the yellow-green ellipse at $\theta=0.2$ plotted in Fig. \ref{R31R22_1}). Focussing on negative values of 
$\theta$ for a moment, the thickness of ellipse decreases as $\theta$ reduces to its minimum value, 
$\theta=-1$. In particular when $\theta$ is precisely $-1$ the ellipse degenerates to a line with slope -1 (see Fig. \ref{R31R22_1}. )

Recall that $\theta=-1$ on the special configuration \eqref{special} that we have focussed attention on in the main text of this paper. In fact the special configuration \eqref{special} lies on 
the blue line (with slope unity) of Fig \ref{R31R22_1}. The fact that the ellipse degenerates to a line at $\theta=-1$ explain why
the configurations \eqref{fpscatnnnn} are always 
either $\mathbb{R}^{2,2}$ or $\mathbb{R}^{2,1}$ but never $\mathbb{R}^{3,1}$.

\begin{figure}[!tbp]
	\centering
	\begin{minipage}[b]{0.45\textwidth}
		\includegraphics[width=8cm]{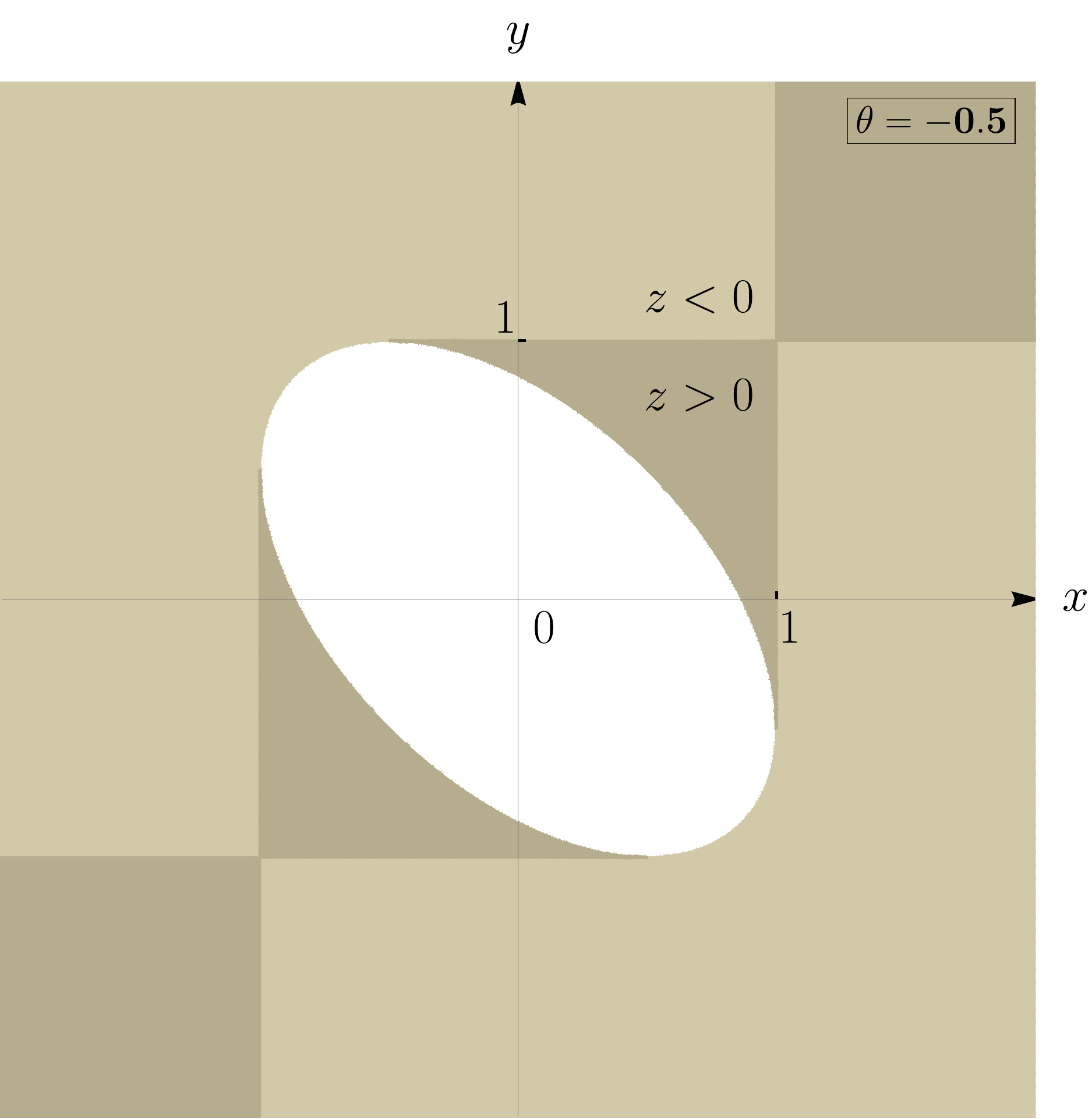}
		\caption{Positivity of $z$.}
	\end{minipage}
	\hfill
	\begin{minipage}[b]{0.45\textwidth}
		\includegraphics[width=8cm]{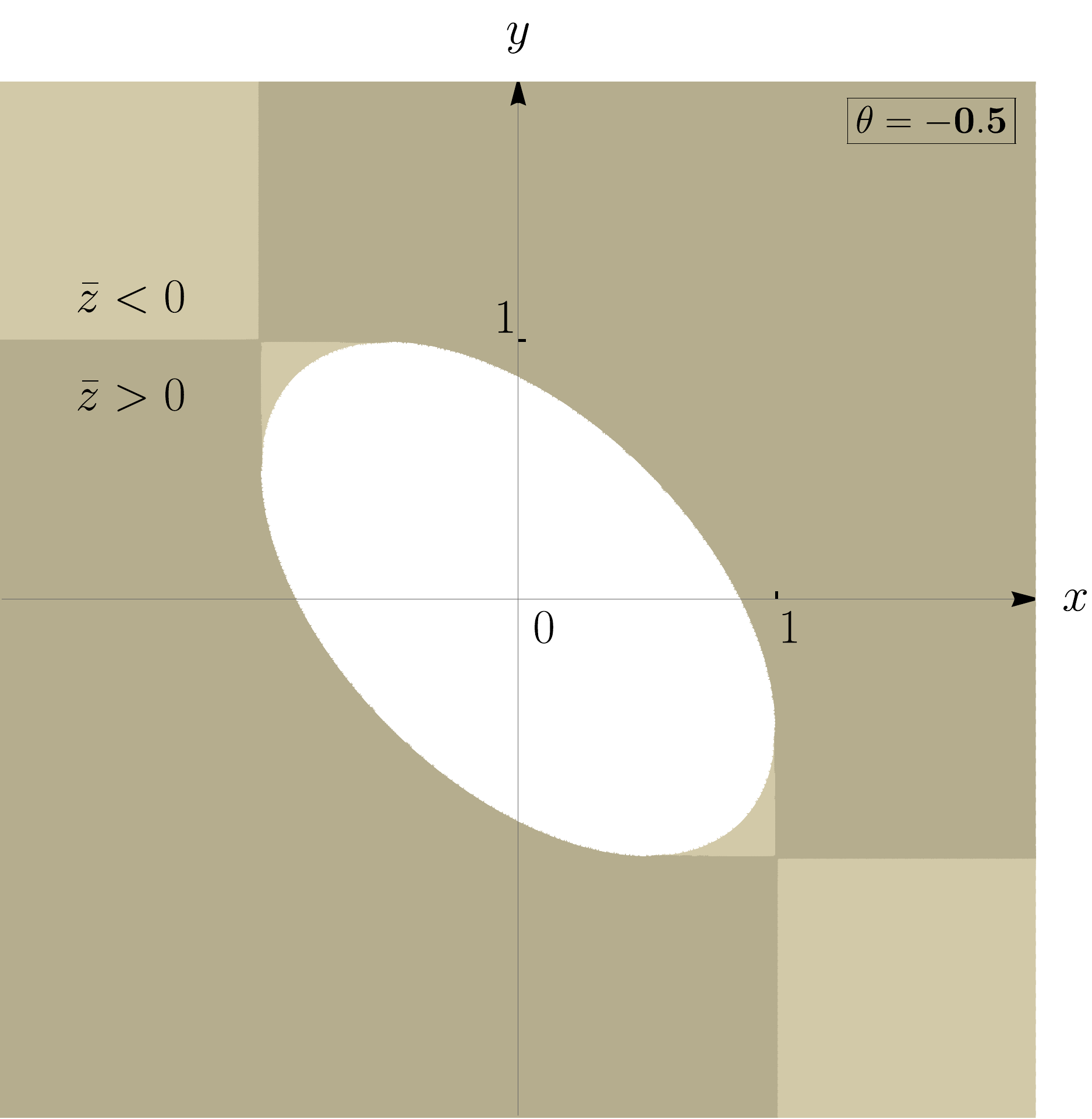}
		\caption{Positivity of $\bar{z}$.}
	\end{minipage}
	\caption{In this plot we have shown the positivity of $z$ and $\bar{z}$ in the $(x,y)$ plane by the dark shaded region. In the light shaded region they are negative. White part is excluded since $z$ and $\bar{z}$ are imaginary in that region.}
	\label{pos_z_zbar}
\end{figure}

\subsection{The sign and ratio of $z$ and ${\bar z}$ in this neighbourhood}

In the previous subsection we have presented a detailed two parameter blow up of the neighbourhood of the Regge point. The small $\theta$ and $\tau$ limit of \eqref{fpscatnnnn} can also be thought of as a restricted blow up of the Regge point. In this subsection we will analyze how generic the configurations \eqref{fpscatnnnn} (more accurately \eqref{special}) are - upto conformal transformations -  in the neighbourhood of the Regge point. 

Let us first recall that, as depicted in Fig. 
\ref{R31R22}, the neighbourhood of the Regge point 
includes points with four different causal configurations. The top right quadrant of 
Fig. \ref{R31R22} ($x>1, y>1$) has the same causal structure as the Causally Regge sheet of 
\eqref{causalrel}. The bottom left quadrant of the same figure,  ($x<1, y<1$), has  the same causal structure as the Causally Scattering sheet of 
\eqref{causalrel}. On the other hand the top left ($x<1, y>1$) and bottom right ($x>1, y<1$) quadrants of Fig. \ref{R31R22} display causal relations that never occur in our special configurations \eqref{fpscatnnnn}. It follows that the special configuration \eqref{fpscatnnnn} lie entirely in the top right and bottom left quadrants of Fig \ref{R31R22}; the remaining quadrants of that figure are not covered by \eqref{fpscatnnnn}.

What parts of the top right and bottom left quadrants of Fig. \ref{R31R22} are covered by
the small $\theta, \tau$ limits of \eqref{fpscatnnnn} (i.e. by \eqref{special})? To 
answer this question we first recall that in the special configuration \eqref{fpscatnnnn}
\begin{itemize} 
	\item The spanning space of the $P_a$ is either $\mathbb{R}^{2,2}$ or $\mathbb{R}^{2,1}$
	\item $z>0$ and ${\bar z}>0$ (i.e. are of Type IIa, see  Appendix \ref{cocr}).
	\item $0 \leq \frac{z}{{\bar z}} \leq 1$
	\end{itemize} 
The first point above immediately tells us that 
points in the interior of the ellipse in Fig. 
\ref{R31R22} are not symmetry related (conformally related) to any of the points \eqref{special}. The second itemized point above tells us that no point in Fig \ref{R31R22} for which either $z$ or 
${\bar z}$ is negative  is conformally related to any of the points \eqref{special}. \footnote{In the language of Appendix \ref{cocr}, this means that  no point of Type IIb, IIc or IId is conformally equivalent to any of the points \eqref{special}}.

Using \eqref{zbzar} it is easy to convince oneself that the regions in which $z$ and ${\bar z}$ are positive are the dark shaded regions in Figs 
\ref{pos_z_zbar}. In particular $z$ and ${\bar z}$ 
are both positive if and only if $xy-\theta$ is positive, and $x^2$ and $y^2$ are both either greater than unity or both less than unity.  
In this case  $(xy -\theta )>\sqrt{
	\left(\theta -xy\right)^2-(x^2-1)(y^2-1) }>0$.	
and so it follows immediately from \eqref{zbzar} 
that in this region $\frac{z}{{\bar z}}<1$ so that 
the third item above is also met. 

It follows all points that are dark shaded in the first of Fig. \ref{pos_z_zbar} are symmetry related to one of the points \eqref{special}. None of the light-shaded points - or the points in the interior of the ellipse- are symmetry related to the points \eqref{special}.

\begin{figure}[H]
	\begin{center}
		\includegraphics[width=12cm]{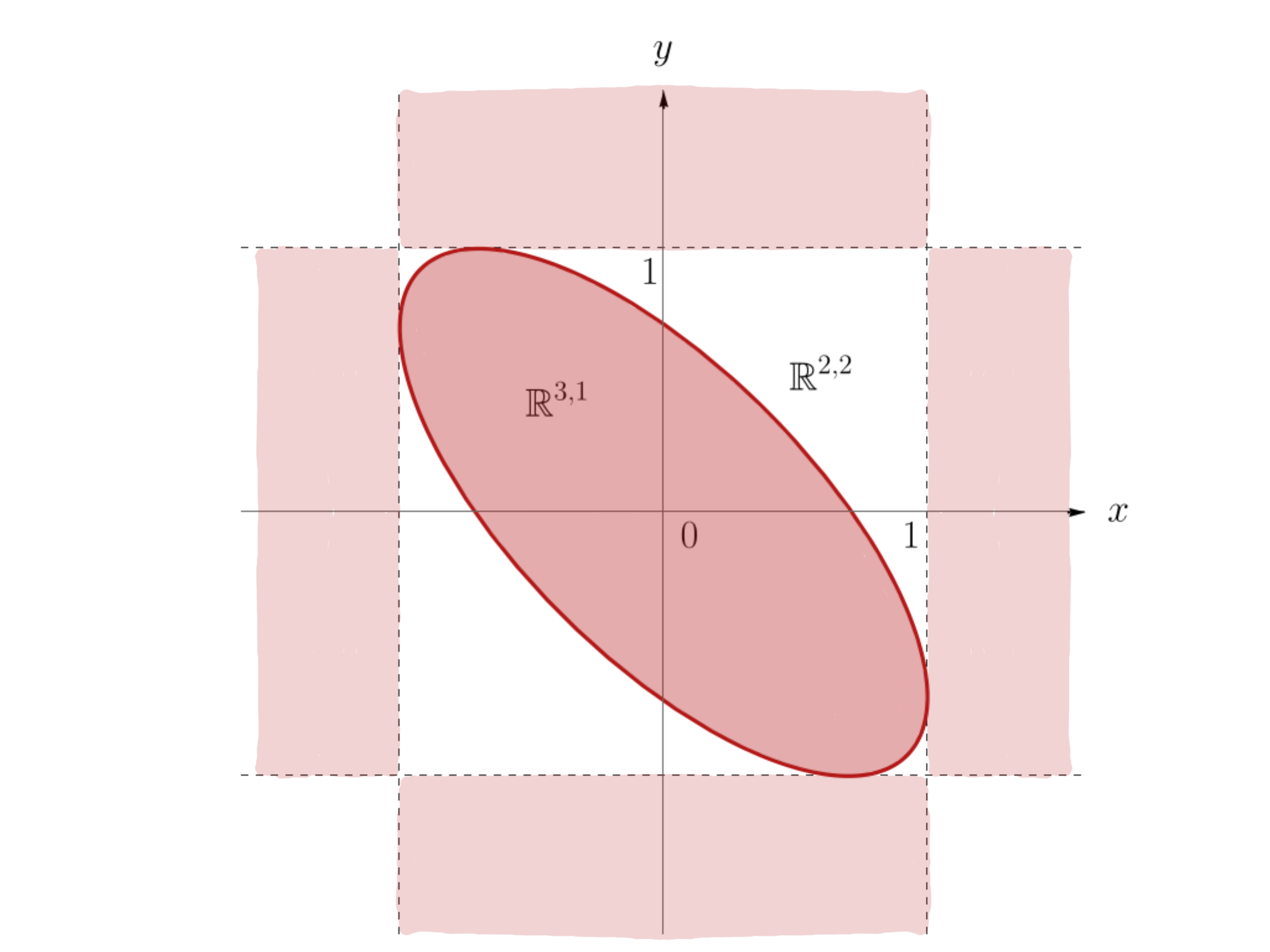}
		\caption{The dark shaded represents the region inside the ${\mathbb R}^{3,1}$ elipse and the light shaded represents the type IId region, where, ${\bar z}$ is negative and ${z}$ is positive.}
		\label{newgrr}
	\end{center}
\end{figure}

As an aside note that $\rho$ is imaginary at every point in Fig. \ref{R31R22} which either lies inside the ${\mathbb R}^{3,1}$ ellipse (these are points of Type I in the language of Appendix \ref{cocr}) or when ${\bar z}$ is negative and ${z}$ is positive (these are  configurations of Type IId in the language of Appendix \ref{cocr}, and are the shaded regions in Fig \ref{newgrr}.) We have carefully checked that in each of the shaded regions in Fig \ref{newgrr}, one of the pairs of points $P_1$ and $P_2$ are spacelike related on the (compact time) manifold \eqref{embed2} while the other pair of points is timelike related on the same manifold, so that the fact that these points have imaginary $\rho$ is consistent with the analysis of Appendix \ref{brtt}. \footnote{We re emphasize that points that this is a distinct question from the causal relation of points in the full covering space $AdS_{D+1}$. Points that are timelike related on the manifold \eqref{embed2} are necessarily timelike related in the full embedding space, but points that are spacelike related in \eqref{embed2} could be either spacelike or timelike separated in the full embedding space. This dichotomy explains the apparent inconsistency between the statements of this paragraph (which refer to the timelike and spacelike separation between points on the space \eqref{embed2}) and causal relations  asserted in Fig. \ref{R31R22}, which refer to causal relations in the full covering $AdS_{D+1}$ space.}

In summary we conclude that the special configuration \eqref{fpscatnnnn} completely covers the Causally Regge sheet of correlators, gives a partial cover of the Causally Scattering sheet (it covers the part of the neighbourhood of the Regge point contained in the dark shaded points in \eqref{fpscatnnnn}).

\section{$i \epsilon$ in position space}

\subsection{The $i \epsilon$ prescription}

Consider a quantum theory with a Hamiltonian  $H$ whose spectrum is bounded from below. A wave function may be evolved either forward or backward in time by the time evolution operation

\begin{equation}\label{evofr}
|\psi(t-t_0) \rangle = e^{-i H (t-t_0)} |\psi(t_0) \rangle. 
\end{equation}
The expression on the RHS of \eqref{evofr} 
consists of a sum of terms with many different frequencies. As the energy of a quantum system is typically unbounded from above, the frequencies that appear in \eqref{evofr} have no bound, and the expression \eqref{evofr} is potentially ill defined. If we are interested in evolving {\it only forward} in time then we can improve the situation. $\eqref{evofr}$ continues 
to be well defined under the replacement 

\begin{equation}\label{reprule} 
t \rightarrow t e^{-i \epsilon} \approx  t-i \epsilon t 
\end{equation}
with $\epsilon >0$. Once we make this replacement, extremely high energy components of \eqref{evofr} are damped out, and \eqref{evofr}
is well defined. This is the $i \epsilon$ prescription that is used to give definite meaning to expressions that are otherwise ill defined in Lorentzian space.

\subsection{Relation to Euclidean space}

More generally, \eqref{evofr} continues to be well defined under the replacement  
$$ t = e^{-i \alpha} t'$$ 
with $t'$ real and $ 0 < \alpha <\pi$. The variable $t'$ at $\alpha=\frac{\pi}{2}$ is 
the so called Euclidean time $\tau$. In other words 
\begin{equation}\label{lorentzeuclid}
t=-i\tau
\end{equation} 
The evolution of the wave function in Euclidean time is given by 
\begin{equation}\label{rwf}
|\psi(\tau-\tau_0) \rangle = e^{- (\tau-\tau_0) H} |\psi(\tau_0)\rangle. 
\end{equation} 
As usual, \eqref{rwf} is well defined only for 
$\tau > \tau_0$. \eqref{rwf} is, of course, extremely well behaved as it is a sum of exponentially decaying rather than oscillating terms. 

One point of view that is sometimes useful to take is the following. The wave function 
\eqref{evofr} is defined starting with the manifestly well defined object \eqref{rwf} and 
then making the replacement 
\begin{equation}\label{ancontrule}
\tau_E = i t e^{-i \epsilon} 
\end{equation} 
\eqref{lorentzeuclid} and \eqref{ancontrule} 
together, of course, reduce to the replacement rule \eqref{reprule}.


\section{Mapping into the $\rho$ plane} \label{stdconfig}

In this appendix we attempt to map the configurations studied in this paper into a possibly more familiar conformal coordinate system - the so called $\rho$ coordinate frame. 

\subsection{The $\rho$ frame} 

Consider the coordinates $z$ and $\bar z$ in $\mathbb{R}^{1,1}$ defined by
\begin{equation}\label{defzbarz}
w=x-t, ~~~~{\bar w}= x+t
\end{equation}
$t$ is Lorentzian time.
\footnote{$w$ and ${\bar w}$ are the standard $\sigma_+$ 
	and $\sigma_-$ configurations of a 2 D CFT.}
\footnote{After continuation to Euclidean time, $t =-i\tau$ 
	find  
	$$w=x+i \tau, ~~~~{\bar z}= w-i\tau;$$
	note that $w$ and ${\bar w}$ are now complex conjugates of each other.}

\begin{figure}[H]
	\begin{center}
		\includegraphics[width=10cm]{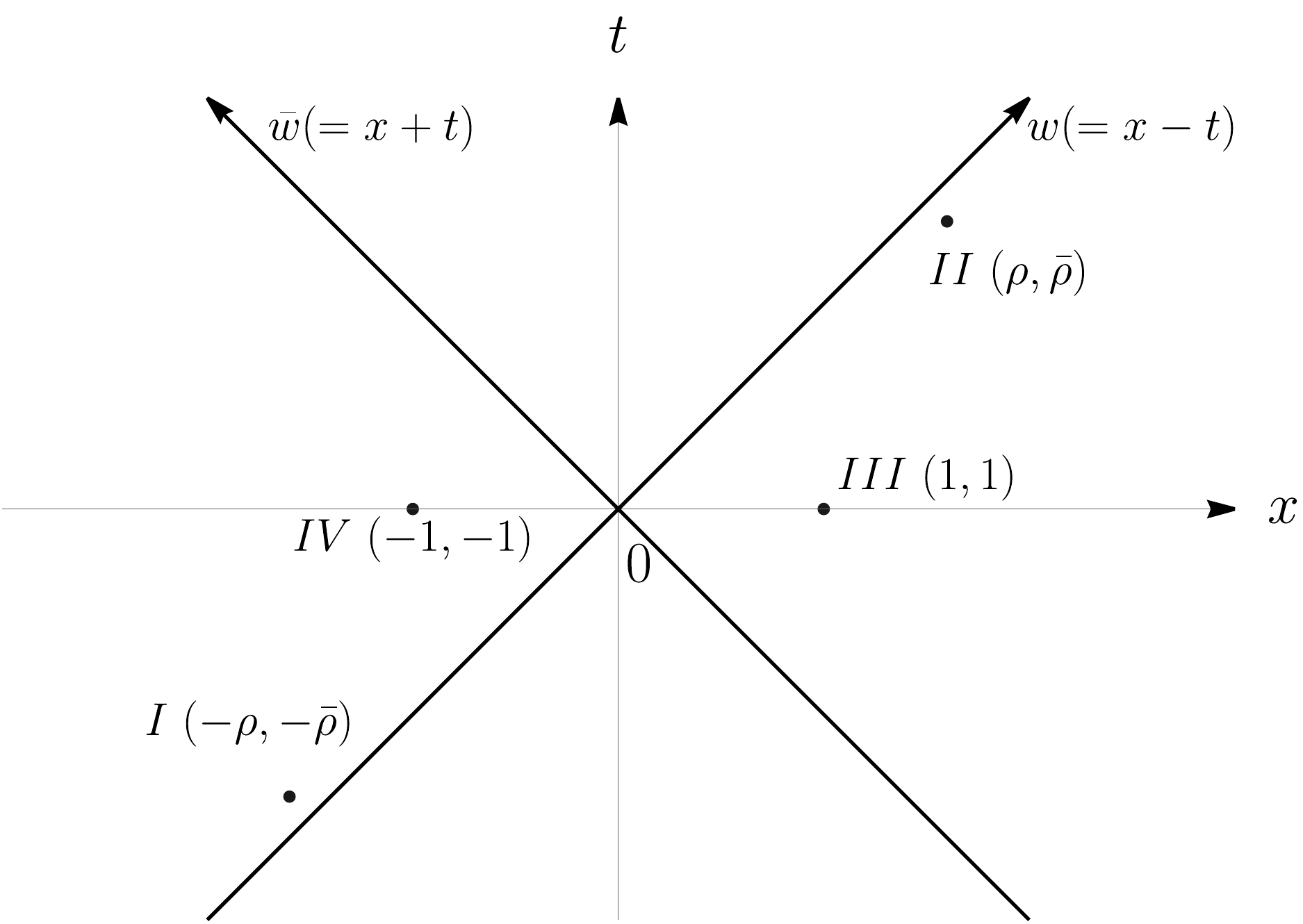}
		\caption{A schematic of the four insertion points on the $\rho$ plane. We have drawn both the $x$ and $t$ as well as the $w$ and 
		${\bar w}$ axes on this graph. The values of coordinates are given in $w$ and ${\bar w}$.}
		\label{xt_pt}
	\end{center}
\end{figure}

Now consider the following standard insertions
(insertions are specified by $(w, {\bar w})$, shown in Fig. \eqref{xt_pt}) 
and their corresponding cross ratios 
\begin{equation} \begin{split} \label{labn} 
&~~~I~:~~~(-\rho , -{\bar \rho})\\
&~~II~:~~~(\rho, {\bar \rho})\\
&~III~:~~~(1,1)\\
&~~IV~:~~(-1, -1)\\
& z(\rho)=\frac{z_{21} z_{34}}{z_{24}z_{31}}= \frac{4 \rho}{(\rho+1)^2} , ~~~
{\bar z}({\bar \rho})= \frac{{\bar z}_{21} {\bar z}_{34}}{{\bar z}_{24} {\bar z}_{31}} = \frac{4 {\bar \rho}}{({\bar \rho}+1)^2}\\
\end{split}
\end{equation} 
These insertions are depicted schematically in the Figure \eqref{xt_pt}. 

(we will consider the $i \epsilon$ corrected cross ratios shortly).

Note that the $\rho$ plane is a double cover of 
the $z$ plane; similarly the ${\bar \rho}$ plane 
is a double cover of the ${\bar z}$ plane. Any given value of $z$ \footnote{Except $z=1$, see below} corresponds to two  distinct values of $\rho$.  In fact if  $\rho$ is one solution to the equation
\begin{equation}\label{zfrh}
\frac{4 \rho}{(\rho+1)^2}=z
\end{equation} 
then $\frac{1}{\rho}$ is a second solution to the same equation. Identical remarks apply to ${\bar \rho}$ and ${\bar z}$.

In what follows we will need the $i \epsilon$ corrections to \eqref{labn}. This is achieved 
by making the following replacements in \eqref{labn}

\begin{equation}\label{repall}
\rho \rightarrow \frac{ \left( \rho + {\bar \rho} \right) }{2} -
\frac{\left( {\bar \rho} - \rho \right) }{2} e^{- i \epsilon} , ~~~~{\bar \rho} \rightarrow \frac{ \left(\rho + {\bar \rho} \right) }{2} +
\frac{\left( {\bar \rho} - \rho \right) }{2} e^{- i \epsilon}
\end{equation}
(\eqref{repall} simply implements \eqref{reprule}).

\begin{figure}[H]
	\begin{center}
		\includegraphics[width=10cm]{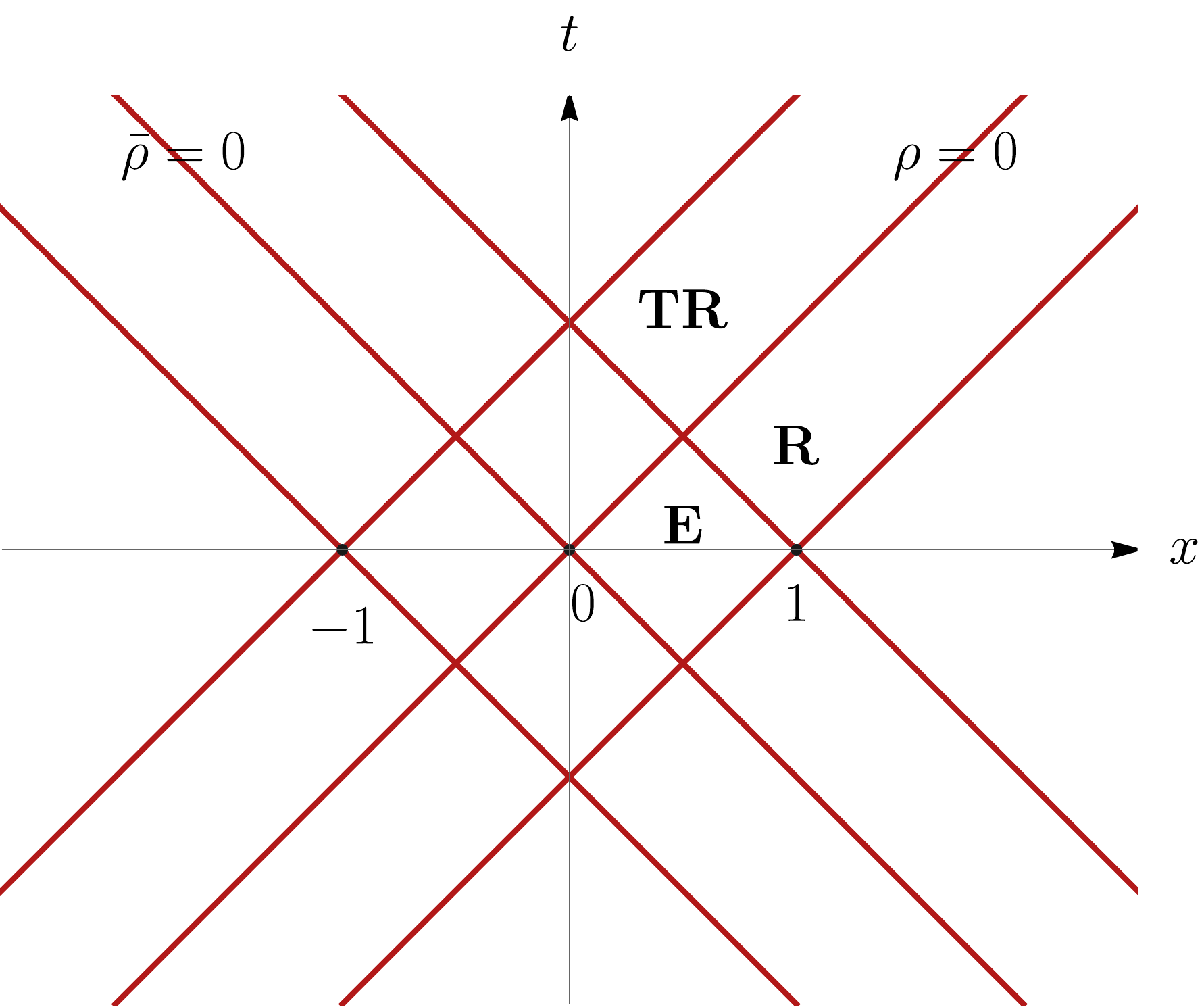}
		\caption{Specification of {\bf E, R,} and {\bf TR} regions in $(t,x)$ plane.}
		\label{xt0}
	\end{center}
\end{figure}
From a causal point of view the $(\rho , {\bar \rho})$ plane has 16 inequivalent regions. This comes about as follows. As depicted in Fig. \ref{xt0}, the $\rho$ plane has  3 interesting lightcone lines with slope unity. As the operator $II$  goes through these lines (moving from down to up), it cuts these  lines  - the right-moving lightcones of operators $III$, $I$ and $IV$. In a similar way the plane has 3 interesting lightcone lines of slope $-1$ (again see Fig. \ref{xt0}). As the operator II moves from bottom to top, it cuts these lines - the lightcones of operators $IV$, $I$ and $III$. This grid of lightcones divides the plane into a criss cross of $(3+1)\times(3+1)=16$ distinct causal regions, 
as depicted in Fig. \ref{xt0}.

In this section we will only explore 3 of these causal regions. The first of these is the diamond marked ${\bf E}$ (for Euclidean) in Fig. \ref{rhthord} defined by the conditions 
$ 0 < \rho <1$ and $0< {\bar \rho} <1$.  In the region ${\bf E}$ all operators are spacelike separated with respect to each other. In other 
words the causal relations between operators in the region ${\bf E}$ is the same as that in the 
Causally Euclidean configurations of \eqref{causalrel}. 

The second region we consider is the half strip marked ${\bf R}$ (for Regge) in Fig. \ref{rhthord} defined by the conditions $0<\rho< 1$ together with ${\bar \rho}>1$. The causal relations between operators in this region is the same as that of the Causally Regge configurations in 
\eqref{causalrel}. 

Finally, the third region we consider is the half strip marked ${\bf TR}$ (for Timelike Regge) in Fig. \ref{rhthord} defined by the conditions $-1<\rho< 0$ together with ${\bar \rho}>1$. The causal relations between operators in this region
is in between that of the Causally Regge and the Causally scattering sheets (it is not identical to that of the Causally Scattering sheet because the operators $III$ and $IV$ are spacelike separated with respect to each other in the region ${\bf TR}$ unlike in the Causally Scattering region).

\subsection{Mapping into the $\rho$ plane: qualitative comments}

Provided $\rho>0$ it from \eqref{zfrh} that
$0<z<1$. Similarly whenever ${\bar \rho}>0$ 
$0<{\bar z}<1$. It follows that both in the regions ${\bf E}$ and the region ${\bf R}$
\begin{equation}\label{brhh}
0<{ z}<1, ~~~~0<{\bar z}<1
\end{equation}
This is precisely the range over which the cross ratios in the Causally Euclidean and Causally Regge regions of \eqref{fpscatnnnn} vary (see \eqref{zineq}). So it seems very plausible that the Causally Euclidean and Causally Regge regions of 
\eqref{fpscatnnnn} map into the regions 
${\bf E}$ and ${\bf R}$ of \eqref{labn}.

In the region ${\bf TR}$, on the other hand we have 
\begin{equation}\label{brhhtr}
z<0, ~~~~0<{\bar z}<1
\end{equation}
However the $z$ cross ratio for the insertions \eqref{fpscatnnnn} is never negative. This is another demonstration of the fact that the  Causally Scattering region of \eqref{fpscatnnnn} cannot map into the region 
${\bf TR}$ of \eqref{labn} (this follows more elementarily, of course, from causal considerations, as we have explained above). In fact the causally scattering region simply has no analogue in 
the coordinate charge \eqref{labn}.
\footnote{A very rough analogy might be the following. If \eqref{fpscatnnnn} is like the ingoing Eddington Finklestein coordinate system
	in a black hole, \eqref{labn} is the analogue of the outgoing Eddington Finklestein coordinate 
	system. Just like the two EF coordinates agree on the exterior of the event horizon but continue to 
	different regions of spacetime for $r< r_H$, the 
	two conformal coordinate patches \eqref{fpscatnnnn} and \eqref{labn} agree on 
	the Causally Euclidean i.e. ${\bf E}$ and Causally Regge i.e. ${\bf R}$ regions, but have 
	different continuations beyond $z=0$.}

If it is indeed the case that the Causally Euclidean region maps to ${\bf E}$ while the 
Causally Regge region maps to ${\bf R}$ then it should also be the case that the transition from 
${\bf E}$ to ${\bf R}$ involves circling counterclockwise around the branch cut at ${\bar z}=1$, as was the case for \eqref{fpscatnnnn}
(see around \eqref{barzexp}). It is easy to directly verify that this is indeed the case. 
Indeed in the neighbourhood of the transition region the Minkowski time is positive which implies that 
$$\rho \rightarrow \rho + i \epsilon, ~~~~
{\bar \rho} \rightarrow {\bar \rho} -i \epsilon$$
In particular
\begin{equation}\label{actualform}
{\bar z}= \frac{4 ({\bar \rho} -i \epsilon) }{({\bar \rho}-i \epsilon +1)^2}, ~~~~{\bar \rho}=\rho_0 e^{\tau}
\end{equation} 
An analysis very similar to that around \eqref{barzexp} will convince the reader that 
the transition from the region ${\bar E}$ to 
${\bar R}$ - which is the transition from ${\bar \rho}<1$ to ${\bar \rho}>1$ takes us along a path in cross ratio space that circles counter clockwise around the branch cut at ${\bar z}=1$ .

\subsection{Mapping into the $\rho$ plane : quantitative formulae}

The quantitative map from \eqref{fpscatnnnn} to 
\eqref{labn} is obtained by equating the $z$ and ${\bar z}$ cross ratios of the two configurations, i.e. by imposing the equations
\begin{equation}\label{maph} \begin{split} 
&\frac{4 \rho}{(\rho+1)^2}=\frac{1-\cos(\theta-\tau)}{2} \\
&\frac{4 {\bar \rho}}{({\bar \rho}+1)^2}=\frac{1-\cos(\theta+\tau)}{2} \\
\end{split}
\end{equation} 
In the ${\bf E}$ region the solution to these equations is given by 
\begin{equation}\label{solbfe}\begin{split}
&\rho= \tan ^2\left(\frac{\tau -\theta }{4}\right)\\
&{\bar \rho}= \cot ^2\left(\frac{\tau+\theta }{4}\right)\\
\end{split} 
\end{equation} 

\begin{figure}[H]
	\begin{center}
		\includegraphics[width=10cm]{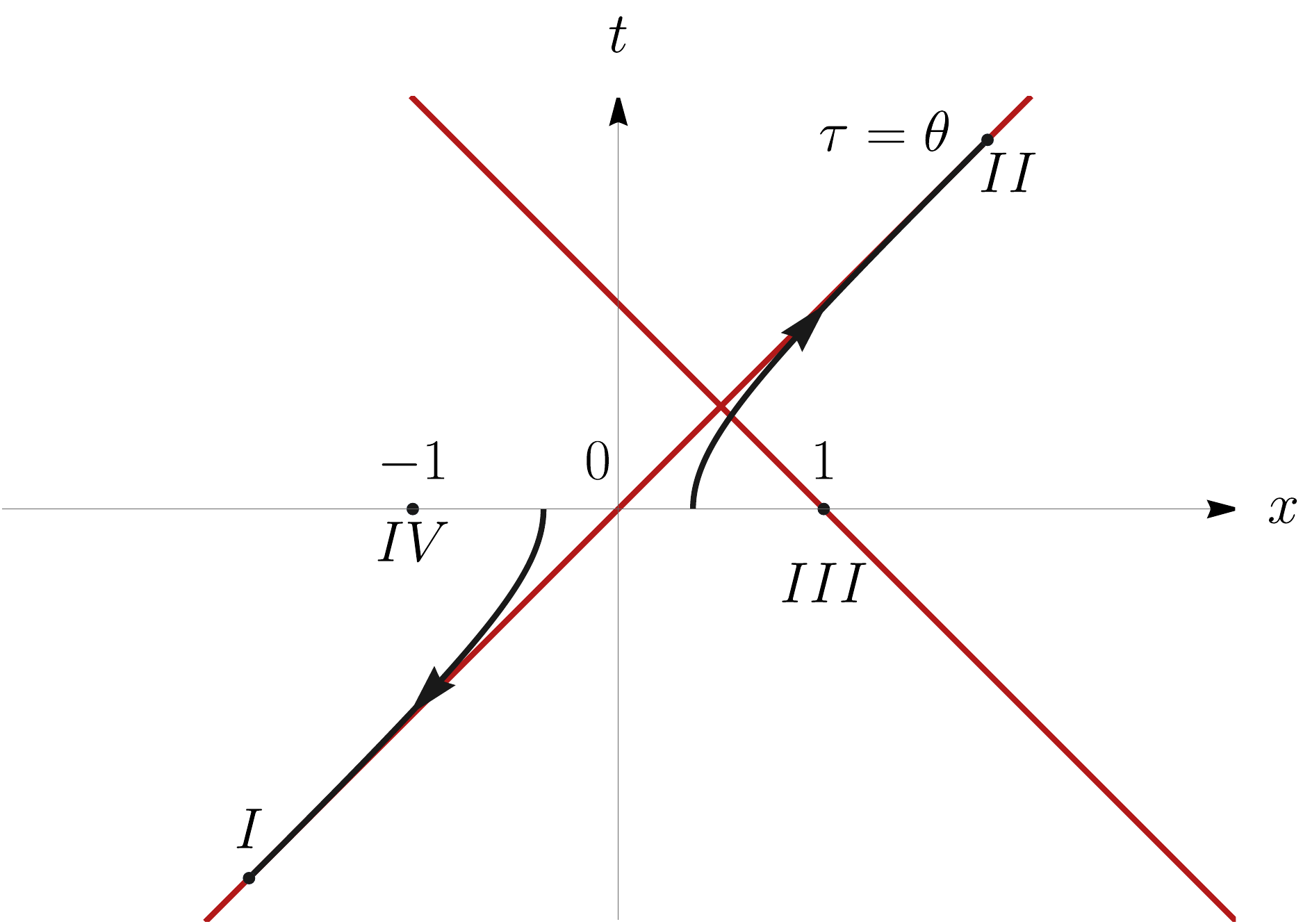}
		\caption{The image of the part of the path of subsection \ref{pcrs} that has a map in the $\rho$, ${\bar \rho}$ plane. The right black curve depicts the insertion position of opertor II. The trajectory is one of decreasing $\tau$; it starts at $\tau=\pi$ and ends at $\tau =\theta$.}
		\label{xt1}
	\end{center}
\end{figure}

It is interesting to investigate how the equivalent insertion points in the $\rho$, ${\bar \rho}$ plane evolve as we move along the path in cross ratio space described in \eqref{pcrs}. Recall that this path is obtained by varying $\tau$ from $\pi$ to $0$ at fixed $t$. As depicted in Fig \ref{xt1}, at $\tau =\pi$ the corresponding insertion for operator II in the $(\rho, {\bar \rho})$ plane starts out at the point on the $x$ axis, $x=\cot^2 \left( \frac{\pi+\theta}{4}\right), t=0$. This point lies in the ${\bf E}$ region of Fig \ref{rhthord}. 	

As $\tau$ decreases the insertion point 
follows the path depicted on the rightmost curve in 
Fig. \ref{xt1}. This path cuts the left-moving lightcone of the operator III at ${\bar \rho}=1$, $\rho=\frac{1-\sin\theta}{1+\sin\theta}$. It then 
moves into the region ${\bf R}$ of Fig. \ref{rhthord}. 
As $\tau$ decreases further the path approaches nearer and nearer to the right moving lightcone of operator I, cutting it when $\tau=\theta$ at the point 
$x=t=\frac{1}{2}\cot^2\left(\frac{\theta}{2}\right)$.

\begin{figure}[H]
	\begin{center}
		\includegraphics[width=10cm]{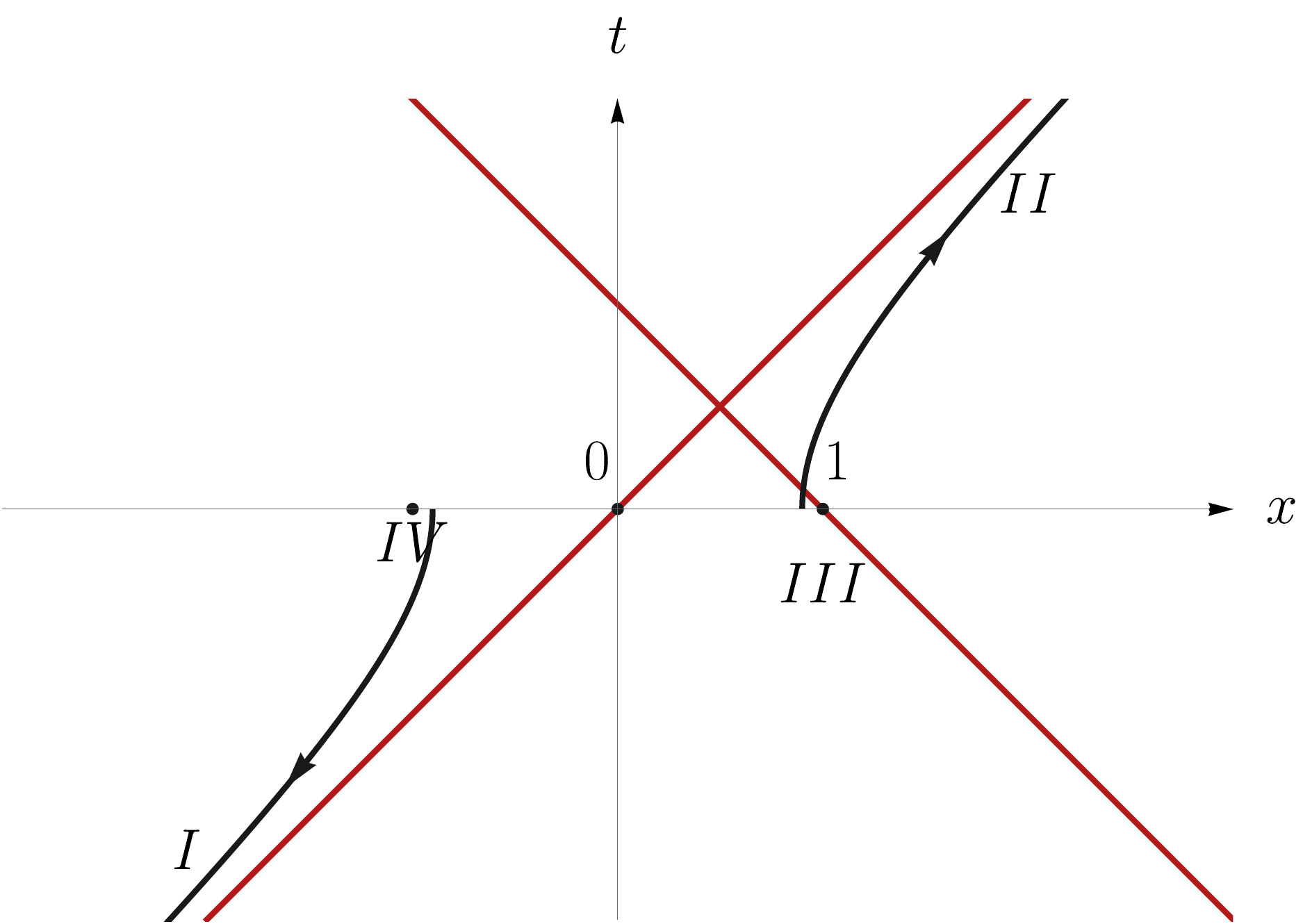}
		\caption{The image of the trajectory of subsection \ref{pcrs} - with a small fixed value of $\theta$ - onto the $\rho$ plane.}
		\label{xt2}
	\end{center}
\end{figure}

In Fig \ref{xt2} we depict the image of the analogous trajectory at a smaller value of $\theta$. Note that the path starts much nearer to the point $x=1$ on the 
$x$ axis and then runs along the lightcone of the origin for a much longer `time', intersecting it only 
at late times (and far distances) $x =t \approx \frac{2}{\theta^2}$. 



\subsection{Summary of the mapping}

Let us summarize. The point $\tau=\pi$ in \eqref{fpscatnnnn} maps to
a point on the $x$ axis $(x= \tan^2 \left( \frac{\pi - \theta}{4}\right), t=0)$ in the $\rho$ plane. Note that the $x$ coordinate of this insertion is less than unity.  As $\tau$ decreases, the our $\rho$ point leaves the $x$ axis, moving to positive values of the time $t$ in the $\rho$ plane (see the trajectory of the point II in Fig. \ref{xt1} ).  When $\tau$ decreases to $\pi-\theta$, the trajectory in the $\rho$ plane cuts the lightcone ${\bar \rho}=1$ (the red line which is oriented at angle $-\frac{\pi}{4}$ to the $x$ axis in Fig \ref{xt1}). As $\tau$ further increased the trajectory approaches the lightcone $\rho=0$ (the red line oriented at an angle $\frac{\pi}{4}$ with the $x$ axis in Fig. \ref{xt1}), hitting it 
at $\tau=\theta$. As we have pointed out, the subsequent evolution of the trajectory of this subsection (the part between $\tau =\theta$ and 
$\tau =0$, i.e. the part of our trajectory on the Conformally Scattering sheet) has no image in the 
$\rho$ plane.

Let us now return to the part of the fixed 
$\theta$ trajectory of this subsection that lies on the Causally Regge sheet. As we have seen above this part does have an image on the $\rho$ plane. When we choose the fixed value of $\theta$ to be small, the image of this trajectory on the $\rho$ plane is qualitatively (though not quantitatively) similar to a familiar trajectory on this plane, namely the path traced out by boosting a point on the $x$ axis, i.e. the trajectory often discussed in the study of the Regge or chaos limits.

\section{Example of bulk Regge scaling}\label{reggecomp}

In this Appendix we explicitly evaluate the bulk
integral \eqref{genformoo} in the Regge limit in the special case that $N$ is a constant (let's take it to be unity).

At leading order in the small $\theta$ limit the bulk integral simplifies to 
\begin{equation}\label{numu}
I=\int_{\mathbb{H}_{D-1}} d^{D-1}y f_i(a, y_0, y_i)
\end{equation} 
where 
\begin{equation}\label{fozA}  
f_i(a, y_0, y_i)= \int 
\frac{dU d V }{(U+a ~ y_0+i\epsilon)^{a_1} (-U+~ y_1 +i\epsilon)^{a_2} (V +a~ ~ y_0+i\epsilon)^{a_3} 
	(-V- ~ y_1+i\epsilon)^{a_4}}
\end{equation} 
As we have explained in the main text, the integral over $U$ and $V$ is easily evaluated. 
Using a change of variable, $U+a~y_0 = \tilde{u}$ and $V+a~y_0 = \tilde{v}$, the above integral becomes,
\begin{equation}\label{fozAS1}  
f_i(a, y_0, y_i)= \int 
\frac{d\tilde{u}~ d\tilde{v} }{(\tilde{u}+i\epsilon)^{a_1} (-\tilde{u}+a~y_0+y_1 +i\epsilon)^{a_2} (\tilde{v}+i\epsilon)^{a_3} 
	(-\tilde{v}+a~y_0 -y_1+i\epsilon)^{a_4}}
\end{equation} 
Using a Schwinger parameter representation of the denominators we find that 
\begin{equation}\label{fozAS2a}  
f_i(a, y_0, y_i)= \frac{C_{a_1,a_2,a_3,a_4}}{\left(a~y_0+y_1 +i\epsilon\right)^{a_1+a_2-1}\left(a~y_0-y_1 +i\epsilon\right)^{a_3+a_4-1}}
\end{equation}
where 
\begin{equation} \label{cdef}  
C_{a_1,a_2,a_3,a_4}=\frac{ \Gamma \left(a_1+a_2-1\right) \Gamma \left(a_3+a_4-1\right) }{\Gamma \left(a_1\right) \Gamma \left( a_2 \right)\Gamma\left(a_3\right) \Gamma \left( a_4 \right)}
\end{equation} 
Plugging \eqref{fozAS2} into \eqref{numu}, we find that 
\begin{equation}\label{Iint}
I= \int_{\mathbb{H}_{D-1}} d^{D-1}y\frac{C_{a_1,a_2,a_3,a_4}}{\left(a~y_0+y_1 +i\epsilon\right)^{a_1+a_2-1}\left(a~y_0-y_1 +i\epsilon\right)^{a_3+a_4-1}}
\end{equation} 
In order to perform the integral over $\mathbb{H}_{D-1}$ it is convenient to use the following coordinates: 
\begin{equation}\label{coord}
y^\mu=(y_0,y_1,y_i)= (\cosh r \cosh \theta,\cosh r \sinh \theta, \sinh r ~\hat{n}_i(\phi^a))
\end{equation}
in terms of which 
\begin{equation}\label{int1}
\begin{split}
&I= \int \frac{\sinh^{D-3} r}{(\cosh r)^{-3+\sum a_i}} dr~d\theta~d\Omega_{D-3}\frac{C_{a_1,a_2,a_3,a_4}}{\left(a \cosh\theta+\sinh\theta +i\epsilon\right)^{a_1+a_2-1}\left(a\cosh\theta-\sinh\theta +i\epsilon\right)^{a_3+a_4-1}}\\
&= C_{a_1,a_2,a_3,a_4}\Omega_{D-3}\int dr \frac{\sinh^{D-3} r}{(\cosh r)^{-3+\sum a_i}}\int \frac{d\theta}{\left(a \cosh\theta+\sinh\theta +i\epsilon\right)^{a_1+a_2-1}\left(a\cosh\theta-\sinh\theta +i\epsilon\right)^{a_3+a_4-1}}
\end{split}
\end{equation}
The integral over $r$ above is a number independent of $a$. Although it will play no role in what follows, for completeness we record the value of this constant. 

\begin{equation}\label{int2}
\begin{split}
&M \equiv \int_0^\infty dr \frac{\sinh^{D-3} r}{\cosh^{4\Delta-3} r} =\frac{\Gamma \left(\frac{D}{2}-1\right) \Gamma \left(2 \Delta -\frac{D}{2}\right)}{2~ \Gamma (2 \Delta -1)}
\end{split}
\end{equation}
where, $\Delta= \frac{1}{4}\sum_{i=0}^4 a_i$.

It is convenient to club all the constants together, i.e. to define 
\begin{equation}\label{Ndefo}
N=M \Omega_{D-3} C_{a_1,a_2,a_3,a_4}
\end{equation}
in terms of which 
\begin{equation}\label{Ialm}
I= N\frac{1}{\left(a \cosh\theta+\sinh\theta +i\epsilon\right)^{a_1+a_2-1}\left(a\cosh\theta-\sinh\theta +i\epsilon\right)^{a_3+a_4-1}}
\end{equation} 

For ease  the rest of this Appendix we will specialize to the case $a_1+a_2=a_3+a_4=2 \Delta$ (this case is relevant, for instance, to the evaluation of the four point function of 4 operators, each of dimension $\Delta$, caused by a bulk $\phi^4$ interaction). With this specialization \eqref{Ialm} simplifies to 
\begin{equation}\label{int3}
\begin{split}
&I=N\int_{-\infty}^\infty \frac{d\theta}{ \left((a+i \epsilon)^2\cosh^2\theta-\sinh^2\theta \right)^{2\Delta-1}}
\end{split}
\end{equation}
When $a>1$ our integrand has no singularities on the real axis. In this case the $i\epsilon$ in 
\eqref{int3} makes no difference to the integral and can be dropped. When $a<1$, on the other hand, the integrand in \eqref{int3} has two poles on the real axis, located at 
$$\tanh \theta = \pm (a+i \epsilon)$$
In this case the $i \epsilon$ is crucial to the definition of the integral in \eqref{int3}.

In the main text we have argued that the integral $I$ has a branch cut singularity at $a=1$. Moreover we have argued that if we evaluate $I$ for $a>1$ and then analytically continue this result to $a<1$ via the upper half of the complex $a$ plane, then we will obtain the correct result for $I$ for $a<1$. In the rest of this Appendix we will directly verify these claims by explicitly evaluating $I$ separately for $a>1$ and $a<1$. 

\subsection{Exact result for every value of $\mathbf{n}$ when $\mathbf{a>1}$}

In this case, as we have explained above, we can ignore the $i\epsilon$ in the integrand. For this reason Mathematica is able to evaluate the integral; we find 
\begin{equation}\label{int4}
\begin{split}
&\int d\theta \frac{1}{ \left(a^2\cosh^2\theta-\sinh^2\theta\right)^{2\Delta-1}} = \frac{4^{2 \Delta -1} F_1\left(2 \Delta -1;2 \Delta -1,2 \Delta -1;2 \Delta ;\frac{1-a}{a+1},\frac{a+1}{1-a}\right)}{(2 \Delta -1)\left(a^2-1\right)^{2 \Delta -1}}
\end{split}
\end{equation}
Where $F_1$ is the Appell function. While the function $F_1$ may be unfamiliar, it is not particularly complicated, atleast for the values of parameters of relevance to our computation. To illustrate this in \eqref{tab:mytab} below we have listed the specific functional form of $F_1$ for small integer values of $n=2\Delta-1$ in terms of elementary functions


\begin{table}[H]
	\centering
	\Large\addtolength{\tabcolsep}{7pt}
	\begin{tabular}{|l|l|}
		\hline
		$\mathbf{n}$ & $\mathbf{\frac{4^{n}}{n\left(a^2-1\right)^n} F_1\left(n;n,n;n+1;\frac{1-a}{a+1},\frac{a+1}{1-a}\right)}$\\
		\hline
		$1$ &  $\frac{1}{a}\log \left(\frac{a+1}{a-1}\right)$\\
		$2$ &  $\frac{1}{2 a^3}\left[\left(a^2+1\right) \log \left(\frac{a+1}{a-1}\right)-2 a\right]$\\
		$3$ &  $\frac{1}{8 a^5}\left[\left(3 a^4+2 a^2+3\right) \log \left(\frac{a+1}{a-1}\right)-6 a\left(a^2+1\right)\right]$\\
		$\vdots$ &  $\vdots$\\
		\hline
	\end{tabular}
	\caption{Functional form of the integral \eqref{int4} for few integer $n$.}	\label{tab:mytab}
\end{table}

As we see from the examples listed in Table 
\ref{tab:mytab}, the integral $I$ does indeed always have a branch cut singularity at $a=1$ 
as anticipated on general grounds. Indeed the fact that this branch cut is logarithmic in nature is true at (atleast) all integer values of $n=(2 \Delta -1)$ as we see from the formula 
\begin{equation}\label{int5}
\begin{split}
&\lim_{a\rightarrow 1^+}\frac{4^{n} F_1\left(n;n,n;n+1;\frac{1-a}{a+1},\frac{a+1}{1-a}\right)}{n\left(a^2-1\right)^n} = \log\left(\frac{2}{a-1}\right) + \mathcal{O}(a-1)^0.~~~~~~~\forall n\in \mathbb{Z}^+
\end{split}
\end{equation}

We will now check that the analytic continuation of this result - taken through the upper half complex $a$ plane - correctly reproduces the result for $I(a)$ for $a<1$. For simplicity we restrict attention in this part of the Appendix to the especially simple case $n=1$, though we do not think it would be too difficult to generalize the computations presented in the rest of this Appendix to (at least) arbitrary integer values of $n$.

\subsection{Complete analytic structure in a special case $n=1$}
We will now completely analyse $I$ in the case $n=1$. We do this by re evaluating the integral over $\theta$, first for the case $a>1$, but in a manner that easily allows us to generalize to $a<1$.
\begin{equation}
\begin{split}
I=&N\int_{-\infty}^\infty \frac{d\theta}{\left(a \cosh\theta+\sinh\theta +i\epsilon\right)\left(a\cosh\theta-\sinh\theta +i\epsilon\right)}\\
=&2N\int_0^\infty \frac{d\theta}{\left(a \cosh\theta+\sinh\theta +i\epsilon\right)\left(a\cosh\theta-\sinh\theta +i\epsilon\right)}
\end{split}
\end{equation}
Now we will do a change of variable, $e^\theta = w$, which gives,
\begin{equation}\label{spcl}
\begin{split}
&I=2N\int_0^\infty \frac{dw}{w}\frac{1}{\left(a \left(w+\frac{1}{w}\right)+\left(w-\frac{1}{w}\right) +i\epsilon\right)\left(a\left(w+\frac{1}{w}\right)-\left(w-\frac{1}{w}\right) +i\epsilon\right)}\\
&=2N\int_0^\infty \frac{w~dw}{\left((a+1)w^2+a -1 +i \epsilon\right)\left((a-1) w^2+a+1 +i \epsilon\right)}\\
&=N\int_0^\infty \frac{dz}{\left((a+1)z+a -1 +i \epsilon\right)\left((a-1) z+a+1 +i \epsilon\right)}
\end{split}
\end{equation}
In going from the second to the third line in 
\eqref{spcl} we have made the variable change 
$z=w^2$.

\eqref{spcl} applies both to the cases $a>1$ and $a<1$.  In the case $a>1$ the integral \eqref{spcl} is elementary because none of the poles lie on the integration axis. In this case  $\epsilon$ can simply be set to zero and we obtain
\begin{equation}\label{spcl1}
\begin{split}
\frac{I_{>}}{N}&=\frac{1}{a}\log \left(\frac{a+1}{a-1}\right)=-\frac{1}{a}\log \left(\frac{a-1}{a+1}\right)
\end{split}
\end{equation}
As the analytic continuation of $\ln(a-1)$ to 
$a <1$ via the upper half complex plane is 
$$ \ln (a-1) \rightarrow  \ln (1-a) +  i\pi $$
\footnote{This is because the argument of $a-1$ changes continuously from $0$ to $\pi$ as we 
	go from the the real axis with $a>1$ to the real axis with $a<1$ via the upper half of the complex plane.}	
it follows that the analytic continuation of 
$I_>$ to $a<1$ via the upper half plane, $I_<$, 
is given by  
\begin{equation}\label{spcl2}
\begin{split}
\frac{I^\circlearrowright_{<}}{N}&=\frac{1}{a}\left[-i\pi -\log \left(\frac{1-a}{1+a}\right)\right]=
\frac{1}{a}\left[-i\pi +\log \left(\frac{1+a}{1-a}\right)\right]
\end{split}
\end{equation}

We will now directly evaluate the integral \eqref{spcl} for the case $a<1$ and verify that we indeed obtain the result \eqref{spcl2}. 
When $a<1$, the integrand in \eqref{spcl} has  two poles that lie (approximately) on the integration contour (i.e. the positive real axis). These two poles occur at $z=z_\pm$ where
\begin{equation}\label{spcl3}
\begin{split}
z_+ = \frac{1+a}{1-a}+\frac{i \epsilon }{1-a}, ~~~~~~~~~\text{and,}~~~~~~~~~  z_-=\frac{1-a}{a+1}-\frac{i \epsilon }{1+a}.
\end{split}
\end{equation}

The integral \eqref{spcl} may be evaluated by rotating the contour counter-clockwise by an angle $\pi$, i.e. changing the integral from zero to $\infty$ along the positive real axis to zero to $- \infty$ along the negative real axis. In performing this integral we cut the pole at $z=z_+$. 
As the integrand decays like $1/|z|^2$ at infinity, the contribution to the integral from the arc at infinity vanishes, and so it follows that the integral in \eqref{spcl} equals the value of the same integral evaluated along the negative real axis plus the contribution of the pole at $z=z_+$. We now evaluate these two contributions separately. 

The integral along the negative real axis can be evaluated by setting $z=-y$ in \eqref{spcl}. rotated path has no pole in its path (recall $a>0)$) so can be integrated in an elementary manner (and in particular dropping the $i\epsilon$) and yields
\begin{equation}\label{spcl4}
\begin{split}
\frac{I^\leftarrow}{N} &=\int_0^\infty \frac{dy}{\left(-(a+1)y+a -1\right)\left((a-1) -y+a+1 \right)}\\
&= \frac{1}{a}\log \left(\frac{1+a}{1-a}\right).
\end{split}
\end{equation}
The contribution of the pole at $z_+$ is given by 

\begin{equation}\label{spcl5}
\begin{split}
\frac{I^\odot}{N} &= ~2 \pi i ~\text{Res} \left[\frac{1}{\left((a+1)z+a -1\right)\left((a-1) z+a+1\right)}\right]_{z=z_+}\\
&= -\frac{i \pi }{a}.
\end{split}
\end{equation}

It follows that the final answer for $I$ at $a<1$ is 
\begin{equation}\label{spcl6}
\begin{split}
\frac{I_<}{N} &= \frac{I^\leftarrow}{N}+\frac{I^\odot}{N}\\
&= \frac{1}{a}\left[-i \pi+\log \left(\frac{1+a}{1-a}\right)\right] = I^\circlearrowleft_<.
\end{split}
\end{equation}
We have thus verified in this special example that the appropriate analytic continuation of the $a>1$ result for $I$ does indeed reproduce $I$ at $a<1$, as predicted on general grounds in the main text. 

\subsection{Integral at small $a$} \label{isa}

In this subsection we evaluate \eqref{Ialm} at leading order in the small $a$ expansion. 
For small $a$, major contribution to the integral
\eqref{Ialm} comes $\theta\sim a$. So the last integral in \eqref{Ialm} becomes,
\begin{equation}\label{int6}
\begin{split}
&\int \frac{d\theta}{\left(a+\theta +i\epsilon\right)^{a_1+a_2-1}\left(a-\theta +i\epsilon\right)^{a_3+a_4-1}}\\
&=\int \frac{d\theta}{\left(\theta +i\epsilon\right)^{a_1+a_2-1}\left(2a-\theta +i\epsilon\right)^{a_3+a_4-1}}\\
&=\frac{\Gamma(a_1+a_2+a_3+a_4-3)}{\Gamma(a_1+a_2-2)\Gamma(a_3+a_4-2)}\frac{1}{(2a)^{-3+4\sum_i a_i}}
\end{split}
\end{equation}


\section{Singularities of holographic correlators from contact interactions} \label{singcor} 

In this subsection we will demonstrate that the 
classical holographic correlator with insertion points \eqref{fpscatnnnn} - generated by a bulk local contact interaction - is an analytic function of of its parameters ($\tau$ and $\theta$) away from the `bulk point singularity' line $\tau=0$ and also the `lightcone singularity lines' $\tau=\theta$ and $\tau=\pi-\theta$ . The result of this section is a simple specialization of the general results of sections 2 and 3 of \cite{Maldacena:2015iua} to the particular case of \eqref{fpscatnnnn}.

Any correlator obtained from a local contact interaction is a sum of terms of the form 
\begin{equation}\label{sumoffterms}
C= \int_{AdS_{D+1}} d ^{D+1} X \frac{Q(Z_i, P_i, X)}
{\prod_{i=1}^4 (-2P_i.X )^{q_i}}
\end{equation}
Here $P_i$ are the the boundary insertion points, $Z_i$ are the boundary polarizations, $X$ position of the interaction vertex in $AdS_{D+1}$ and $q_i=\Delta_i + n_i$ 
where $\Delta_i$ is the dimension of the operator inserted at $P_i$ and $n_i$ is a positive integer.
\footnote{In the case that the operator $O_i$ 
	is traceless symmetric with $J_i$ indices, 
	$n_i \geq J_i$. We get terms with $n_i>J_i$ when 
	derivatives - from the bulk interaction vertex - hit the propagator.}

Because of potential singularities from the denominator, the expression \eqref{sumoffterms} is not yet completely unambiguous; it needs to be $i \epsilon$ corrected. Note that 
$$-2P_i.X= 2 \cosh r\cos (\tau -\tau_i) + ...$$
where $\tau_i$ is the global time of the 
boundary point $P_i$ and $\tau$ is the global point of the bulk interaction point $X$. The $i \epsilon$ replacement rule \eqref{reprule} instructs us to make the replacement 
\begin{equation}\label{repruu}
-2P_i.X \rightarrow -2P_i.X + i \sin(\tau-\tau_i) (\tau -\tau_i) \epsilon 
\end{equation} 
In particular when  
\begin{equation}\label{slab} 
|\tau -\tau_i|<\pi
\end{equation}
\eqref{repruu} simplifies to 
\begin{equation}\label{effrepru}
-2P_i.X \rightarrow -2P_i.X + i\epsilon
\end{equation} 
For $X$ such that \eqref{slab} is obeyed, 
\eqref{sumoffterms} is modified to 
\begin{equation}\label{sumofftermsmod}
C= \int_{AdS_{D+1}} d ^{D+1} X \frac{Q(Z_i, P_i, X)}
{\prod_{i=1}^4 (-2P_i.X + i \epsilon  )^{q_i}}
\end{equation}

The integral over $X$ in \eqref{sumofftermsmod} is potentially singular when one or more of the four denominator factors in \eqref{sumofftermsmod} vanish - and the contour of integration over $X^\mu$ cannot be modified in the complex $X^M$ plane  (in a manner consistent with Cauchy's theorem, i.e. without crossing a pole) to avoid all singular points. 

\subsection{End point singularities}\label{eps}

The boundary of the integration contour for the integral over the $AdS$ lies at the boundary of $AdS$. The integral is potentially singular when the integrand of 
\eqref{sumofftermsmod} has a pole at a boundary point. This happens at four points on the boundary, namely $X=P_1$, 
$X=P_2$, $X=P_3$ and $X=P_4$.

For generic values of $P_a$ these  potentially problematic points do not in fact lead to a singularity in the integral. Let us, for instance, consider the pole at $X=P_1$. The $i\epsilon$ in \eqref{sumofftermsmod} ensures that the singularity does not really lie on the integration contour, the potentially dangerous contribution from integral in the neighbourhood of this point is proportional to the non-singular residue of this `pole'. This residue is singular only if 
$P_a.P_1=0$ for $a=2,3, 4$. This happens only when the points 
$P_1$ and $P_a$ are lightlike separated on the boundary (i.e), and leads to the usual `lightcone' singularities familiar from the study of conformal field theories. As is familiar - and as we have explained earlier in this section - these light cone singularities lead to branch cuts in correlators, and the correlation function continues to be analytic on the branched cover of cross ratio space. 

In this paper we are interested in the correlator at the points 
\eqref{fpscatnnnn}. Lightcone singularities occur at $\tau=\pi -\theta$ (at which point the pairs of points $(P_1, P_4)$ and $(P_3, P_2)$ are lightlike separated, and the conformal cross ratio ${\bar z}=1$) and also at $\tau=\theta$ (at which point $(P_1, P_2)$ and $(P_3, P_4)$ are lightlike separated and the conformal cross ratio $z=1$). At exactly the points described, 
(namely $\tau=\pi -\theta$ and $\tau=\theta$) we land exactly at the branch points of the correlator which is thus singular. As we have explained, however, we can continue past this singularity by going around it; and so the value of the correlator at in the ranges $\tau \in (0,\theta)$, $\tau \in (\theta, \pi-\theta)$, $(\pi-\theta, \pi)$ are different `boundary values' (across cuts) of the same analytic function.

\subsection{Pinch singularities}

As explained in \cite{Maldacena:2015iua}, a potential singularity at $X=X_0$ (where some subset $\{ P_a \}=S$ obey $P_a.X_0=0$) can be avoided provided we can find a small vector $\delta Q$ so that on the new contour 
$$X_0 \rightarrow X_0 +i \delta Q $$
and 
\begin{equation}\label{condofe}
~~~~\delta Q. X_0=0, ~~~~{\rm and}~~~-2P_a . \delta Q >0, ~~~~{\rm for ~all~P_a~in} ~ S,  
\end{equation} 
The first equation is needed to ensure that 
\begin{equation}\label{stayonads}
(X_0+i\delta Q)^2=-1
\end{equation} 
(i.e. the modified contour is on the $AdS_{D+1}$ sub-manifold - note we work to first order in the small modification $\delta Q$). The second condition is needed to ensure that the modification does not pass through any of the poles in \eqref{sumoffterms}.  

\footnote{To see why this is the case, 
	consider the single variable integral  $\int \frac{dz}{z+i \epsilon}$ 
	along the real axis. The pole in the integrand lies at $z=i\epsilon$. Hence we are free to move the contour of integration in the upper half plane, i.e. to give $z$ a positive imaginary part. In other words the $+i \epsilon$ in the integrand tells us that we are allowed to change the integration contour so that ${\rm Im}(z)$ is positive when ${\rm Re}(z)=0$; however the reverse modification is not allowed. In the same way the integral contour in \eqref{sumofftermsmod} can be modified (without changing the value of the integral) if we ensure that at all $X$ that obey  $P_a . X=0$, $P_a.{\rm Im}(X)>0$. }

The first condition 
\eqref{condofe} tells us that $\delta Q$ is orthogonal to $X$, and so forces $\delta Q$ to lie in an $\mathbb{R}^{D,1}$ in $\mathbb{R}^{D,2}$. As each $P_a$ 
obeys $P_a.X=0$, each $P_a$ lies in this $\mathbb{R}^{D,1}$. Each $P_a$ is associated with a co-dimension one hyperplane in $\mathbb{R}^{D,1}$ that passes through the origin ($P_a$ is the one-form normal to this plane - this is the plane, whose intersection with $X^2=-1$ gives the light sheet that emanates out of $P_a$). The condition $P_a. \delta Q >0$ tells us that $\delta Q$ lies on one side of this hyperplane in $\mathbb{R}^{D,1}$. 

Let us first consider a point $X$ at which the equation $P_a.X=0$ is obeyed for $m$  
values of $a$. Since no collection of three or fewer $P_a$ in \eqref{fpscatnnnn} are linearly dependent (and since $D-1 \geq 3$) the hyperplanes associated with the $m$ $P_a$ slice up the $\mathbb{R}^{D,1}$ into $2^m$ distinct sectors. The positivity condition in \eqref{condofe}
is met provided we choose $\delta Q$ to lie within one of these (the all positive) sector. 
It follows that \eqref{condofe} admits an infinite number of solutions, and the integral 
\eqref{sumofftermsmod} receives regular contributions from all such points. 

If $\tau \neq 0$ then the four $P_i$ span an $\mathbb{R}^{2,2}$ and there are no solutions (in $AdS_{D+1}$) to the simultaneous equation
$P_i.X=0$. 

It follows that the integral \eqref{sumofftermsmod} can only be singular 
at $\tau=0$. At this value of $\tau$ the four 
$P_i$ span an $\mathbb{R}^{2,1}$ and the simultaneous equations  $P_i.X=0$ are obeyed on the $\mathbb{H}_{D-2}$
$$X=\left(0, \cosh r, 0, 0, (\sinh r) {\vec m}
\right)$$
where ${\vec m}$ is any unit vector in $\mathbb{R}^{D-2}$. 
When $\tau=0$ 
\begin{equation}\label{sumofpapp}
P_1+P_2+P_3+P_4=0
\end{equation} 
It follows from \eqref{sumofp} that 
\begin{equation}\label{sumofpcons}
P_1. \delta Q+P_2. \delta Q+P_3. \delta Q+P_4. \delta Q=0
\end{equation} 
As \eqref{sumofp} is clearly inconsistent with 
the condition $P_a.\delta Q>0$ for all $a$ \cite{Maldacena:2015iua}, it 
follows that \eqref{condofe} cannot be obeyed, and so it is not possible to deform the integration contour away from the singular hyperboloid and  the contribution to \eqref{sumofftermsmod} is singular. 

In summary, we have demonstrated that the holographic correlator with insertions at  \eqref{fpscatnnnn} has a pinch singularity at $\tau=0$. It also has endpoint singularities at $\tau=\theta$ and $\tau=\pi-\theta$, but these are branch cut singularities that can be continued around, as we have explained earlier in this section. 

\section{$\rho=0$ branch cuts and UV softening} \label{rous}
 
\subsection{$D=2$}

The branch cut singularity in \eqref{finrestn_TEST} had its origin in the integral
\begin{equation}\label{omegainteg}
\int_0^\infty d\omega ~\left( \omega^{\Delta -4} ~\omega^r ~
e^{-2i\omega \tau } \right) 
\end{equation} 
(see the middle equation in \eqref{nntre}).
The factor of $\omega^r$ in the integrand in \eqref{omegainteg} had its origin in the fact that the bulk $S$ matrix for the corresponding contact term grew like $\omega^r$. Of course, this is an approximation. In any `real' bulk theory we expect this power-law growth of the S matrix to be ameliorated by stringy and quantum effects. As a crude model for the stringy softening, we can follow \cite{Okuda:2010ym, Maldacena:2015iua} and make the replacement $\omega^r \rightarrow \omega^r e^{- l_s^2 \omega^2}$. We have chosen the name $l_s$ suggestively. In weakly coupled string theory $l_s$ will be proportional to the string length. In the stringy context while $l_s$ is constant in the sense that it is independent of energy, it is, however, a function of the scattering angle. We will return to a study of the angular dependence below.

In this section we wish to study the integral
\begin{equation}\label{omegainteg2tin}
{\tilde I}(\tau)= \int_0^\infty  d\omega ~\left( \omega^{\Delta -4+r}  e^{- l_s^2 \omega^2-2i\omega \tau } \right) 
\end{equation} 
In our intermediate analysis, we will find it more convenient to work with the differently normalized integral
\begin{equation}\label{omegainteg2}
I(\tau)= l_s^{\Delta -3+r} \int_0^\infty  d\omega ~\left( \omega^{\Delta -4+r}  e^{- l_s^2 \omega^2-2i\omega \tau } \right) 
\end{equation} 
Of ${\tilde I}(\tau)$ is proportional to $I(\tau)$
\begin{equation}\label{difnor}
{\tilde I}(\tau)= l_s^{-\Delta +3-r} I(\tau); 
\end{equation} 
this proportionality factor will gain physical significance when $l_s$ is a function of the scattering angle.

The change of variables $\omega =\frac{x}{l_s}$ and $y=\frac{\tau}{l_s}$transforms \eqref{omegainteg2} to 
\begin{equation}\label{omegainteg3}
I(y)=
\int_0^\infty  dx  ~\left( x^{\Delta -4+r}  e^{-  x^2-2i x y} \right) 
\end{equation} 
Note that $\tau$, $\rho$ and $y$ are all proportional to each other at fixed $\sigma$, and so the analytic structure of $I(y)$ around $y=0$ is the same as the analytic structure of the RHS of \eqref{finrestn_TEST}.
In the rest of this subsection, we analyze the function $I(y)$. 

First, note that $I(y)$ is a manifestly single-valued function of $y$ that is analytic everywhere in the $y$ complex plane. It is instructive to study this function in various regimes.  

For $|y| \ll 1$ the term $2 i xy $ is a small perturbation of the argument of the exponent in \eqref{omegainteg3}. It follows that the integral in \eqref{omegainteg3} admits a power series expansion in $y$, which can be evaluated by first expanding the integrand in a power series and performing the integral term by term. We find 
\begin{equation}\label{psey}
I(y)=\sum_n a_n y^n~~~~~~~~a_n= \frac{(-2i)^n}{2 n!}\Gamma\left( \frac{\Delta-3+r+n}{2}\right) 
\end{equation} 
At large values of $n$ 
\begin{equation}\label{largena}
a_n \approx \frac{1}{2}\left(-i\sqrt{\frac{2e}{n}}\right)^n
\end{equation}
From which it follows that the expansion \eqref{psey}
is convergent, with an infinite radius of convergence \footnote{See section 3.2 of \cite{Okuda:2010ym} for discussion on hard scattering of string amplitudes whose bulk point integral behave in a similar manner.}. 
Of course the truncation of \eqref{psey} to the first few terms gives a good approximation to $I(y)$ only for all complex values such that $|y|\ll 1$ or 
\begin{equation}\label{ineqontau}
|\tau| \ll l_s 
\end{equation} 

When $|y| \gg 1$, on the other hand, $I(y)$ behaves very differently depending on whether ${\rm Im} (y) >0$ or ${\rm Im}(y)<0$. When ${\rm Im} (y)>0$ the 
damping and oscillations from the factor $e^{2i x y}$ cut off the integral before the factor $e^{-x^2}$ becomes important at all when $y$ is large. In this case the integral is well approximated by dropping the factor $e^{-x^2}$ and so is well approximated by
\begin{equation}\label{approxlargey}
I(y)= \frac{\Gamma\left(\Delta-3+r\right)}{(2i)^{\left(\Delta-3+r\right)}y^{\Delta -3+r}}
\end{equation} 
Note that $y \propto \rho$, so \eqref{approxlargey}
reproduces the singularity visible in \eqref{finrestn_TEST}. Corrections to \eqref{approxlargey} can be systematically computed by expanding the factor of $e^{-x^2}$ in a power seies in 
$x$ and then integrating term by term. This is reasonable as the integral receives its dominant contribution from small values of $x$ when 
${\rm Im y}$ is large and positive. We find 
\begin{equation}\label{approxlargeycor} \begin{split} 
&I(y) = \sum_{n=0}^\infty \frac{(-1)^n}{n!} \frac{\Gamma(\Delta+r+2n-3)}{(2i)^{\Delta+r+2n-3}}\frac{1}{y^{\Delta+r+2n-3}}\\
\end{split}
\end{equation} 
In contrast with the expansion \eqref{psey}, \eqref{approxlargeycor} is an asymptotic rather than a convergent expansion. This mathematical fact reflects the physical fact that the truncation of \eqref{approxlargeycor} to a few terms gives us a 
good approximation to the actual behaviour of the function $I(y)$ only when ${\rm Im}(y)$ is positive rather 
than negative. \footnote{An analogy is the following. 
The perturbation series for the energy spectrum of a harmonic oscillator is asymptotic rather than convergent, reflecting the fact that the first few terms of the expansion give a good approximation to the spectrum of the theory in the case that the quartic term in the potential is positive, but a very bad approximation to the (ill defined, unbounded) spectrum of the theory in the case that the quartic term is negative.}

When $|y| \gg 1$ but ${\rm Im}(y)<0$, $I(y)$ behaves completely differently from the case just examined above. In this case, the factor $e^{2i x y}$ exponentially enhances (instead of cutting off)
the integrand, which continues to grow until it is eventually cut off at of order unity by the factor 
$e^{-x^2}$ (which now plays a crucial role; without this factor the integral would have divergent and so ill defined). In this regime, very, very approximately, 
\begin{equation}\label{Ivo}
|I(y)| \sim e^{-{\rm Im}(y) }
\end{equation} 
(moreover, we expect the phase of $I(y)$ to oscillate rapidly with the real part of $y$). 

We now have a good qualitative picture of the behaviour of $I(y)$ on the complex $y$ plane. This function is well approximated by the first few terms of \eqref{psey} when $|y| \ll 1$, is well approximated by the first few terms of \eqref{approxlargeycor} when $|y| \gg 1$ and ${\rm Im}(y)>0$, but is harder to control (very approximately given by \eqref{Ivo} when 
$|y| \gg 1$ and ${\rm Im}(y)<0$.

In summary, the  softening of the integral induced by the factor $e^{-x^2}$ in $I(y)$ impacts the functional dependence of \eqref{finrestn_TEST} in the complex  $\rho$ differently at different values of $\rho$ (recall $\rho$ is proportional to $y$). First, it modifies the function at small values of $\rho$ to smoothen out the $\rho=0$ ($y=0$) singularity in \eqref{finrestn_TEST}. Second, it has a negligible impact on the value of the function at large values of $|\rho|$ provided ${\rm Im}(\rho)>0$; in this regime  \eqref{finrestn_TEST} continues to be a good approximation to a regulated function.  When 
${\rm Im}(\rho)<0$, on the other hand, $I(y)$ is very different from \eqref{finrestn_TEST} even at large values of $|y|$ (the function behaves rather wildly in this region, growing very large in modulus and also oscillating very rapidly). It follows that \eqref{finrestn_TEST} cannot be used to reliably compute the discontinuity around $\rho=0$ even at large $\rho$. In fact $I(y)$ is an everywhere analytic and single valued function of $y$, that does not have a branch cut.

\subsubsection{Angular dependence} 

Returning to the function ${\tilde I}(\tau)$ and restricting our attention to the upper half plane, we 
see that ${\tilde I}(\tau)$ interpolates between two cases, 

\begin{equation} \label{actI}
{\tilde I}(\tau) = 
\begin{cases}
\frac{\Gamma\left( \frac{\Delta-3+r}{2}\right)}{2l_s^{\Delta +r -3}} & \hspace{1cm}\tau \ll l_s\\
\frac{\Gamma\left(\Delta-3+r\right)}{(2i)^{\left(\Delta-3+r\right)} \tau^{\Delta -3+r}} &\hspace{1cm}\tau \gg l_s
\end{cases}
\end{equation}

Note that $l_s$ controls the exponential damping of the scattering amplitude in units in which 
the radius of AdS is unity (as has been assumed through this paper) i.e. $l_s$ in the formulae above is really 
$\frac{l_s}{R}$ where $R$ is the AdS radius. 

 We have mentioned above that the parameter $l_s$ is, in general, a function of the scattering angle. If, for instance, we follow \cite{Okuda:2010ym} and use the textbook formula for the 
 fixed angle high energy behaviour of the classical string S matrix we obtain
 \begin{equation}\label{lsang}
 l_s^2= \frac{\alpha'}{2R^2}  \left(-\sin^2\left(\frac{\theta}{2}\right)\log \left[\sin^2\left(\frac{\theta}{2}\right)\right]- \cos^2\left(\frac{\theta}{2}\right)\log \left[\cos^2\left(\frac{\theta}{2}\right)\right] \right) 
 \end{equation}
 In the small scattering angle limit this formula reduces to 
 \begin{equation}\label{lsangs}
 l_s= \frac{\sqrt{\alpha'}}{2R} \theta\log|\theta|+\cdots
 \end{equation}
 
Plugging \eqref{lsangs} into \eqref{actualform} we obtain  (see\cite{Okuda:2010ym})\footnote{See \cite{Dodelson:2019ddi} for related discussion in Mellin amplitude.}
\begin{equation}\label{interpnn}
{\tilde I}(\tau) = 
\begin{cases}
\frac{1}{2}\Gamma\left( \frac{\Delta-3+r}{2}\right)\left(\frac{2R}{\sqrt{\alpha'}}\right)^{\Delta +r -3}\frac{1}{\left(\theta\log|\theta|\right)^{\Delta +r -3}} & \hspace{1cm}\tau \ll \frac{\sqrt{\alpha'}}{2R} \theta\log|\theta|\\
\frac{\Gamma\left(\Delta-3+r\right)}{(2i)^{\left(\Delta-3+r\right)} }\frac{1}{\tau^{\Delta+r-3}} &\hspace{1cm}\tau \gg \frac{\sqrt{\alpha'}}{2R} \theta\log|\theta|
\end{cases}
\end{equation}

In \eqref{interpnn} have reinstated $R$, the radius of $AdS$, (recall that through out this paper we have worked in units in which $R$ is unity).
Note that the logarithm of the angle in the first of \eqref{interpnn}, together with interpolation between singular behaviours in $\tau$ and $\theta$ lends  ${\tilde I}(\tau)$ an intricate analytic structure as a function of $\theta$ and $\tau$, i.e. of $\rho$ and $\sigma$. This observation suggests that the holographic correlator resulting from full string interactions has a much more interesting analytic structure than the correlators obtained from local bulk contact interactions that we have studied in the bulk of this paper. We hope to return to the study of correlators holographically generated by string interactions in future work.

\subsection{$D>2$}\label{rgtt}

When $D>2$ the singular part of \eqref{gsingnn} is generated by the integral 
\begin{equation}\label{omegainteg4}
J(y)=
\int_0^\infty  d \zeta \sinh^{D-3} \zeta 
\int_0^\infty  dx  ~\left( x^{\Delta -4+r}  e^{-  x^2-2i x y \cosh \zeta} \right). 
\end{equation} 
Performing the integral over $x$ we obtain
\begin{equation}\label{zepa}
J(y)=
\int_0^\infty  d \zeta \sinh^{D-3} \zeta~ I(y \cosh \zeta)
\end{equation}
It is easy to approximately perform the integral in \eqref{zepa} in two limits. When $|y| \gg 1$  and 
${\rm Im}(y)>0$ the argument of the function $I$ in \eqref{zepa} shares these two properties, so we can use the approximation \eqref{approxlargey} for $I(\cosh \zeta y)$ to find 
\begin{equation}\label{zepamod}
\begin{split}
&J(y)\approx 
\frac{\Gamma\left(\Delta-3+r\right)}{(2i)^{\left(\Delta-3+r\right)}y^{\Delta -3+r}} 
\int_0^\infty  d \zeta \frac{\sinh^{D-3} \zeta}{
\cosh^{\Delta -3+r} \zeta} \\
&=N_{D, \Delta} \frac{\Gamma\left(\Delta-3+r\right)}{(2i)^{\left(\Delta-3+r\right)}} ~ \frac{1}{y^{\Delta -3+r}}
\end{split}
\end{equation}
where $N_{D, \Delta}$ is defined in \eqref{gsingnn2}. 
The approximation \eqref{zepa} is easily improved by by inserting \eqref{approxlargeycor} rather than \eqref{approxlargey} into \eqref{zepa} but we will not bother to do so here. 

At small $|y|$, on the other hand, the dominant contribution to the integral in \eqref{zepa} comes from $\zeta$ such that $y \cosh \zeta$ less than or of order unity. Working in a very crude approximation one can replace 
$I(y \cosh \zeta)$ by $I(0) ~\theta( 1-y \cosh \zeta )$. Making this replacement we find that for $|y|\ll 1$
\begin{equation}\label{zepaimp}
J(y) \sim  \frac{1}{2(D-3)}\Gamma\left( \frac{\Delta-3+r}{2}\right) ~ \frac{1}{y^{D-3}} 
\end{equation}
The approximation \eqref{zepaimp} is very crude (even the overall coefficient on the RHS is not reliable but the form of the $y$ dependence is). We will not attempt to improve this approximation here, but leave this as an exercise for the interested reader.

The main qualitative point is that the function $J(y)$ roughly similar to $I(y)$ in the previous section with one key difference. While $I(y)$ interpolates from a constant to a rapid decay $\propto \frac{1}{y^{\Delta -3+r}}$ as $y$ increases from zero to infinity, 
$J(y)$ interpolates from the weaker power 
$\frac{1}{y^{D-3}}$ \cite{Maldacena:2015iua} to the stronger power (more rapid decay) $\frac{1}{y^{\Delta -3+r}}$ as $y$ varies over the same range. Most of the  comments in the previous subsection about the analytic properties of $I(y)$ also hold for $J(y)$ with small modifications. In particular it is a single valued function of $y$ everywhere in the complex plane.

\section{Massive higher spin particles} \label{mhsp}

\subsection{Multiple powers of $\rho$}

In this Appendix we outline some of the complications that arise when attempting to generalize the analysis of this paper to massive higher spin particles. We focus on the simplest case, namely that of massive vector particles. The complications that arise all have their origin in a familiar fact; namely that the S matrices of `longitudinal' polarizations of massive particles grow faster with energy in the high energy limit than those of transverse polarizations. 

Quantitatively, the complications arise as follows. As we have 
explained in the main text (\eqref{polgnn}) the bulk to boundary propagator for a vector operator of dimension $\Delta$ 
is given by 
\begin{equation}\label{polgnnapp} 
\left(1- \frac{1}{\Delta} \right) \frac{Z_A^\perp}{(P.X)^\Delta} 
+  \nabla_A \left( \frac{Z.X}{\Delta (P.X)^\Delta}\right) 
\end{equation} 
Applying \eqref{ident} on \eqref{polgnnapp} and working to leading order we find the wave form \eqref{scatwavesprelmain}.

Let us first proceed by supposing \eqref{nntrespin} (which we reproduce here for convenience) 
\begin{equation}\label{nntrevapp}
\begin{split} 
G_{\rm sing}&=-\frac{2\pi^3 \left( \prod_a \tilde{\mathcal{C}}_{\Delta_a, 1}\right)}{ \sqrt{\sigma(1-\sigma)}}
\int_{\mathbb{H}_{D-2}} \sqrt{g_{D-2}} ~d^{D-2} X \int d\omega \omega^{\Delta -4}
e^{i\omega P.X } {\cal S}\left(\omega \right)\\
\Delta&= \sum_i \Delta_i \\
\tilde{\mathcal{C}}_{\Delta_a, 1}&=\frac{\mathcal{C}_{\Delta_a,1}}{2^{\Delta+1} i^{\Delta}\Gamma(\Delta)}
\end{split} 
\end{equation} 
continues to capture all relevant singularities (we will see later this is untrue) where ${\cal S}(\omega)$ is the flat space S matrix for the waves \eqref{polgnnapp}. 

Let us suppose that the bulk interaction term is of $r^{th}$ order in derivatives. Let us decompose the $S$ matrix in \eqref{nntrevapp} as follows 
\begin{equation}\label{smatrixdecomp}
{\cal S}(\epsilon_i, k_i)=\sum_{m=0}^4 {\cal S}_m(\epsilon_i, k_i)
\end{equation}  
where $S_n$ is the S matrix of $n$ `longitudinal' polarization (i.e. the polarization 
proportional to $k_i$) in \eqref{scatwavesprelmain} and $4-n$ transverse polarizations (i.e. the polarizations proportional to $Z_i^\perp$).
\footnote{$S_n$ can be defined more formally as follows. We formally modify \eqref{scatwavesprelmain} to 
	included a new counting variable $\theta$ (which is set to unity at the end of the computation) as
	\begin{equation}\label{modep}
	\epsilon_M = \left( 1- \frac{1}{\Delta_i} \right) 
	(Z_i^\perp)_M + i \theta k^i_M(Z_i.X_0) 
	\end{equation}
	${\cal S}_n$ is the part of the S matrix which scales like 
	$\theta^n$.}. When the scattering matrix results from an $r$ derivative bulk interaction term, the quantity ${\cal S}_m$ scales with the 
overall energy scale of the scattering momenta 
\eqref{forvec} like 
\begin{equation}\label{overscal} 
{\cal S}_m \sim \omega^{r+m}
\end{equation} 
(the dependence of the RHS of \eqref{overscal} on $m$  follows from the extra factor of momentum in  \eqref{scatwavesprelmain}).
Note that the scattering waves \eqref{scatwavesprelmain} that produce ${\cal S}_n$ are not canonically normalized (more about this in the next subsection)\footnote{The fact that these waves do not have standard normalization is of no qualitative importance for the transverse wave - whose normalization factor is independent of $\omega$. However this point is of crucial importance for the longitudinal mode. As we discuss in more detail in the next subsection, the norm of this mode in \eqref{scatwavesprelmain} is $\omega$ dependent, and goes to zero in the limit $\omega \to 0$, cancelling the singular behaviour of longitudinal mode scattering amplitudes at high energy.}.

Plugging \eqref{smatrixdecomp} into \eqref{nntrevapp} we find that one contribution to the singularities in 
$\rho$ of the correlator is given by
\begin{equation}\label{singcorv}\begin{split} 
G_{\rm sing}&=\sum_{m=0}^4 G^m_{\rm sing}\\
G^m_{\rm sing}&=i\left( 2\pi^3 \left(\tilde{\mathcal{C}}_{\Delta_a,1}\right)^4\right)\Gamma (\Delta +r +m-3)e^{-\frac{i(\Delta +r+m)}{2}} ~~~
\frac{\sqrt{1-\sigma}^{\left( \Delta+r+m-4 \right)}   }{\sigma^{\frac{\Delta + r+m-2}{2}} \rho^{\Delta+r+m-3}}\\
&~~\times  \int d \omega_{D-3}  d\zeta~ \frac{\sinh^{D-3} \zeta}{ \cosh^{\Delta +r+m-3 } \zeta} 
~{\hat {\cal S}}_m (X_0)\\ 
{\hat {\cal S}}_m &= \frac{ {\cal S}_m}{\omega^{r+m} }\\
 X_0&=(0, \cosh \zeta, 0, 0, \sinh \zeta {\hat n}_i), ~~~~i=1 \ldots D-2\\
\end{split} 
\end{equation} 
and ${\cal S}_m$ is the part of the S matrix of the un normalized waves \eqref{scatwavesprelmain} involving 
$m$ longitudinal and $4-m$ transverse modes, where 
the scattering takes place for the waves 
\eqref{scatwavesprelmain}, with the scattering momenta 
given by \eqref{forvec} with $\omega$ set to unity.
The dependence of the power of the $\rho$ singularity on $m$ in \eqref{singcorv} is a consequence of \eqref{overscal}. 

Had \eqref{singcorv} accurately captured the coefficients of all the $\rho$ singularities that appear in that formula, we could have used it to establish that if any of the S matrices that appear as coefficients of different powers of $\rho$ grow faster than $s^2$ in the Regge limit then the correlator 
continued to the Causally Regge sheet would violate the chaos bound. Unfortunately \eqref{singcorv} is not complete. The problem is that the term with, say, $m=4$ in \eqref{smatrixdecomp} has its origin in the overlap of four longitudinal modes. The $m=4$ term in \eqref{smatrixdecomp} 
does indeed capture the most singular term that arises from this overlap. However, the overlap of four longitudinal modes could also produce lower order singularities (the coefficients of these singularities would `see' the fact that the bulk to boundary propagator does not quite yield a plane wave, and that the elevator is not quite a flat space). These lower order singularities can, in principle, modify the coefficients of $\frac{1}{\rho^{\Delta + r+m-3}}$ for 
$m \leq 3$. Similarly, subleading corrections from the overlap of 3 longitudinal and one transverse polarization can modify the coefficients of $\frac{1}{\rho^{\Delta + r+m-3}}$ for $m \leq 2$.
Unlikely as it seems, these correction terms could, in principle, cancel that rapid Regge growth of the S matrix from a particular contact term, allowing it to give rise to a correlator that obeys the chaos bound even though the S matrix in question violates the CRG conjecture. While it seems very likely that this will in fact happen
\footnote{ The reader might hope that the contamination from subleading corrections could be `quarantined' in a class of index structures, allowing us to read off the scattering of less than a maximal number of longitudinal polarizations from the other index structures. We were, however,  unable to show this will always be the case. Note that the fact that the  terms in 
	$G^m_{\rm sing}$ listed in \eqref{singcor} obeys the equations 
	\begin{equation} \begin{split} \label{invdist}
	&\nabla^{M_1} \nabla^{M_2} \dots \nabla^{M_{m+1}} 
	G_{M_1 M_2 \ldots M_4} =0  \\
	&\nabla^{M_1} \nabla^{M_2} \dots \nabla^{M_m} 
	G_{M_1 M_2 \ldots M_4} \neq 0  \\
	\end{split}
	\end{equation} 
	does not help us, as the nothing we can see prevents the subleading corrections to flat space scattering from obeying the same equations.} a clear argument for the connection between the chaos bound and  massive higher spin scattering requires further work.

\subsection{Normalizations and connections to scattering}

We have indicated above that the additional power of $\omega$ in the longitudinal waves is essentially a reflection of the 
enhanced $\omega$ scaling of the scattering of longitudinal waves as compared to transverse waves. In this brief subsection we explain this connection more clearly. 

Through this paper we have been slightly imprecise in  discussing the flat space  ${\cal S}$ matrix of the wave \eqref{scatwavesprelmain} in the case of massive particles (even massive scalars). For massive particles \eqref{scatwavesprelmain} are free solutions of the relevant flat space bulk wave equations only in the strict $\omega \to \infty$ limit (because the momenta of the scattering waves obey $k^2=0$ rather than $k^2=-m^2$). As the singularity in the flat space correlator arises from the $\omega \to \infty$ part of frequency space, this imprecision was unimportant in the case of scalars.  

In the case of massive vectors the imprecision extends also to the polarizations. A polarization proportional to $k_M$ is not quite transverse and so not quite allowed except in the $\omega \to \infty$ limit. In order to see this more clearly we could, for instance, arbitrarily replace the scattering waves \eqref{scatwavesprelmain} with 
\begin{equation}\label{scatwaves} 
A^i_M= 
\left(  \left( 1- \frac{1}{\Delta_i} \right) 
(Z_i^\perp)_M + \frac{i m_i}{\Delta_i} \epsilon^\parallel_i(Z_i.X_0) \right)
e^{i {\tilde k}_i.x}
\end{equation} 
where
\begin{equation}\label{tpte}
\begin{split}
&{\tilde k}_1=(\omega_1, 0, p_1, 0 )~~~~~~~~~~~~~~~~~~~~~~~~~~~~~~~~~\epsilon^\parallel_1=\frac{1}{m_1} ( p_1, 0, \omega_1, 0)\\
& {\tilde k}_3=(\omega_3, 0, -p_3, 0 )~~~~~~~~~~~~~~~~~~~~~~~~~~~~~~~\epsilon^\parallel_3=\frac{1}{m_3} ( p_3, 0, -\omega_3, 0)\\
&{\tilde k}_2=-(\omega_2, 0, p_2 \cos \theta , p_2 \sin \theta)~~~~~~~~~~~~~~~~~\epsilon^\parallel_2=\frac{1}{m_2} ( p_2, 0, \omega_2 \cos \theta, \omega_2 \sin \theta)\\
& {\tilde k}_4=(-\omega_4, 0, -p_4 \cos \theta, -p_4 \sin \theta )~~~~~~~~~~~~\epsilon^\parallel_4=\frac{1}{m_4} ( p_4, 0, -\omega_4\cos \theta, -\omega_4 \sin \theta )\\
&p_1=p_3=p, ~~~~p_2=p_4=p'~~~~\omega_i=\sqrt{m_i^2+p_i^2}\\
& \sqrt{m_1^2+p^2}+\sqrt{m_3^2+p^2}=
\sqrt{m_2^2+(p')^2}+\sqrt{m_4^2+(p')^2}
\end{split}
\end{equation} 
Note that the waves \eqref{scatwaves} are identical to 
the waves \eqref{scatwavesprelmain} in the large $\omega$ limit that is relevant for generating the singularity of the correlators. Unlike the waves \eqref{scatwavesprelmain}, however, \eqref{scatwaves} are 
genuine solutions of the free flat space massive massive vector equation with mass $m$, \footnote{
	In \eqref{tpte} we have corrected the momenta to ensure that they obey the true mass shell condition $k^2=-m^2$, and have simultaneously modified the expression for $\epsilon_i^\parallel$ to make sure that 
	the equation $\epsilon^\parallel.k=0$ continues to be obeyed. Note that we have chosen our modifications to ensure that scattering continues to take place in the centre of mass frame, as was the case in for the momenta \eqref{forvec}. We have also ensured that the 
	normalization vectors $\epsilon_i^\parallel$ are all 
	normalized. This is what results in the factor $\frac{1}{m_i}$ in the expressions for $\epsilon_i^\parallel$ in \eqref{tpte}, and the compensating factors of $m_i$ in the coefficient of 
	$\epsilon^\parallel_i$ in \eqref{scatwaves}.
	Note, in particular,
	that the normalization of the part of the scattering wave proportional to $\epsilon^\parallel$ in \eqref{scatwaves} is proportional to $m_i$. This is what allows the scattering amplitudes of longitudinal polarizations (which come with a single factor of $\frac{1}{m_i}$ for each factor of $m_i$) to be well defined even in the limit $m_i \to 0$. }
and so the ${\cal S}$ matrix of these waves is genuinely well defined. The quantity ${\cal S}$ that enters formulae such as \eqref{singcorv} should really be thought of as the S matrix of the true scattering waves \eqref{scatwaves}.

The point we want to make is the following. In the expression \eqref{scatwavesprelmain} the extra factor of $\omega$ in the longitudinal polarization comes from the fact that the polarization is proportional to $k^\mu$. The reader might have suspected that this extra scaling with $\omega$ is a consequence of using an $\omega$ dependent normalization for this wave. This is not the case. In \eqref{tpte} makes clear that the polarization that appears in \eqref{scatwavesprelmain} differs from the properly normalized polarization $\epsilon_1^\parallel$ by the $\omega$ independent factor $m_i$. The additional growth with $\omega$ of this S matrix is a dynamical fact and not an issue of normalization.

The arbitrary replacement \eqref{tpte} (or any other convenient replacement) would be sufficient to determine the leading order singularity in $\rho$. The crude replacement \eqref{tpte} would not, however, be sufficient to capture the subleading $\rho$ singularities that arise in longitudinal scattering, and a more delicate analysis would be needed to capture these effects.

\providecommand{\href}[2]{#2}\begingroup\raggedright\endgroup

\end{document}